\documentclass[runningheads]{llncs}
\pdfoutput=1
\usepackage[T1]{fontenc}
\usepackage[utf8]{inputenc}
\usepackage{textcomp}
\usepackage[cmex10]{amsmath}
\usepackage{amssymb}
\usepackage{latexsym}
\usepackage{stmaryrd}
\usepackage{color}
\usepackage{wrapfig}
\renewcommand{\subsubsection}[1]{\paragraph{\rm {\bf #1.}}}
\newcommand{\myparagraph}[1]{
	\smallskip\noindent{\em #1}}

\usepackage{subfig}

\usepackage{hyperref}
\hypersetup{%
    final=true,
    colorlinks=true,
    urlcolor=black,
    linkcolor=black,
    citecolor=black,
    pdfborder={0 0 0},
    pdfauthor={Timothy Bourke \& Robert J. van Glabbeek \& Peter Höfner},
    pdftitle={A mechanized proof of loop freedom of the (untimed) AODV
    routing protocol},
    pdfsubject={Interactive proof of a routing protocol},
    pdfkeywords={AODV, protocol verification, Isabelle/HOL},
}
\usepackage{breakurl} % added by Rob; affects latex -> dvi -> ps -> pdf route only

\usepackage[final]{graphicx}
\usepackage[inline]{enumitem}
\newlist{inparaenum}{enumerate*}{1}
\setlist[inparaenum,1]{label=(\arabic*),ref=\arabic*}

\usepackage{mathpartir}
\mprset{sep=1.10em}  % hor. sep. between premises

\usepackage{isabelle, isabellesym}
\isabellestyle{tim}
\usepackage{relsize}
\renewcommand{\isasymAnd}{\isamath{{\mathsmaller{\bigwedge}}}}

\definecolor{gray}{rgb}{0.6,0.6,0.6}
\newcommand{\gray}[1]{\textcolor{gray}{#1}}

\newcommand{\snip}[4]
  {\expandafter\newcommand\csname snippet--#1\endcsname{#4}}
\newcommand{\snippet}[1]{\csname snippet--#1\endcsname}
\newcommand{\msnippet}[1]{\mbox{\csname snippet--#1\endcsname}}

\snip{seqp_choice}{0}{0}{\isa{p\isactrlsub {\isadigit{1}}\hspace{.12em}{\isasymoplus}\hspace{.2em}p\isactrlsub {\isadigit{2}}}}
\snip{seqp_call}{0}{0}{\isa{call{\isacharparenleft}pn{\isacharparenright}}}

\snip{lseqp_guard}{0}{0}{\isa{\gray{{\isacharbraceleft}l{\isacharbraceright}}{\isasymlangle}g{\isasymrangle}\hspace{.2em}p}}
\snip{lseqp_assign}{0}{0}{\isa{\gray{{\isacharbraceleft}l{\isacharbraceright}}{\isasymlbrakk}u{\isasymrbrakk}\hspace{.2em}p}}
\snip{lseqp_ucast}{0}{0}{\isa{\gray{{\isacharbraceleft}l{\isacharbraceright}}unicast{\isacharparenleft}\selip,\ \selmsg{\isacharparenright}{\isachardot}p\hspace{.24em}{\isasymtriangleright}\hspace{.19em}q}}
\snip{lseqp_bcast}{0}{0}{\isa{\gray{{\isacharbraceleft}l{\isacharbraceright}}broadcast{\isacharparenleft}\selmsg{\isacharparenright}{\isachardot}p}}
\snip{lseqp_gcast}{0}{0}{\isa{\gray{{\isacharbraceleft}l{\isacharbraceright}}groupcast{\isacharparenleft}\selips,\ \selmsg{\isacharparenright}{\isachardot}p}}
\snip{lseqp_send}{0}{0}{\isa{\gray{{\isacharbraceleft}l{\isacharbraceright}}send{\isacharparenleft}\selmsg{\isacharparenright}{\isachardot}p}}
\snip{lseqp_deliver}{0}{0}{\isa{\gray{{\isacharbraceleft}l{\isacharbraceright}}deliver{\isacharparenleft}\seldata{\isacharparenright}{\isachardot}p}}
\snip{lseqp_receive}{0}{0}{\isa{\gray{{\isacharbraceleft}l{\isacharbraceright}}receive{\isacharparenleft}\updmsg{\isacharparenright}{\isachardot}p}}

\snip{seqp_guard}{0}{0}{\isa{{\isasymlangle}g{\isasymrangle}\ p}}
\snip{seqp_assign}{0}{0}{\isa{{\isasymlbrakk}u{\isasymrbrakk}\ p}}
\snip{seqp_ucast}{0}{0}{\isa{unicast{\isacharparenleft}\selip,\ \selmsg{\isacharparenright}\ {\isachardot}\ p\ {\isasymtriangleright}\ q}}
\snip{seqp_bcast}{0}{0}{\isa{broadcast{\isacharparenleft}\selmsg{\isacharparenright}\ {\isachardot}\ p}}
\snip{seqp_gcast}{0}{0}{\isa{groupcast{\isacharparenleft}\selips,\ \selmsg{\isacharparenright}\ {\isachardot}\ p}}
\snip{seqp_send}{0}{0}{\isa{send{\isacharparenleft}\selmsg{\isacharparenright}\ {\isachardot}\ p}}
\snip{seqp_deliver}{0}{0}{\isa{deliver{\isacharparenleft}\seldata{\isacharparenright}\ {\isachardot}\ p}}
\snip{seqp_receive}{0}{0}{\isa{receive{\isacharparenleft}\updmsg{\isacharparenright}\ {\isachardot}\ p}}

\snip{seqp_guard_map}{0}{0}{\isa{{\isacharprime}s\ {\isasymRightarrow}\ {\isacharprime}s\ set}}
\snip{seqp_assign_map}{0}{0}{\isa{{\isacharprime}s\ {\isasymRightarrow}\ {\isacharprime}s}}
\snip{seqp_cast_ip_map}{0}{0}{\isa{{\isacharprime}s\ {\isasymRightarrow}\ ip}}
\snip{seqp_cast_msg_map}{0}{0}{\isa{{\isacharprime}s\ {\isasymRightarrow}\ {\isacharprime}m}}
\snip{seqp_cast_ip_set_map}{0}{0}{\isa{{\isacharprime}s\ {\isasymRightarrow}\ ip\ set}}

\snip{gamma_fun}{0}{0}{\isa{{\isasymGamma}{\isasymColon}{\isacharprime}p\ {\isasymRightarrow}\ {\isacharparenleft}{\isacharprime}s,\ {\isacharprime}m,\ {\isacharprime}p,\ {\isacharprime}l{\isacharparenright}\ seqp}}
\snip{wellformed_gamma}{0}{0}{\isa{wellformed\ {\isasymGamma}}}

\snip{broadcastT}{0}{0}{\isa{{\isacharparenleft}{\isacharparenleft}{\isasymxi},\ \gray{{\isacharbraceleft}l{\isacharbraceright}}broadcast{\isacharparenleft}\selmsg{\isacharparenright}{\isachardot}p{\isacharparenright},\ broadcast\ {\isacharparenleft}\selmsg\ {\isasymxi}{\isacharparenright},\ {\isacharparenleft}{\isasymxi},\ p{\isacharparenright}{\isacharparenright}{\isasymin}\seqpsos\ {\isasymGamma}}}
\snip{groupcastT}{0}{0}{\isa{{\isacharparenleft}{\isacharparenleft}{\isasymxi},\ \gray{{\isacharbraceleft}l{\isacharbraceright}}groupcast{\isacharparenleft}\selips,\ \selmsg{\isacharparenright}\ {\isachardot}\ p{\isacharparenright},\ groupcast\ {\isacharparenleft}\selips\ {\isasymxi}{\isacharparenright}\ {\isacharparenleft}\selmsg\ {\isasymxi}{\isacharparenright},\ {\isacharparenleft}{\isasymxi},\ p{\isacharparenright}{\isacharparenright}{\isasymin}\seqpsos\ {\isasymGamma}}}
\snip{unicastT}{0}{0}{\isa{{\isacharparenleft}{\isacharparenleft}{\isasymxi},\ \gray{{\isacharbraceleft}l{\isacharbraceright}}unicast{\isacharparenleft}\selip,\ \selmsg{\isacharparenright}\ {\isachardot}\ p\ {\isasymtriangleright}\ q{\isacharparenright},\ unicast\ {\isacharparenleft}\selip\ {\isasymxi}{\isacharparenright}\ {\isacharparenleft}\selmsg\ {\isasymxi}{\isacharparenright},\ {\isacharparenleft}{\isasymxi},\ p{\isacharparenright}{\isacharparenright}{\isasymin}\seqpsos\ {\isasymGamma}}}
\snip{notunicastT}{0}{0}{\isa{{\isacharparenleft}{\isacharparenleft}{\isasymxi},\ \gray{{\isacharbraceleft}l{\isacharbraceright}}unicast{\isacharparenleft}\selip,\ \selmsg{\isacharparenright}\ {\isachardot}\ p\ {\isasymtriangleright}\ q{\isacharparenright},\ {\isasymnot}unicast\ {\isacharparenleft}\selip\ {\isasymxi}{\isacharparenright},\ {\isacharparenleft}{\isasymxi},\ q{\isacharparenright}{\isacharparenright}{\isasymin}\seqpsos\ {\isasymGamma}}}
\snip{sendT}{0}{0}{\isa{{\isacharparenleft}{\isacharparenleft}{\isasymxi},\ \gray{{\isacharbraceleft}l{\isacharbraceright}}send{\isacharparenleft}\selmsg{\isacharparenright}\ {\isachardot}\ p{\isacharparenright},\ send\ {\isacharparenleft}\selmsg\ {\isasymxi}{\isacharparenright},\ {\isacharparenleft}{\isasymxi},\ p{\isacharparenright}{\isacharparenright}{\isasymin}\seqpsos\ {\isasymGamma}}}
\snip{deliverT}{0}{0}{\isa{{\isacharparenleft}{\isacharparenleft}{\isasymxi},\ \gray{{\isacharbraceleft}l{\isacharbraceright}}deliver{\isacharparenleft}\seldata{\isacharparenright}\ {\isachardot}\ p{\isacharparenright},\ deliver\ {\isacharparenleft}\seldata\ {\isasymxi}{\isacharparenright},\ {\isacharparenleft}{\isasymxi},\ p{\isacharparenright}{\isacharparenright}{\isasymin}\seqpsos\ {\isasymGamma}}}
\snip{receiveT}{0}{0}{\isa{{\isacharparenleft}{\isacharparenleft}{\isasymxi},\ \gray{{\isacharbraceleft}l{\isacharbraceright}}receive{\isacharparenleft}\updmsg{\isacharparenright}\ {\isachardot}\ p{\isacharparenright},\ receive\ msg,\ {\isacharparenleft}\updmsg\ msg\ {\isasymxi},\ p{\isacharparenright}{\isacharparenright}{\isasymin}\seqpsos\ {\isasymGamma}}}
\snip{assignT}{0}{0}{\isa{{\isacharparenleft}{\isacharparenleft}{\isasymxi},\ \gray{{\isacharbraceleft}l{\isacharbraceright}}{\isasymlbrakk}u{\isasymrbrakk}\ p{\isacharparenright},\ {\isasymtau},\ {\isacharparenleft}u\ {\isasymxi},\ p{\isacharparenright}{\isacharparenright}{\isasymin}\seqpsos\ {\isasymGamma}}}
\snip{assignT'}{0}{0}{\isa{\mbox{}\inferrule{\mbox{{\isasymxi}{\isacharprime}\ {\isacharequal}\ u\ {\isasymxi}}}{\mbox{{\isacharparenleft}{\isacharparenleft}{\isasymxi},\ \gray{{\isacharbraceleft}l{\isacharbraceright}}{\isasymlbrakk}u{\isasymrbrakk}\ p{\isacharparenright},\ {\isasymtau},\ {\isacharparenleft}{\isasymxi}{\isacharprime},\ p{\isacharparenright}{\isacharparenright}{\isasymin}\seqpsos\ {\isasymGamma}}}}}
\snip{callT}{0}{0}{\isa{\mbox{}\inferrule{\mbox{{\isacharparenleft}{\isacharparenleft}{\isasymxi},\ {\isasymGamma}\ pn{\isacharparenright},\ a,\ {\isacharparenleft}{\isasymxi}{\isacharprime},\ p{\isacharprime}{\isacharparenright}{\isacharparenright}{\isasymin}\seqpsos\ {\isasymGamma}}}{\mbox{{\isacharparenleft}{\isacharparenleft}{\isasymxi},\ call{\isacharparenleft}pn{\isacharparenright}{\isacharparenright},\ a,\ {\isacharparenleft}{\isasymxi}{\isacharprime},\ p{\isacharprime}{\isacharparenright}{\isacharparenright}{\isasymin}\seqpsos\ {\isasymGamma}}}}}
\snip{choiceT1}{0}{0}{\isa{\mbox{}\inferrule{\mbox{{\isacharparenleft}{\isacharparenleft}{\isasymxi},\ p{\isacharparenright},\ a,\ {\isacharparenleft}{\isasymxi}{\isacharprime},\ p{\isacharprime}{\isacharparenright}{\isacharparenright}{\isasymin}\seqpsos\ {\isasymGamma}}}{\mbox{{\isacharparenleft}{\isacharparenleft}{\isasymxi},\ p\ {\isasymoplus}\ q{\isacharparenright},\ a,\ {\isacharparenleft}{\isasymxi}{\isacharprime},\ p{\isacharprime}{\isacharparenright}{\isacharparenright}{\isasymin}\seqpsos\ {\isasymGamma}}}}}
\snip{choiceT2}{0}{0}{\isa{\mbox{}\inferrule{\mbox{{\isacharparenleft}{\isacharparenleft}{\isasymxi},\ q{\isacharparenright},\ a,\ {\isacharparenleft}{\isasymxi}{\isacharprime},\ q{\isacharprime}{\isacharparenright}{\isacharparenright}{\isasymin}\seqpsos\ {\isasymGamma}}}{\mbox{{\isacharparenleft}{\isacharparenleft}{\isasymxi},\ p\ {\isasymoplus}\ q{\isacharparenright},\ a,\ {\isacharparenleft}{\isasymxi}{\isacharprime},\ q{\isacharprime}{\isacharparenright}{\isacharparenright}{\isasymin}\seqpsos\ {\isasymGamma}}}}}
\snip{guardT}{0}{0}{\isa{\mbox{}\inferrule{\mbox{{\isasymxi}{\isacharprime}{\isasymin}g\ {\isasymxi}}}{\mbox{{\isacharparenleft}{\isacharparenleft}{\isasymxi},\ \gray{{\isacharbraceleft}l{\isacharbraceright}}{\isasymlangle}g{\isasymrangle}\ p{\isacharparenright},\ {\isasymtau},\ {\isacharparenleft}{\isasymxi}{\isacharprime},\ p{\isacharparenright}{\isacharparenright}{\isasymin}\seqpsos\ {\isasymGamma}}}}}

\snip{parleft}{0}{0}{\isa{\mbox{}\inferrule{\mbox{{\isacharparenleft}s,\ a,\ s{\isacharprime}{\isacharparenright}{\isasymin}S}\\\ \mbox{{\isasymAnd}m{\isachardot}\ a\ {\isasymnoteq}\ receive\ m}}{\mbox{{\isacharparenleft}{\isacharparenleft}s,\ t{\isacharparenright},\ a,\ {\isacharparenleft}s{\isacharprime},\ t{\isacharparenright}{\isacharparenright}{\isasymin}\parpsos\ S\ T}}}}
\snip{parright}{0}{0}{\isa{\mbox{}\inferrule{\mbox{{\isacharparenleft}t,\ a,\ t{\isacharprime}{\isacharparenright}{\isasymin}T}\\\ \mbox{{\isasymAnd}m{\isachardot}\ a\ {\isasymnoteq}\ send\ m}}{\mbox{{\isacharparenleft}{\isacharparenleft}s,\ t{\isacharparenright},\ a,\ {\isacharparenleft}s,\ t{\isacharprime}{\isacharparenright}{\isacharparenright}{\isasymin}\parpsos\ S\ T}}}}
\snip{parboth}{0}{0}{\isa{\mbox{}\inferrule{\mbox{{\isacharparenleft}s,\ receive\ m,\ s{\isacharprime}{\isacharparenright}{\isasymin}S}\\\ \mbox{{\isacharparenleft}t,\ send\ m,\ t{\isacharprime}{\isacharparenright}{\isasymin}T}}{\mbox{{\isacharparenleft}{\isacharparenleft}s,\ t{\isacharparenright},\ {\isasymtau},\ {\isacharparenleft}s{\isacharprime},\ t{\isacharprime}{\isacharparenright}{\isacharparenright}{\isasymin}\parpsos\ S\ T}}}}

\snip{par_comp}{0}{0}{\isa{s\ {\isasymlangle}{\isasymlangle}\ t\ {\isacharequal}\ {\isasymlparr}init\ {\isacharequal}\ init\ s\ {\isasymtimes}\ init\ t,\ trans\ {\isacharequal}\ \parpsos\ {\isacharparenleft}trans\ s{\isacharparenright}\ {\isacharparenleft}trans\ t{\isacharparenright}{\isasymrparr}}}
\snip{par_comp_term}{0}{0}{\isa{s\ {\isasymlangle}{\isasymlangle}\ t}}
\snip{par_comp_type}{0}{0}{\isa{{\isacharparenleft}{\isacharprime}s\ {\isasymtimes}\ {\isacharprime}p,\ {\isacharprime}m\ seq{\isacharunderscore}action{\isacharparenright}\ automaton}}

\snip{node_bcast}{0}{0}{\isa{\mbox{}\inferrule{\mbox{{\isacharparenleft}s,\ broadcast\ m,\ s{\isacharprime}{\isacharparenright}{\isasymin}S}}{\mbox{{\isacharparenleft}\NodeS\ i\ s\ R,\ R{\isacharcolon}{\isacharasterisk}cast{\isacharparenleft}m{\isacharparenright},\ \NodeS\ i\ s{\isacharprime}\ R{\isacharparenright}{\isasymin}\nodesos\ S}}}}
\snip{node_gcast}{0}{0}{\isa{\mbox{}\inferrule{\mbox{{\isacharparenleft}s,\ groupcast\ D\ m,\ s{\isacharprime}{\isacharparenright}{\isasymin}S}}{\mbox{{\isacharparenleft}\NodeS\ i\ s\ R,\ {\isacharparenleft}R\ {\isasyminter}\ D{\isacharparenright}{\isacharcolon}{\isacharasterisk}cast{\isacharparenleft}m{\isacharparenright},\ \NodeS\ i\ s{\isacharprime}\ R{\isacharparenright}{\isasymin}\nodesos\ S}}}}
\snip{node_ucast}{0}{0}{\isa{\mbox{}\inferrule{\mbox{{\isacharparenleft}s,\ unicast\ d\ m,\ s{\isacharprime}{\isacharparenright}{\isasymin}S}\\\ \mbox{d{\isasymin}R}}{\mbox{{\isacharparenleft}\NodeS\ i\ s\ R,\ {\isacharbraceleft}d{\isacharbraceright}{\isacharcolon}{\isacharasterisk}cast{\isacharparenleft}m{\isacharparenright},\ \NodeS\ i\ s{\isacharprime}\ R{\isacharparenright}{\isasymin}\nodesos\ S}}}}
\snip{node_notucast}{0}{0}{\isa{\mbox{}\inferrule{\mbox{{\isacharparenleft}s,\ {\isasymnot}unicast\ d,\ s{\isacharprime}{\isacharparenright}{\isasymin}S}\\\ \mbox{d\ {\isasymnotin}\ R}}{\mbox{{\isacharparenleft}\NodeS\ i\ s\ R,\ {\isasymtau},\ \NodeS\ i\ s{\isacharprime}\ R{\isacharparenright}{\isasymin}\nodesos\ S}}}}
\snip{node_deliver}{0}{0}{\isa{\mbox{}\inferrule{\mbox{{\isacharparenleft}s,\ deliver\ d,\ s{\isacharprime}{\isacharparenright}{\isasymin}S}}{\mbox{{\isacharparenleft}\NodeS\ i\ s\ R,\ i{\isacharcolon}deliver{\isacharparenleft}d{\isacharparenright},\ \NodeS\ i\ s{\isacharprime}\ R{\isacharparenright}{\isasymin}\nodesos\ S}}}}
\snip{node_receive}{0}{0}{\isa{\mbox{}\inferrule{\mbox{{\isacharparenleft}s,\ receive\ m,\ s{\isacharprime}{\isacharparenright}{\isasymin}S}}{\mbox{{\isacharparenleft}\NodeS\ i\ s\ R,\ {\isacharbraceleft}i{\isacharbraceright}{\isasymnot}{\isasymemptyset}{\isacharcolon}arrive{\isacharparenleft}m{\isacharparenright},\ \NodeS\ i\ s{\isacharprime}\ R{\isacharparenright}{\isasymin}\nodesos\ S}}}}
\snip{node_tau}{0}{0}{\isa{\mbox{}\inferrule{\mbox{{\isacharparenleft}s,\ {\isasymtau},\ s{\isacharprime}{\isacharparenright}{\isasymin}S}}{\mbox{{\isacharparenleft}\NodeS\ i\ s\ R,\ {\isasymtau},\ \NodeS\ i\ s{\isacharprime}\ R{\isacharparenright}{\isasymin}\nodesos\ S}}}}
\snip{node_arrive}{0}{0}{\isa{{\isacharparenleft}\NodeS\ i\ s\ R,\ {\isasymemptyset}{\isasymnot}{\isacharbraceleft}i{\isacharbraceright}{\isacharcolon}arrive{\isacharparenleft}m{\isacharparenright},\ \NodeS\ i\ s\ R{\isacharparenright}{\isasymin}\nodesos\ S}}
\snip{node_connect1}{0}{0}{\isa{{\isacharparenleft}\NodeS\ i\ s\ R,\ connect{\isacharparenleft}i,\ i{\isacharprime}{\isacharparenright},\ \NodeS\ i\ s\ {R\ {\isasymunion}\ {\isacharbraceleft}i{\isacharprime}{\isacharbraceright}}{\isacharparenright}{\isasymin}\nodesos\ S}}
\snip{node_connect2}{0}{0}{\isa{{\isacharparenleft}\NodeS\ i\ s\ R,\ connect{\isacharparenleft}i{\isacharprime},\ i{\isacharparenright},\ \NodeS\ i\ s\ {R\ {\isasymunion}\ {\isacharbraceleft}i{\isacharprime}{\isacharbraceright}}{\isacharparenright}{\isasymin}\nodesos\ S}}
\snip{node_disconnect1}{0}{0}{\isa{{\isacharparenleft}\NodeS\ i\ s\ R,\ disconnect{\isacharparenleft}i,\ i{\isacharprime}{\isacharparenright},\ \NodeS\ i\ s\ {\isacharparenleft}R\ {\isacharminus}\ {\isacharbraceleft}i{\isacharprime}{\isacharbraceright}{\isacharparenright}{\isacharparenright}{\isasymin}\nodesos\ S}}
\snip{node_disconnect2}{0}{0}{\isa{{\isacharparenleft}\NodeS\ i\ s\ R,\ disconnect{\isacharparenleft}i{\isacharprime},\ i{\isacharparenright},\ \NodeS\ i\ s\ {\isacharparenleft}R\ {\isacharminus}\ {\isacharbraceleft}i{\isacharprime}{\isacharbraceright}{\isacharparenright}{\isacharparenright}{\isasymin}\nodesos\ S}}
\snip{node_connect_other}{0}{0}{\isa{\mbox{}\inferrule{\mbox{i\ {\isasymnoteq}\ i{\isacharprime}}\\\ \mbox{i\ {\isasymnoteq}\ i{\isacharprime}{\isacharprime}}}{\mbox{{\isacharparenleft}\NodeS\ i\ s\ R,\ connect{\isacharparenleft}i{\isacharprime},\ i{\isacharprime}{\isacharprime}{\isacharparenright},\ \NodeS\ i\ s\ R{\isacharparenright}{\isasymin}\nodesos\ S}}}}
\snip{node_disconnect_other}{0}{0}{\isa{\mbox{}\inferrule{\mbox{i\ {\isasymnoteq}\ i{\isacharprime}}\\\ \mbox{i\ {\isasymnoteq}\ i{\isacharprime}{\isacharprime}}}{\mbox{{\isacharparenleft}\NodeS\ i\ s\ R,\ disconnect{\isacharparenleft}i{\isacharprime},\ i{\isacharprime}{\isacharprime}{\isacharparenright},\ \NodeS\ i\ s\ R{\isacharparenright}{\isasymin}\nodesos\ S}}}}

\snip{node_comp}{0}{0}{\isa{{\isasymlangle}i\ {\isacharcolon}\ np\ {\isacharcolon}\ \Ri{\isasymrangle}\ {\isacharequal}\ {\isasymlparr}init\ {\isacharequal}\ {\isacharbraceleft}\NodeS\ i\ s\ \Ri\ {\isacharbar}\ s{\isasymin}init\ np{\isacharbraceright},\ trans\ {\isacharequal}\ \nodesos\ {\isacharparenleft}trans\ np{\isacharparenright}{\isasymrparr}}}
\snip{node_comp_term}{0}{0}{\isa{{\isasymlangle}i\ {\isacharcolon}\ S\ {\isacharcolon}\ \Ri{\isasymrangle}}}
\snip{node_comp_type}{0}{0}{\isa{{\isacharparenleft}{\isacharprime}a\ net{\isacharunderscore}state,\ {\isacharprime}b\ node{\isacharunderscore}action{\isacharparenright}\ automaton}}

\snip{pnet_cast1}{0}{0}{\isa{\mbox{}\inferrule{\mbox{{\isacharparenleft}s,\ R{\isacharcolon}{\isacharasterisk}cast{\isacharparenleft}m{\isacharparenright},\ s{\isacharprime}{\isacharparenright}{\isasymin}S}\\\ \mbox{{\isacharparenleft}t,\ H{\isasymnot}K{\isacharcolon}arrive{\isacharparenleft}m{\isacharparenright},\ t{\isacharprime}{\isacharparenright}{\isasymin}T}\\\ \mbox{H\ {\isasymsubseteq}\ R}\\\ \mbox{K\ {\isasyminter}\ R\ {\isacharequal}\ {\isasymemptyset}}}{\mbox{{\isacharparenleft}\subnets{s}{t},\ R{\isacharcolon}{\isacharasterisk}cast{\isacharparenleft}m{\isacharparenright},\ \subnets{s{\isacharprime}}{t{\isacharprime}}{\isacharparenright}{\isasymin}\pnetsos\ S\ T}}}}
\snip{pnet_cast2}{0}{0}{\isa{\mbox{}\inferrule{\mbox{{\isacharparenleft}s,\ H{\isasymnot}K{\isacharcolon}arrive{\isacharparenleft}m{\isacharparenright},\ s{\isacharprime}{\isacharparenright}{\isasymin}S}\\\ \mbox{{\isacharparenleft}t,\ R{\isacharcolon}{\isacharasterisk}cast{\isacharparenleft}m{\isacharparenright},\ t{\isacharprime}{\isacharparenright}{\isasymin}T}\\\ \mbox{H\ {\isasymsubseteq}\ R}\\\ \mbox{K\ {\isasyminter}\ R\ {\isacharequal}\ {\isasymemptyset}}}{\mbox{{\isacharparenleft}\subnets{s}{t},\ R{\isacharcolon}{\isacharasterisk}cast{\isacharparenleft}m{\isacharparenright},\ \subnets{s{\isacharprime}}{t{\isacharprime}}{\isacharparenright}{\isasymin}\pnetsos\ S\ T}}}}
\snip{pnet_arrive}{0}{0}{\isa{\mbox{}\inferrule{\mbox{{\isacharparenleft}s,\ H{\isasymnot}K{\isacharcolon}arrive{\isacharparenleft}m{\isacharparenright},\ s{\isacharprime}{\isacharparenright}{\isasymin}S}\\\ \mbox{{\isacharparenleft}t,\ H{\isacharprime}{\isasymnot}K{\isacharprime}{\isacharcolon}arrive{\isacharparenleft}m{\isacharparenright},\ t{\isacharprime}{\isacharparenright}{\isasymin}T}}{\mbox{{\isacharparenleft}\subnets{s}{t},\ {\isacharparenleft}H\ {\isasymunion}\ H{\isacharprime}{\isacharparenright}{\isasymnot}{\isacharparenleft}K\ {\isasymunion}\ K{\isacharprime}{\isacharparenright}{\isacharcolon}arrive{\isacharparenleft}m{\isacharparenright},\ \subnets{s{\isacharprime}}{t{\isacharprime}}{\isacharparenright}{\isasymin}\pnetsos\ S\ T}}}}
\snip{pnet_deliver1}{0}{0}{\isa{\mbox{}\inferrule{\mbox{{\isacharparenleft}s,\ i{\isacharcolon}deliver{\isacharparenleft}d{\isacharparenright},\ s{\isacharprime}{\isacharparenright}{\isasymin}S}}{\mbox{{\isacharparenleft}\subnets{s}{t},\ i{\isacharcolon}deliver{\isacharparenleft}d{\isacharparenright},\ \subnets{s{\isacharprime}}{t}{\isacharparenright}{\isasymin}\pnetsos\ S\ T}}}}
\snip{pnet_deliver2}{0}{0}{\isa{\mbox{}\inferrule{\mbox{{\isacharparenleft}t,\ i{\isacharcolon}deliver{\isacharparenleft}d{\isacharparenright},\ t{\isacharprime}{\isacharparenright}{\isasymin}T}}{\mbox{{\isacharparenleft}\subnets{s}{t},\ i{\isacharcolon}deliver{\isacharparenleft}d{\isacharparenright},\ \subnets{s}{t{\isacharprime}}{\isacharparenright}{\isasymin}\pnetsos\ S\ T}}}}
\snip{pnet_tau1}{0}{0}{\isa{\mbox{}\inferrule{\mbox{{\isacharparenleft}s,\ {\isasymtau},\ s{\isacharprime}{\isacharparenright}{\isasymin}S}}{\mbox{{\isacharparenleft}\subnets{s}{t},\ {\isasymtau},\ \subnets{s{\isacharprime}}{t}{\isacharparenright}{\isasymin}\pnetsos\ S\ T}}}}
\snip{pnet_tau2}{0}{0}{\isa{\mbox{}\inferrule{\mbox{{\isacharparenleft}t,\ {\isasymtau},\ t{\isacharprime}{\isacharparenright}{\isasymin}T}}{\mbox{{\isacharparenleft}\subnets{s}{t},\ {\isasymtau},\ \subnets{s}{t{\isacharprime}}{\isacharparenright}{\isasymin}\pnetsos\ S\ T}}}}
\snip{pnet_connect}{0}{0}{\isa{\mbox{}\inferrule{\mbox{{\isacharparenleft}s,\ connect{\isacharparenleft}i,\ i{\isacharprime}{\isacharparenright},\ s{\isacharprime}{\isacharparenright}{\isasymin}S}\\\ \mbox{{\isacharparenleft}t,\ connect{\isacharparenleft}i,\ i{\isacharprime}{\isacharparenright},\ t{\isacharprime}{\isacharparenright}{\isasymin}T}}{\mbox{{\isacharparenleft}\subnets{s}{t},\ connect{\isacharparenleft}i,\ i{\isacharprime}{\isacharparenright},\ \subnets{s{\isacharprime}}{t{\isacharprime}}{\isacharparenright}{\isasymin}\pnetsos\ S\ T}}}}
\snip{pnet_disconnect}{0}{0}{\isa{\mbox{}\inferrule{\mbox{{\isacharparenleft}s,\ disconnect{\isacharparenleft}i,\ i{\isacharprime}{\isacharparenright},\ s{\isacharprime}{\isacharparenright}{\isasymin}S}\\\ \mbox{{\isacharparenleft}t,\ disconnect{\isacharparenleft}i,\ i{\isacharprime}{\isacharparenright},\ t{\isacharprime}{\isacharparenright}{\isasymin}T}}{\mbox{{\isacharparenleft}\subnets{s}{t},\ disconnect{\isacharparenleft}i,\ i{\isacharprime}{\isacharparenright},\ \subnets{s{\isacharprime}}{t{\isacharprime}}{\isacharparenright}{\isasymin}\pnetsos\ S\ T}}}}

\snip{pnet}{0}{0}{\isa{pnet}}
\snip{pnet_node_term}{0}{0}{\isa{{\isasymlangle}i{\isacharsemicolon}\ \Ri{\isasymrangle}}}
\snip{pnet_node_type}{0}{0}{\isa{net{\isacharunderscore}tree}}
\snip{pnet_par_term}{0}{0}{\isa{p\isactrlsub {\isadigit{1}}\hspace{-.1em}\parallelcomp{}p\isactrlsub {\isadigit{2}}}}
\snip{pnet_par_type}{0}{0}{\isa{net{\isacharunderscore}tree}}
\snip{net_tree}{0}{0}{\isa{net{\isacharunderscore}tree}}
\snip{net_tree_ips}{0}{0}{\isa{net{\isacharunderscore}tree{\isacharunderscore}ips}}

\snip{act_arrive}{0}{0}{\isa{H{\isasymnot}K{\isacharcolon}arrive{\isacharparenleft}m{\isacharparenright}}}
\snip{act_not_arrive}{0}{0}{\isa{{\isasymemptyset}{\isasymnot}{\isacharbraceleft}i{\isacharbraceright}{\isacharcolon}arrive{\isacharparenleft}m{\isacharparenright}}}
\snip{act_connect}{0}{0}{\isa{connect{\isacharparenleft}i,\ i{\isacharprime}{\isacharparenright}}}
\snip{act_disconnect}{0}{0}{\isa{disconnect{\isacharparenleft}i,\ i{\isacharprime}{\isacharparenright}}}
\snip{act_receive}{0}{0}{\isa{receive\ m}}
\snip{act_cast}{0}{0}{\isa{R{\isacharcolon}{\isacharasterisk}cast{\isacharparenleft}m{\isacharparenright}}}

\snip{net_state}{0}{0}{\isa{{\isacharprime}s\ net{\isacharunderscore}state}}
\snip{net_state_nodes}{0}{0}{\isa{\NodeS\ i\ s\ R}}
\snip{net_state_subnets}{0}{0}{\isa{\subnets{s\isactrlsub {\isadigit{1}}}{s\isactrlsub {\isadigit{2}}}}}

\snip{wf_net_tree}{0}{0}{\isa{wf{\isacharunderscore}net{\isacharunderscore}tree\ {\isacharparenleft}p{\isadigit{1}}{\isachardot}{\isadigit{0}}\parallelcomp{}p{\isadigit{2}}{\isachardot}{\isadigit{0}}{\isacharparenright}\ {\isacharequal}\ {\isacharparenleft}net{\isacharunderscore}tree{\isacharunderscore}ips\ p{\isadigit{1}}{\isachardot}{\isadigit{0}}\ {\isasyminter}\ net{\isacharunderscore}tree{\isacharunderscore}ips\ p{\isadigit{2}}{\isachardot}{\isadigit{0}}\ {\isacharequal}\ {\isasymemptyset}\ {\isasymand}\ wf{\isacharunderscore}net{\isacharunderscore}tree\ p{\isadigit{1}}{\isachardot}{\isadigit{0}}\ {\isasymand}\ wf{\isacharunderscore}net{\isacharunderscore}tree\ p{\isadigit{2}}{\isachardot}{\isadigit{0}}{\isacharparenright}\isasep\isanewline%
wf{\isacharunderscore}net{\isacharunderscore}tree\ {\isasymlangle}i{\isacharsemicolon}\ R{\isasymrangle}\ {\isacharequal}\ True}}
\snip{pnet_rules}{0}{0}{\isa{pnet\ np\ {\isasymlangle}i{\isacharsemicolon}\ \Ri{\isasymrangle}\ {\isacharequal}\ {\isasymlangle}i\ {\isacharcolon}\ np\ i\ {\isacharcolon}\ \Ri{\isasymrangle}\isasep\isanewline%
pnet\ np\ {\isacharparenleft}p\isactrlsub {\isadigit{1}}\parallelcomp{}p\isactrlsub {\isadigit{2}}{\isacharparenright}\ {\isacharequal}\ {\isasymlparr}init\ {\isacharequal}\ {\isacharbraceleft}\subnets{s\isactrlsub {\isadigit{1}}}{s\isactrlsub {\isadigit{2}}}\ {\isacharbar}\ s\isactrlsub {\isadigit{1}}{\isasymin}init\ {\isacharparenleft}pnet\ np\ p\isactrlsub {\isadigit{1}}{\isacharparenright}\ {\isasymand}\ s\isactrlsub {\isadigit{2}}{\isasymin}init\ {\isacharparenleft}pnet\ np\ p\isactrlsub {\isadigit{2}}{\isacharparenright}{\isacharbraceright},\ trans\ {\isacharequal}\ \pnetsos\ {\isacharparenleft}trans\ {\isacharparenleft}pnet\ np\ p\isactrlsub {\isadigit{1}}{\isacharparenright}{\isacharparenright}\ {\isacharparenleft}trans\ {\isacharparenleft}pnet\ np\ p\isactrlsub {\isadigit{2}}{\isacharparenright}{\isacharparenright}{\isasymrparr}}}

\snip{pnet1_lhs}{0}{0}{\isa{pnet\ np\ {\isasymlangle}i{\isacharsemicolon}\ \Ri{\isasymrangle}}}
\snip{pnet1_rhs}{0}{0}{\isa{{\isasymlangle}i\ {\isacharcolon}\ np\ i\ {\isacharcolon}\ \Ri{\isasymrangle}}}
\snip{pnet2_lhs}{0}{0}{\isa{pnet\ np\ {\isacharparenleft}p\isactrlsub {\isadigit{1}}\parallelcomp{}p\isactrlsub {\isadigit{2}}{\isacharparenright}}}
\snip{pnet2_rhs}{0}{0}{\isa{{\isasymlparr}init\ {\isacharequal}\ {\isacharbraceleft}\subnets{s\isactrlsub {\isadigit{1}}}{s\isactrlsub {\isadigit{2}}}\ {\isacharbar}\ s\isactrlsub {\isadigit{1}}{\isasymin}init\ {\isacharparenleft}pnet\ np\ p\isactrlsub {\isadigit{1}}{\isacharparenright}\ {\isasymand}\ s\isactrlsub {\isadigit{2}}{\isasymin}init\ {\isacharparenleft}pnet\ np\ p\isactrlsub {\isadigit{2}}{\isacharparenright}{\isacharbraceright},}\\ && \isa{\phantom{\isasymlparr}trans {\isacharequal}\ \pnetsos\ {\isacharparenleft}trans\ {\isacharparenleft}pnet\ np\ p\isactrlsub {\isadigit{1}}{\isacharparenright}{\isacharparenright}\ {\isacharparenleft}trans\ {\isacharparenleft}pnet\ np\ p\isactrlsub {\isadigit{2}}{\isacharparenright}{\isacharparenright}{\isasymrparr}}}

\snip{opnet1_lhs}{0}{0}{\isa{opnet\ onp\ {\isasymlangle}i{\isacharsemicolon}\ \Ri{\isasymrangle}}}
\snip{opnet1_rhs}{0}{0}{\isa{{\isasymlangle}i\ {\isacharcolon}\ onp\ i\ {\isacharcolon}\ \Ri{\isasymrangle}\isactrlsub o}}
\snip{opnet2_lhs}{0}{0}{\isa{opnet\ onp\ {\isacharparenleft}p\isactrlsub {\isadigit{1}}\parallelcomp{}p\isactrlsub {\isadigit{2}}{\isacharparenright}}}
\snip{opnet2_rhs}{0}{0}{\isa{{\isasymlparr}init\ {\isacharequal}\ {\isacharbraceleft}{\isacharparenleft}{\isasymsigma},\ \subnets{s\isactrlsub {\isadigit{1}}}{s\isactrlsub {\isadigit{2}}}{\isacharparenright}\ {\isacharbar}\ {\isacharparenleft}{\isasymsigma},\ s\isactrlsub {\isadigit{1}}{\isacharparenright}{\isasymin}init\ {\isacharparenleft}opnet\ onp\ p\isactrlsub {\isadigit{1}}{\isacharparenright}}\\ && \isa{\phantom{\isasymlparr init\ = \ {\isacharbraceleft}{\isacharparenleft}{\isasymsigma}{\isacharcomma}\ \subnets{s\isactrlsub {\isadigit{1}}}{s\isactrlsub {\isadigit{2}}}{\isacharparenright}\ {\isacharbar}}{\isasymand}\ {\isacharparenleft}{\isasymsigma},\ s\isactrlsub {\isadigit{2}}{\isacharparenright}{\isasymin}init\ {\isacharparenleft}opnet\ onp\ p\isactrlsub {\isadigit{2}}{\isacharparenright}}\\ && \isa{\phantom{\isasymlparr init\ = \ {\isacharbraceleft}{\isacharparenleft}{\isasymsigma}{\isacharcomma}\ \subnets{s\isactrlsub {\isadigit{1}}}{s\isactrlsub {\isadigit{2}}}{\isacharparenright}\ {\isacharbar}}{\isasymand}\ net{\isacharunderscore}ips\ s\isactrlsub {\isadigit{1}}\ {\isasyminter}\ net{\isacharunderscore}ips\ s\isactrlsub {\isadigit{2}}\ {\isacharequal}\ {\isasymemptyset}{\isacharbraceright},}\\ && \isa{\phantom{\isasymlparr}trans {\isacharequal}\ \opnetsos\ {\isacharparenleft}trans\ {\isacharparenleft}opnet\ onp\ p\isactrlsub {\isadigit{1}}{\isacharparenright}{\isacharparenright}\ {\isacharparenleft}trans\ {\isacharparenleft}opnet\ onp\ p\isactrlsub {\isadigit{2}}{\isacharparenright}{\isacharparenright}{\isasymrparr}}}

\snip{oclosed'}{0}{0}{\isa{oclosed\ A}}
\snip{oclosed_term}{0}{0}{\isa{A{\isasymlparr}trans\ {\isacharcolon}{\isacharequal}\ \ocnetsos\ {\isacharparenleft}trans\ A{\isacharparenright}{\isasymrparr}}}

\snip{netglobal}{0}{0}{\isa{netglobal\ P\ {\isacharequal}\ {\isasymlambda}s{\isachardot}\ P\ {\isacharparenleft}default\ aodv{\isacharunderscore}init\ {\isacharparenleft}netlift\ fst\ s{\isacharparenright}{\isacharparenright}}}

\snip{default}{0}{0}{\isa{default\ df\ f\ {\isacharequal}\ {\isacharparenleft}{\isasymlambda}i{\isachardot}\ \textsf{case}\ f\ i\ \textsf{of}\ None\ {\isasymRightarrow}\ df\ i\ {\isacharbar}\ Some\ s\ {\isasymRightarrow}\ s{\isacharparenright}}}
\snip{default_lhs}{0}{0}{\isa{default\ df\ f}}
\snip{default_rhs}{0}{0}{\isa{{\isasymlambda}i{\isachardot}\ \textsf{case}\ f\ i\ \textsf{of}\ None\ {\isasymRightarrow}\ df\ i\ {\isacharbar}\ Some\ s\ {\isasymRightarrow}\ s}}

\snip{netlift1_lhs}{0}{0}{\isa{netlift\ sr\ {\isacharparenleft}\NodeS\ i\ s\ R{\isacharparenright}}}
\snip{netlift1_rhs}{0}{0}{\isa{{\isacharbrackleft}i\ {\isasymmapsto}\ fst\ {\isacharparenleft}sr\ s{\isacharparenright}{\isacharbrackright}}}
\snip{netlift2_lhs}{0}{0}{\isa{netlift\ sr\ {\isacharparenleft}\subnets{s}{t}{\isacharparenright}}}
\snip{netlift2_rhs}{0}{0}{\isa{netlift\ sr\ s\ \listconcat\ netlift\ sr\ t}}

\snip{netliftl1_lhs}{0}{0}{\isa{netliftl\ sr\ {\isacharparenleft}\NodeS\ i\ s\ R{\isacharparenright}}}
\snip{netliftl1_rhs}{0}{0}{\isa{\NodeS\ i\ {{\isacharparenleft}snd\ {\isacharparenleft}sr\ s{\isacharparenright}{\isacharparenright}}\ R}}
\snip{netliftl2_lhs}{0}{0}{\isa{netliftl\ sr\ {\isacharparenleft}\subnets{s}{t}{\isacharparenright}}}
\snip{netliftl2_rhs}{0}{0}{\isa{\subnets{{\isacharparenleft}netliftl\ sr\ s{\isacharparenright}}{{\isacharparenleft}netliftl\ sr\ t{\isacharparenright}}}}

\snip{netgmap1_lhs}{0}{0}{\isa{netgmap\ sr\ {\isacharparenleft}\NodeS\ i\ s\ R{\isacharparenright}}}
\snip{netgmap1_rhs}{0}{0}{\isa{{\isacharparenleft}{\isacharbrackleft}i\ {\isasymmapsto}\ fst\ {\isacharparenleft}sr\ s{\isacharparenright}{\isacharbrackright},\ \NodeS\ i\ {{\isacharparenleft}snd\ {\isacharparenleft}sr\ s{\isacharparenright}{\isacharparenright}}\ R{\isacharparenright}}}
\snip{netgmap2_lhs}{0}{0}{\isa{netgmap\ sr\ {\isacharparenleft}\subnets{s\isactrlsub {\isadigit{1}}}{s\isactrlsub {\isadigit{2}}}{\isacharparenright}}}
\snip{netgmap2_rhs}{0}{0}{\isa{\textsf{let}\ {\isacharparenleft}{\isasymsigma}\isactrlsub {\isadigit{1}},\ ss{\isacharparenright}\ {\isacharequal}\ netgmap\ sr\ s\isactrlsub {\isadigit{1}}{\isacharsemicolon}}\\&&\isa{\phantom{let\ }{\isacharparenleft}{\isasymsigma}\isactrlsub {\isadigit{2}},\ tt{\isacharparenright}\ {\isacharequal}\ netgmap\ sr\ s\isactrlsub {\isadigit{2}}\ }\\&&\isa{\textsf{in}\ {\isacharparenleft}{\isasymsigma}\isactrlsub {\isadigit{1}}\ \listconcat\ {\isasymsigma}\isactrlsub {\isadigit{2}},\ \subnets{ss}{tt}{\isacharparenright}}}
\snip{initmissing_def}{0}{0}{\isa{initmissing\ {\isasymsigma}\ {\isacharequal}\ {\isacharparenleft}{\isasymlambda}i{\isachardot}\ \textsf{case}\ fst\ {\isasymsigma}\ i\ \textsf{of}\ None\ {\isasymRightarrow}\ aodv{\isacharunderscore}init\ i\ {\isacharbar}\ Some\ s\ {\isasymRightarrow}\ s,\ snd\ {\isasymsigma}{\isacharparenright}}}
\snip{initmissing_lhs}{0}{0}{\isa{initmissing\ {\isasymsigma}}}
\snip{initmissing_rhs}{0}{0}{\isa{{\isacharparenleft}{\isasymlambda}i{\isachardot}\ \textsf{case}\ fst\ {\isasymsigma}\ i\ \textsf{of}\ None\ {\isasymRightarrow}\ aodv{\isacharunderscore}init\ i\ {\isacharbar}\ Some\ s\ {\isasymRightarrow}\ s,\ snd\ {\isasymsigma}{\isacharparenright}}}

\snip{onl}{0}{0}{\isa{onl\ {\isasymGamma}\ P\ {\isacharequal}\ {\isasymlambda}{\isacharparenleft}{\isasymxi},\ p{\isacharparenright}{\isachardot}\ {\isasymforall}l{\isasymin}labels\ {\isasymGamma}\ p{\isachardot}\ P\ {\isacharparenleft}{\isasymxi},\ l{\isacharparenright}}}
\snip{onl_lhs}{0}{0}{\isa{onl\ {\isasymGamma}\ P}}
\snip{onl_rhs}{0}{0}{\isa{{\isasymlambda}{\isacharparenleft}{\isasymxi},\ p{\isacharparenright}{\isachardot}\ {\isasymforall}l{\isasymin}labels\ {\isasymGamma}\ p{\isachardot}\ P\ {\isacharparenleft}{\isasymxi},\ l{\isacharparenright}}}

\snip{cnet_connect}{0}{0}{\isa{\mbox{}\inferrule{\mbox{{\isacharparenleft}s,\ connect{\isacharparenleft}i,\ i{\isacharprime}{\isacharparenright},\ s{\isacharprime}{\isacharparenright}{\isasymin}S}}{\mbox{{\isacharparenleft}s,\ connect{\isacharparenleft}i,\ i{\isacharprime}{\isacharparenright},\ s{\isacharprime}{\isacharparenright}{\isasymin}\cnetsos\ S}}}}
\snip{cnet_disconnect}{0}{0}{\isa{\mbox{}\inferrule{\mbox{{\isacharparenleft}s,\ disconnect{\isacharparenleft}i,\ i{\isacharprime}{\isacharparenright},\ s{\isacharprime}{\isacharparenright}{\isasymin}S}}{\mbox{{\isacharparenleft}s,\ disconnect{\isacharparenleft}i,\ i{\isacharprime}{\isacharparenright},\ s{\isacharprime}{\isacharparenright}{\isasymin}\cnetsos\ S}}}}
\snip{cnet_cast}{0}{0}{\isa{\mbox{}\inferrule{\mbox{{\isacharparenleft}s,\ R{\isacharcolon}{\isacharasterisk}cast{\isacharparenleft}m{\isacharparenright},\ s{\isacharprime}{\isacharparenright}{\isasymin}S}}{\mbox{{\isacharparenleft}s,\ {\isasymtau},\ s{\isacharprime}{\isacharparenright}{\isasymin}\cnetsos\ S}}}}
\snip{cnet_tau}{0}{0}{\isa{\mbox{}\inferrule{\mbox{{\isacharparenleft}s,\ {\isasymtau},\ s{\isacharprime}{\isacharparenright}{\isasymin}S}}{\mbox{{\isacharparenleft}s,\ {\isasymtau},\ s{\isacharprime}{\isacharparenright}{\isasymin}\cnetsos\ S}}}}
\snip{cnet_deliver}{0}{0}{\isa{\mbox{}\inferrule{\mbox{{\isacharparenleft}s,\ i{\isacharcolon}deliver{\isacharparenleft}d{\isacharparenright},\ s{\isacharprime}{\isacharparenright}{\isasymin}S}}{\mbox{{\isacharparenleft}s,\ i{\isacharcolon}deliver{\isacharparenleft}d{\isacharparenright},\ s{\isacharprime}{\isacharparenright}{\isasymin}\cnetsos\ S}}}}
\snip{cnet_newpkt}{0}{0}{\isa{\mbox{}\inferrule{\mbox{{\isacharparenleft}s,\ {\isacharbraceleft}i{\isacharbraceright}{\isasymnot}K{\isacharcolon}arrive{\isacharparenleft}newpkt\ {\isacharparenleft}d,\ di{\isacharparenright}{\isacharparenright},\ s{\isacharprime}{\isacharparenright}{\isasymin}S}}{\mbox{{\isacharparenleft}s,\ i{\isacharcolon}newpkt{\isacharparenleft}d,\ di{\isacharparenright},\ s{\isacharprime}{\isacharparenright}{\isasymin}\cnetsos\ S}}}}

\snip{closed}{0}{0}{\isa{closed\ {\isacharparenleft}pnet\ {\isacharparenleft}{\isasymlambda}i{\isachardot}\ paodv\ i\ {\isasymlangle}{\isasymlangle}\ qmsg{\isacharparenright}\ any{\isacharparenright}}}
\snip{closed'}{0}{0}{\isa{closed\ A}}
\snip{closed_term}{0}{0}{\isa{A{\isasymlparr}trans\ {\isacharcolon}{\isacharequal}\ \cnetsos\ {\isacharparenleft}trans\ A{\isacharparenright}{\isasymrparr}}}

\snip{closed_type}{0}{0}{\isa{{\isacharparenleft}{\isacharparenleft}{\isacharparenleft}state\ {\isasymtimes}\ {\isacharparenleft}state,\ msg,\ pseqp,\ pseqp\ label{\isacharparenright}\ seqp{\isacharparenright}\ {\isasymtimes}\ msg\ list\ {\isasymtimes}\ {\isacharparenleft}msg\ list,\ msg,\ unit,\ unit\ label{\isacharparenright}\ seqp{\isacharparenright}\ net{\isacharunderscore}state,\ msg\ node{\isacharunderscore}action{\isacharparenright}\ automaton}}
\snip{update_type}{0}{0}{\isa{rt\ {\isasymRightarrow}\ ip\ {\isasymRightarrow}\ r\ {\isasymRightarrow}\ rt}}
\snip{invalidate_type}{0}{0}{\isa{rt\ {\isasymRightarrow}\ {\isacharparenleft}ip\ {\isasymrightharpoonup}\ sqn{\isacharparenright}\ {\isasymRightarrow}\ rt}}
\snip{nhop_type}{0}{0}{\isa{rt\ {\isasymRightarrow}\ ip\ {\isasymrightharpoonup}\ ip}}
\snip{addpreRT_type}{0}{0}{\isa{rt\ {\isasymRightarrow}\ ip\ {\isasymRightarrow}\ ip\ set\ {\isasymrightharpoonup}\ rt}}
\snip{state_type}{0}{0}{\isa{state}}
\snip{sigma_type}{0}{0}{\isa{ip\ {\isasymRightarrow}\ state}}
\snip{r_type}{0}{0}{\isa{sqn~{\isasymtimes}~k~{\isasymtimes}~f~{\isasymtimes}~nat~{\isasymtimes}~ip~{\isasymtimes}~ip~set}}

\snip{obroadcastT}{0}{0}{\isa{\mbox{}\inferrule{\mbox{{\isasymsigma}{\isacharprime}\ i\ {\isacharequal}\ {\isasymsigma}\ i}}{\mbox{{\isacharparenleft}{\isacharparenleft}{\isasymsigma},\ \gray{{\isacharbraceleft}l{\isacharbraceright}}broadcast{\isacharparenleft}\selmsg{\isacharparenright}\ {\isachardot}\ p{\isacharparenright},\ broadcast\ {\isacharparenleft}\selmsg\ {\isacharparenleft}{\isasymsigma}\ i{\isacharparenright}{\isacharparenright},\ {\isacharparenleft}{\isasymsigma}{\isacharprime},\ p{\isacharparenright}{\isacharparenright}{\isasymin}\oseqpsos\ {\isasymGamma}\ i}}}}
\snip{ogroupcastT}{0}{0}{\isa{\mbox{}\inferrule{\mbox{{\isasymsigma}{\isacharprime}\ i\ {\isacharequal}\ {\isasymsigma}\ i}}{\mbox{{\isacharparenleft}{\isacharparenleft}{\isasymsigma},\ \gray{{\isacharbraceleft}l{\isacharbraceright}}groupcast{\isacharparenleft}\selips,\ \selmsg{\isacharparenright}\ {\isachardot}\ p{\isacharparenright},\ groupcast\ {\isacharparenleft}\selips\ {\isacharparenleft}{\isasymsigma}\ i{\isacharparenright}{\isacharparenright}\ {\isacharparenleft}\selmsg\ {\isacharparenleft}{\isasymsigma}\ i{\isacharparenright}{\isacharparenright},\ {\isacharparenleft}{\isasymsigma}{\isacharprime},\ p{\isacharparenright}{\isacharparenright}{\isasymin}\oseqpsos\ {\isasymGamma}\ i}}}}
\snip{ounicastT}{0}{0}{\isa{\mbox{}\inferrule{\mbox{{\isasymsigma}{\isacharprime}\ i\ {\isacharequal}\ {\isasymsigma}\ i}}{\mbox{{\isacharparenleft}{\isacharparenleft}{\isasymsigma},\ \gray{{\isacharbraceleft}l{\isacharbraceright}}unicast{\isacharparenleft}\selip,\ \selmsg{\isacharparenright}\ {\isachardot}\ p\ {\isasymtriangleright}\ q{\isacharparenright},\ unicast\ {\isacharparenleft}\selip\ {\isacharparenleft}{\isasymsigma}\ i{\isacharparenright}{\isacharparenright}\ {\isacharparenleft}\selmsg\ {\isacharparenleft}{\isasymsigma}\ i{\isacharparenright}{\isacharparenright},\ {\isacharparenleft}{\isasymsigma}{\isacharprime},\ p{\isacharparenright}{\isacharparenright}{\isasymin}\oseqpsos\ {\isasymGamma}\ i}}}}
\snip{onotunicastT}{0}{0}{\isa{\mbox{}\inferrule{\mbox{{\isasymsigma}{\isacharprime}\ i\ {\isacharequal}\ {\isasymsigma}\ i}}{\mbox{{\isacharparenleft}{\isacharparenleft}{\isasymsigma},\ \gray{{\isacharbraceleft}l{\isacharbraceright}}unicast{\isacharparenleft}\selip,\ \selmsg{\isacharparenright}\ {\isachardot}\ p\ {\isasymtriangleright}\ q{\isacharparenright},\ {\isasymnot}unicast\ {\isacharparenleft}\selip\ {\isacharparenleft}{\isasymsigma}\ i{\isacharparenright}{\isacharparenright},\ {\isacharparenleft}{\isasymsigma}{\isacharprime},\ q{\isacharparenright}{\isacharparenright}{\isasymin}\oseqpsos\ {\isasymGamma}\ i}}}}
\snip{osendT}{0}{0}{\isa{\mbox{}\inferrule{\mbox{{\isasymsigma}{\isacharprime}\ i\ {\isacharequal}\ {\isasymsigma}\ i}}{\mbox{{\isacharparenleft}{\isacharparenleft}{\isasymsigma},\ \gray{{\isacharbraceleft}l{\isacharbraceright}}send{\isacharparenleft}\selmsg{\isacharparenright}\ {\isachardot}\ p{\isacharparenright},\ send\ {\isacharparenleft}\selmsg\ {\isacharparenleft}{\isasymsigma}\ i{\isacharparenright}{\isacharparenright},\ {\isacharparenleft}{\isasymsigma}{\isacharprime},\ p{\isacharparenright}{\isacharparenright}{\isasymin}\oseqpsos\ {\isasymGamma}\ i}}}}
\snip{odeliverT}{0}{0}{\isa{\mbox{}\inferrule{\mbox{{\isasymsigma}{\isacharprime}\ i\ {\isacharequal}\ {\isasymsigma}\ i}}{\mbox{{\isacharparenleft}{\isacharparenleft}{\isasymsigma},\ \gray{{\isacharbraceleft}l{\isacharbraceright}}deliver{\isacharparenleft}\seldata{\isacharparenright}\ {\isachardot}\ p{\isacharparenright},\ deliver\ {\isacharparenleft}\seldata\ {\isacharparenleft}{\isasymsigma}\ i{\isacharparenright}{\isacharparenright},\ {\isacharparenleft}{\isasymsigma}{\isacharprime},\ p{\isacharparenright}{\isacharparenright}{\isasymin}\oseqpsos\ {\isasymGamma}\ i}}}}
\snip{oreceiveT}{0}{0}{\isa{\mbox{}\inferrule{\mbox{{\isasymsigma}{\isacharprime}\ i\ {\isacharequal}\ \updmsg\ msg\ {\isacharparenleft}{\isasymsigma}\ i{\isacharparenright}}}{\mbox{{\isacharparenleft}{\isacharparenleft}{\isasymsigma},\ \gray{{\isacharbraceleft}l{\isacharbraceright}}receive{\isacharparenleft}\updmsg{\isacharparenright}\ {\isachardot}\ p{\isacharparenright},\ receive\ msg,\ {\isacharparenleft}{\isasymsigma}{\isacharprime},\ p{\isacharparenright}{\isacharparenright}{\isasymin}\oseqpsos\ {\isasymGamma}\ i}}}}
\snip{oassignT}{0}{0}{\isa{\mbox{}\inferrule{\mbox{{\isasymsigma}{\isacharprime}\ i\ {\isacharequal}\ u\ {\isacharparenleft}{\isasymsigma}\ i{\isacharparenright}}}{\mbox{{\isacharparenleft}{\isacharparenleft}{\isasymsigma},\ \gray{{\isacharbraceleft}l{\isacharbraceright}}{\isasymlbrakk}u{\isasymrbrakk}\ p{\isacharparenright},\ {\isasymtau},\ {\isacharparenleft}{\isasymsigma}{\isacharprime},\ p{\isacharparenright}{\isacharparenright}{\isasymin}\oseqpsos\ {\isasymGamma}\ i}}}}
\snip{ocallT}{0}{0}{\isa{\mbox{}\inferrule{\mbox{{\isacharparenleft}{\isacharparenleft}{\isasymsigma},\ {\isasymGamma}\ pn{\isacharparenright},\ a,\ {\isacharparenleft}{\isasymsigma}{\isacharprime},\ p{\isacharprime}{\isacharparenright}{\isacharparenright}{\isasymin}\oseqpsos\ {\isasymGamma}\ i}}{\mbox{{\isacharparenleft}{\isacharparenleft}{\isasymsigma},\ call{\isacharparenleft}pn{\isacharparenright}{\isacharparenright},\ a,\ {\isacharparenleft}{\isasymsigma}{\isacharprime},\ p{\isacharprime}{\isacharparenright}{\isacharparenright}{\isasymin}\oseqpsos\ {\isasymGamma}\ i}}}}
\snip{ochoiceT1}{0}{0}{\isa{\mbox{}\inferrule{\mbox{{\isacharparenleft}{\isacharparenleft}{\isasymsigma},\ p{\isacharparenright},\ a,\ {\isacharparenleft}{\isasymsigma}{\isacharprime},\ p{\isacharprime}{\isacharparenright}{\isacharparenright}{\isasymin}\oseqpsos\ {\isasymGamma}\ i}}{\mbox{{\isacharparenleft}{\isacharparenleft}{\isasymsigma},\ p\ {\isasymoplus}\ q{\isacharparenright},\ a,\ {\isacharparenleft}{\isasymsigma}{\isacharprime},\ p{\isacharprime}{\isacharparenright}{\isacharparenright}{\isasymin}\oseqpsos\ {\isasymGamma}\ i}}}}
\snip{ochoiceT2}{0}{0}{\isa{\mbox{}\inferrule{\mbox{{\isacharparenleft}{\isacharparenleft}{\isasymsigma},\ q{\isacharparenright},\ a,\ {\isacharparenleft}{\isasymsigma}{\isacharprime},\ q{\isacharprime}{\isacharparenright}{\isacharparenright}{\isasymin}\oseqpsos\ {\isasymGamma}\ i}}{\mbox{{\isacharparenleft}{\isacharparenleft}{\isasymsigma},\ p\ {\isasymoplus}\ q{\isacharparenright},\ a,\ {\isacharparenleft}{\isasymsigma}{\isacharprime},\ q{\isacharprime}{\isacharparenright}{\isacharparenright}{\isasymin}\oseqpsos\ {\isasymGamma}\ i}}}}
\snip{oguardT}{0}{0}{\isa{\mbox{}\inferrule{\mbox{{\isasymsigma}{\isacharprime}\ i{\isasymin}g\ {\isacharparenleft}{\isasymsigma}\ i{\isacharparenright}}}{\mbox{{\isacharparenleft}{\isacharparenleft}{\isasymsigma},\ \gray{{\isacharbraceleft}l{\isacharbraceright}}{\isasymlangle}g{\isasymrangle}\ p{\isacharparenright},\ {\isasymtau},\ {\isacharparenleft}{\isasymsigma}{\isacharprime},\ p{\isacharparenright}{\isacharparenright}{\isasymin}\oseqpsos\ {\isasymGamma}\ i}}}}

\snip{oassignT_prem_1}{0}{0}{\isa{{\isasymsigma}{\isacharprime}\ i\ {\isacharequal}\ u\ {\isacharparenleft}{\isasymsigma}\ i{\isacharparenright}}}
\snip{ounicastT_prem_1}{0}{0}{\isa{{\isasymsigma}{\isacharprime}\ i\ {\isacharequal}\ {\isasymsigma}\ i}}

\snip{oparleft}{0}{0}{\isa{\mbox{}\inferrule{\mbox{{\isacharparenleft}{\isacharparenleft}{\isasymsigma},\ s{\isacharparenright},\ a,\ {\isacharparenleft}{\isasymsigma}{\isacharprime},\ s{\isacharprime}{\isacharparenright}{\isacharparenright}{\isasymin}S}\\\ \mbox{{\isasymAnd}m{\isachardot}\ a\ {\isasymnoteq}\ receive\ m}}{\mbox{{\isacharparenleft}{\isacharparenleft}{\isasymsigma},\ s,\ t{\isacharparenright},\ a,\ {\isasymsigma}{\isacharprime},\ s{\isacharprime},\ t{\isacharparenright}{\isasymin}\oparpsos\ i\ S\ T}}}}
\snip{oparright}{0}{0}{\isa{\mbox{}\inferrule{\mbox{{\isacharparenleft}t,\ {\isacharparenleft}a,\ t{\isacharprime}{\isacharparenright}{\isacharparenright}{\isasymin}T}\\\ \mbox{{\isasymAnd}m{\isachardot}\ a\ {\isasymnoteq}\ send\ m}\\\ \mbox{{\isasymsigma}{\isacharprime}\ i\ {\isacharequal}\ {\isasymsigma}\ i}}{\mbox{{\isacharparenleft}{\isacharparenleft}{\isasymsigma},\ s,\ t{\isacharparenright},\ a,\ {\isasymsigma}{\isacharprime},\ s,\ t{\isacharprime}{\isacharparenright}{\isasymin}\oparpsos\ i\ S\ T}}}}
\snip{oparboth}{0}{0}{\isa{\mbox{}\inferrule{\mbox{{\isacharparenleft}{\isacharparenleft}{\isasymsigma},\ s{\isacharparenright},\ receive\ m,\ {\isacharparenleft}{\isasymsigma}{\isacharprime},\ s{\isacharprime}{\isacharparenright}{\isacharparenright}{\isasymin}S}\\\ \mbox{{\isacharparenleft}t,\ send\ m,\ t{\isacharprime}{\isacharparenright}{\isasymin}T}}{\mbox{{\isacharparenleft}{\isacharparenleft}{\isasymsigma},\ {\isacharparenleft}s,\ t{\isacharparenright}{\isacharparenright},\ {\isasymtau},\ {\isacharparenleft}{\isasymsigma}{\isacharprime},\ {\isacharparenleft}s{\isacharprime},\ t{\isacharprime}{\isacharparenright}{\isacharparenright}{\isacharparenright}{\isasymin}\oparpsos\ i\ S\ T}}}}

\snip{oparcomp}{0}{0}{\begin{isabelle}%
s\ {\isasymlangle}{\isasymlangle}\isactrlbsub i\isactrlesub \ t\ {\isacharequal}\isanewline
{\isasymlparr}init\ {\isacharequal}\ extg\ {\isacharbackquote}\ {\isacharparenleft}init\ s\ {\isasymtimes}\ init\ t{\isacharparenright},\isanewline
\isaindent{\ \ \ }trans\ {\isacharequal}\ \oparpsos\ i\ {\isacharparenleft}trans\ s{\isacharparenright}\ {\isacharparenleft}trans\ t{\isacharparenright}{\isasymrparr}%
\end{isabelle}}
\snip{opar_comp'}{0}{0}{\mbox{\isa{s\ {\isasymlangle}{\isasymlangle}\isactrlbsub i\isactrlesub \ t\ {\isacharequal}\ {\isasymlparr}}} & \mbox{\isa{init\ {\isacharequal}\ {\isacharbraceleft}{\isacharparenleft}{\isasymsigma},\ (s\isactrlsub l,\ t\isactrlsub l){\isacharparenright}\ {\isacharbar}\ {\isacharparenleft}{\isasymsigma},\ s\isactrlsub l{\isacharparenright}{\isasymin}init\ s\ {\isasymand}\ t\isactrlsub l{\isasymin}init\ t{\isacharbraceright},}}\\ & \mbox{\isa{trans\ {\isacharequal}\ \oparpsos\ i\ {\isacharparenleft}trans\ s{\isacharparenright}\ {\isacharparenleft}trans\ t{\isacharparenright}{\isasymrparr}}}}
\snip{oparcomp_term}{0}{0}{\isa{s\ {\isasymlangle}{\isasymlangle}\isactrlbsub i\isactrlesub \ t{\isasymColon}{\isacharparenleft}{\isacharparenleft}nat\ {\isasymRightarrow}\ {\isacharprime}a{\isacharparenright}\ {\isasymtimes}\ {\isacharprime}b\ {\isasymtimes}\ {\isacharprime}d,\ {\isacharprime}c\ seq{\isacharunderscore}action{\isacharparenright}\ automaton}}

\snip{onode_bcast}{0}{0}{\isa{\mbox{}\inferrule{\mbox{{\isacharparenleft}{\isacharparenleft}{\isasymsigma},\ s{\isacharparenright},\ broadcast\ m,\ {\isasymsigma}{\isacharprime},\ s{\isacharprime}{\isacharparenright}{\isasymin}S}}{\mbox{{\isacharparenleft}{\isacharparenleft}{\isasymsigma},\ \NodeS\ i\ s\ R{\isacharparenright},\ R{\isacharcolon}{\isacharasterisk}cast{\isacharparenleft}m{\isacharparenright},\ {\isacharparenleft}{\isasymsigma}{\isacharprime},\ \NodeS\ i\ s{\isacharprime}\ R{\isacharparenright}{\isacharparenright}{\isasymin}\onodesos\ S}}}}
\snip{onode_gcast}{0}{0}{\isa{\mbox{}\inferrule{\mbox{{\isacharparenleft}{\isacharparenleft}{\isasymsigma},\ s{\isacharparenright},\ groupcast\ D\ m,\ {\isasymsigma}{\isacharprime},\ s{\isacharprime}{\isacharparenright}{\isasymin}S}}{\mbox{{\isacharparenleft}{\isacharparenleft}{\isasymsigma},\ \NodeS\ i\ s\ R{\isacharparenright},\ {\isacharparenleft}R\ {\isasyminter}\ D{\isacharparenright}{\isacharcolon}{\isacharasterisk}cast{\isacharparenleft}m{\isacharparenright},\ {\isacharparenleft}{\isasymsigma}{\isacharprime},\ \NodeS\ i\ s{\isacharprime}\ R{\isacharparenright}{\isacharparenright}{\isasymin}\onodesos\ S}}}}
\snip{onode_ucast}{0}{0}{\isa{\mbox{}\inferrule{\mbox{{\isacharparenleft}{\isacharparenleft}{\isasymsigma},\ s{\isacharparenright},\ unicast\ d\ m,\ {\isasymsigma}{\isacharprime},\ s{\isacharprime}{\isacharparenright}{\isasymin}S}\\\ \mbox{d{\isasymin}R}}{\mbox{{\isacharparenleft}{\isacharparenleft}{\isasymsigma},\ \NodeS\ i\ s\ R{\isacharparenright},\ {\isacharbraceleft}d{\isacharbraceright}{\isacharcolon}{\isacharasterisk}cast{\isacharparenleft}m{\isacharparenright},\ {\isacharparenleft}{\isasymsigma}{\isacharprime},\ \NodeS\ i\ s{\isacharprime}\ R{\isacharparenright}{\isacharparenright}{\isasymin}\onodesos\ S}}}}
\snip{onode_notucast}{0}{0}{\isa{\mbox{}\inferrule{\mbox{{\isacharparenleft}{\isacharparenleft}{\isasymsigma},\ s{\isacharparenright},\ {\isasymnot}unicast\ d,\ {\isasymsigma}{\isacharprime},\ s{\isacharprime}{\isacharparenright}{\isasymin}S}\\\ \mbox{d\ {\isasymnotin}\ R}\\\ \mbox{{\isasymforall}j{\isachardot}\ j\ {\isasymnoteq}\ i\ {\isasymlongrightarrow}\ {\isasymsigma}{\isacharprime}\ j\ {\isacharequal}\ {\isasymsigma}\ j}}{\mbox{{\isacharparenleft}{\isacharparenleft}{\isasymsigma},\ \NodeS\ i\ s\ R{\isacharparenright},\ {\isasymtau},\ {\isacharparenleft}{\isasymsigma}{\isacharprime},\ \NodeS\ i\ s{\isacharprime}\ R{\isacharparenright}{\isacharparenright}{\isasymin}\onodesos\ S}}}}
\snip{onode_deliver}{0}{0}{\isa{\mbox{}\inferrule{\mbox{{\isacharparenleft}{\isacharparenleft}{\isasymsigma},\ s{\isacharparenright},\ deliver\ d,\ {\isasymsigma}{\isacharprime},\ s{\isacharprime}{\isacharparenright}{\isasymin}S}\\\ \mbox{{\isasymforall}j{\isachardot}\ j\ {\isasymnoteq}\ i\ {\isasymlongrightarrow}\ {\isasymsigma}{\isacharprime}\ j\ {\isacharequal}\ {\isasymsigma}\ j}}{\mbox{{\isacharparenleft}{\isacharparenleft}{\isasymsigma},\ \NodeS\ i\ s\ R{\isacharparenright},\ i{\isacharcolon}deliver{\isacharparenleft}d{\isacharparenright},\ {\isacharparenleft}{\isasymsigma}{\isacharprime},\ \NodeS\ i\ s{\isacharprime}\ R{\isacharparenright}{\isacharparenright}{\isasymin}\onodesos\ S}}}}
\snip{onode_receive}{0}{0}{\isa{\mbox{}\inferrule{\mbox{{\isacharparenleft}{\isacharparenleft}{\isasymsigma},\ s{\isacharparenright},\ receive\ m,\ {\isacharparenleft}{\isasymsigma}{\isacharprime},\ s{\isacharprime}{\isacharparenright}{\isacharparenright}{\isasymin}S}}{\mbox{{\isacharparenleft}{\isacharparenleft}{\isasymsigma},\ \NodeS\ i\ s\ R{\isacharparenright},\ {\isacharbraceleft}i{\isacharbraceright}{\isasymnot}{\isasymemptyset}{\isacharcolon}arrive{\isacharparenleft}m{\isacharparenright},\ {\isacharparenleft}{\isasymsigma}{\isacharprime},\ \NodeS\ i\ s{\isacharprime}\ R{\isacharparenright}{\isacharparenright}{\isasymin}\onodesos\ S}}}}
\snip{onode_tau}{0}{0}{\isa{\mbox{}\inferrule{\mbox{{\isacharparenleft}{\isacharparenleft}{\isasymsigma},\ s{\isacharparenright},\ {\isasymtau},\ {\isacharparenleft}{\isasymsigma}{\isacharprime},\ s{\isacharprime}{\isacharparenright}{\isacharparenright}{\isasymin}S}\\\ \mbox{{\isasymforall}j {\isasymnoteq} i{\isachardot}\ {\isasymsigma}{\isacharprime}\ j\ {\isacharequal}\ {\isasymsigma}\ j}}{\mbox{{\isacharparenleft}{\isacharparenleft}{\isasymsigma},\ \NodeS\ i\ s\ R{\isacharparenright},\ {\isasymtau},\ {\isacharparenleft}{\isasymsigma}{\isacharprime},\ \NodeS\ i\ s{\isacharprime}\ R{\isacharparenright}{\isacharparenright}{\isasymin}\onodesos\ S}}}}
\snip{onode_arrive}{0}{0}{\isa{\mbox{}\inferrule{\mbox{{\isasymsigma}{\isacharprime}\ i\ {\isacharequal}\ {\isasymsigma}\ i}}{\mbox{{\isacharparenleft}{\isacharparenleft}{\isasymsigma},\ \NodeS\ i\ s\ R{\isacharparenright},\ {\isasymemptyset}{\isasymnot}{\isacharbraceleft}i{\isacharbraceright}{\isacharcolon}arrive{\isacharparenleft}m{\isacharparenright},\ {\isacharparenleft}{\isasymsigma}{\isacharprime},\ \NodeS\ i\ s\ R{\isacharparenright}{\isacharparenright}{\isasymin}\onodesos\ S}}}}
\snip{onode_connect1}{0}{0}{\isa{\mbox{}\inferrule{\mbox{{\isasymsigma}{\isacharprime}\ i\ {\isacharequal}\ {\isasymsigma}\ i}}{\mbox{{\isacharparenleft}{\isacharparenleft}{\isasymsigma},\ \NodeS\ i\ s\ R{\isacharparenright},\ connect{\isacharparenleft}i,\ i{\isacharprime}{\isacharparenright},\ {\isacharparenleft}{\isasymsigma}{\isacharprime},\ \NodeS\ i\ s\ {R\ {\isasymunion}\ {\isacharbraceleft}i{\isacharprime}{\isacharbraceright}}{\isacharparenright}{\isacharparenright}{\isasymin}\onodesos\ S}}}}
\snip{onode_connect2}{0}{0}{\isa{\mbox{}\inferrule{\mbox{{\isasymsigma}{\isacharprime}\ i\ {\isacharequal}\ {\isasymsigma}\ i}}{\mbox{{\isacharparenleft}{\isacharparenleft}{\isasymsigma},\ \NodeS\ i\ s\ R{\isacharparenright},\ connect{\isacharparenleft}i{\isacharprime},\ i{\isacharparenright},\ {\isacharparenleft}{\isasymsigma}{\isacharprime},\ \NodeS\ i\ s\ {R\ {\isasymunion}\ {\isacharbraceleft}i{\isacharprime}{\isacharbraceright}}{\isacharparenright}{\isacharparenright}{\isasymin}\onodesos\ S}}}}
\snip{onode_disconnect1}{0}{0}{\isa{\mbox{}\inferrule{\mbox{{\isasymsigma}{\isacharprime}\ i\ {\isacharequal}\ {\isasymsigma}\ i}}{\mbox{{\isacharparenleft}{\isacharparenleft}{\isasymsigma},\ \NodeS\ i\ s\ R{\isacharparenright},\ disconnect{\isacharparenleft}i,\ i{\isacharprime}{\isacharparenright},\ {\isacharparenleft}{\isasymsigma}{\isacharprime},\ \NodeS\ i\ s\ {\isacharparenleft}R\ {\isacharminus}\ {\isacharbraceleft}i{\isacharprime}{\isacharbraceright}{\isacharparenright}{\isacharparenright}{\isacharparenright}{\isasymin}\onodesos\ S}}}}
\snip{onode_disconnect2}{0}{0}{\isa{\mbox{}\inferrule{\mbox{{\isasymsigma}{\isacharprime}\ i\ {\isacharequal}\ {\isasymsigma}\ i}}{\mbox{{\isacharparenleft}{\isacharparenleft}{\isasymsigma},\ \NodeS\ i\ s\ R{\isacharparenright},\ disconnect{\isacharparenleft}i{\isacharprime},\ i{\isacharparenright},\ {\isacharparenleft}{\isasymsigma}{\isacharprime},\ \NodeS\ i\ s\ {\isacharparenleft}R\ {\isacharminus}\ {\isacharbraceleft}i{\isacharprime}{\isacharbraceright}{\isacharparenright}{\isacharparenright}{\isacharparenright}{\isasymin}\onodesos\ S}}}}
\snip{onode_connect_other}{0}{0}{\isa{\mbox{}\inferrule{\mbox{i\ {\isasymnoteq}\ i{\isacharprime}}\\\ \mbox{i\ {\isasymnoteq}\ i{\isacharprime}{\isacharprime}}\\\ \mbox{{\isasymsigma}{\isacharprime}\ i\ {\isacharequal}\ {\isasymsigma}\ i}}{\mbox{{\isacharparenleft}{\isacharparenleft}{\isasymsigma},\ \NodeS\ i\ s\ R{\isacharparenright},\ connect{\isacharparenleft}i{\isacharprime},\ i{\isacharprime}{\isacharprime}{\isacharparenright},\ {\isacharparenleft}{\isasymsigma}{\isacharprime},\ \NodeS\ i\ s\ R{\isacharparenright}{\isacharparenright}{\isasymin}\onodesos\ S}}}}
\snip{onode_disconnect_other}{0}{0}{\isa{\mbox{}\inferrule{\mbox{i\ {\isasymnoteq}\ i{\isacharprime}}\\\ \mbox{i\ {\isasymnoteq}\ i{\isacharprime}{\isacharprime}}\\\ \mbox{{\isasymsigma}{\isacharprime}\ i\ {\isacharequal}\ {\isasymsigma}\ i}}{\mbox{{\isacharparenleft}{\isacharparenleft}{\isasymsigma},\ \NodeS\ i\ s\ R{\isacharparenright},\ disconnect{\isacharparenleft}i{\isacharprime},\ i{\isacharprime}{\isacharprime}{\isacharparenright},\ {\isacharparenleft}{\isasymsigma}{\isacharprime},\ \NodeS\ i\ s\ R{\isacharparenright}{\isacharparenright}{\isasymin}\onodesos\ S}}}}

\snip{onode_tau_prem_2}{0}{0}{\isa{{\isasymforall}j {\isasymnoteq} i{\isachardot}\ {\isasymsigma}{\isacharprime}\ j\ {\isacharequal}\ {\isasymsigma}\ j}}

\snip{onode_comp}{0}{0}{\isa{{\isasymlangle}i\ {\isacharcolon}\ onp\ {\isacharcolon}\ \Ri{\isasymrangle}\isactrlsub o\ {\isacharequal}\ {\isasymlparr}init\ {\isacharequal}\ {\isacharbraceleft}{\isacharparenleft}{\isasymsigma},\ \NodeS\ i\ s\ \Ri{\isacharparenright}\ {\isacharbar}\ {\isacharparenleft}{\isasymsigma},\ s{\isacharparenright}{\isasymin}init\ onp{\isacharbraceright},\ trans\ {\isacharequal}\ \onodesos\ {\isacharparenleft}trans\ onp{\isacharparenright}{\isasymrparr}}}
\snip{onode_comp_term}{0}{0}{\isa{{\isasymlangle}i\ {\isacharcolon}\ S\ {\isacharcolon}\ R{\isasymrangle}\isactrlsub o{\isasymColon}{\isacharparenleft}{\isacharparenleft}nat\ {\isasymRightarrow}\ {\isacharprime}a{\isacharparenright}\ {\isasymtimes}\ {\isacharprime}b\ net{\isacharunderscore}state,\ {\isacharprime}c\ node{\isacharunderscore}action{\isacharparenright}\ automaton}}

\snip{opnet_cast1}{0}{0}{\isa{\mbox{}\inferrule{\mbox{{\isacharparenleft}{\isacharparenleft}{\isasymsigma},\ s{\isacharparenright},\ R{\isacharcolon}{\isacharasterisk}cast{\isacharparenleft}m{\isacharparenright},\ {\isasymsigma}{\isacharprime},\ s{\isacharprime}{\isacharparenright}{\isasymin}S}\\\ \mbox{{\isacharparenleft}{\isacharparenleft}{\isasymsigma},\ t{\isacharparenright},\ H{\isasymnot}K{\isacharcolon}arrive{\isacharparenleft}m{\isacharparenright},\ {\isasymsigma}{\isacharprime},\ t{\isacharprime}{\isacharparenright}{\isasymin}T}\\\ \mbox{H\ {\isasymsubseteq}\ R}\\\ \mbox{K\ {\isasyminter}\ R\ {\isacharequal}\ {\isasymemptyset}}}{\mbox{{\isacharparenleft}{\isacharparenleft}{\isasymsigma},\ \subnets{s}{t}{\isacharparenright},\ R{\isacharcolon}{\isacharasterisk}cast{\isacharparenleft}m{\isacharparenright},\ {\isacharparenleft}{\isasymsigma}{\isacharprime},\ \subnets{s{\isacharprime}}{t{\isacharprime}}{\isacharparenright}{\isacharparenright}{\isasymin}\opnetsos\ S\ T}}}}
\snip{opnet_cast2}{0}{0}{\isa{\mbox{}\inferrule{\mbox{{\isacharparenleft}{\isacharparenleft}{\isasymsigma},\ s{\isacharparenright},\ H{\isasymnot}K{\isacharcolon}arrive{\isacharparenleft}m{\isacharparenright},\ {\isasymsigma}{\isacharprime},\ s{\isacharprime}{\isacharparenright}{\isasymin}S}\\\ \mbox{{\isacharparenleft}{\isacharparenleft}{\isasymsigma},\ t{\isacharparenright},\ R{\isacharcolon}{\isacharasterisk}cast{\isacharparenleft}m{\isacharparenright},\ {\isasymsigma}{\isacharprime},\ t{\isacharprime}{\isacharparenright}{\isasymin}T}\\\ \mbox{H\ {\isasymsubseteq}\ R}\\\ \mbox{K\ {\isasyminter}\ R\ {\isacharequal}\ {\isasymemptyset}}}{\mbox{{\isacharparenleft}{\isacharparenleft}{\isasymsigma},\ \subnets{s}{t}{\isacharparenright},\ R{\isacharcolon}{\isacharasterisk}cast{\isacharparenleft}m{\isacharparenright},\ {\isacharparenleft}{\isasymsigma}{\isacharprime},\ \subnets{s{\isacharprime}}{t{\isacharprime}}{\isacharparenright}{\isacharparenright}{\isasymin}\opnetsos\ S\ T}}}}
\snip{opnet_arrive}{0}{0}{\isa{\mbox{}\inferrule{\mbox{{\isacharparenleft}{\isacharparenleft}{\isasymsigma},\ s{\isacharparenright},\ H{\isasymnot}K{\isacharcolon}arrive{\isacharparenleft}m{\isacharparenright},\ {\isacharparenleft}{\isasymsigma}{\isacharprime},\ s{\isacharprime}{\isacharparenright}{\isacharparenright}{\isasymin}S}\\\ \mbox{{\isacharparenleft}{\isacharparenleft}{\isasymsigma},\ t{\isacharparenright},\ H{\isacharprime}{\isasymnot}K{\isacharprime}{\isacharcolon}arrive{\isacharparenleft}m{\isacharparenright},\ {\isacharparenleft}{\isasymsigma}{\isacharprime},\ t{\isacharprime}{\isacharparenright}{\isacharparenright}{\isasymin}T}}{\mbox{{\isacharparenleft}{\isacharparenleft}{\isasymsigma},\ \subnets{s}{t}{\isacharparenright},\ {\isacharparenleft}H\ {\isasymunion}\ H{\isacharprime}{\isacharparenright}{\isasymnot}{\isacharparenleft}K\ {\isasymunion}\ K{\isacharprime}{\isacharparenright}{\isacharcolon}arrive{\isacharparenleft}m{\isacharparenright},\ {\isacharparenleft}{\isasymsigma}{\isacharprime},\ \subnets{s{\isacharprime}}{t{\isacharprime}}{\isacharparenright}{\isacharparenright}{\isasymin}\opnetsos\ S\ T}}}}
\snip{opnet_deliver1}{0}{0}{\isa{\mbox{}\inferrule{\mbox{{\isacharparenleft}{\isacharparenleft}{\isasymsigma},\ s{\isacharparenright},\ i{\isacharcolon}deliver{\isacharparenleft}d{\isacharparenright},\ {\isasymsigma}{\isacharprime},\ s{\isacharprime}{\isacharparenright}{\isasymin}S}}{\mbox{{\isacharparenleft}{\isacharparenleft}{\isasymsigma},\ \subnets{s}{t}{\isacharparenright},\ i{\isacharcolon}deliver{\isacharparenleft}d{\isacharparenright},\ {\isacharparenleft}{\isasymsigma}{\isacharprime},\ \subnets{s{\isacharprime}}{t}{\isacharparenright}{\isacharparenright}{\isasymin}\opnetsos\ S\ T}}}}
\snip{opnet_deliver2}{0}{0}{\isa{\mbox{}\inferrule{\mbox{{\isacharparenleft}{\isacharparenleft}{\isasymsigma},\ t{\isacharparenright},\ i{\isacharcolon}deliver{\isacharparenleft}d{\isacharparenright},\ {\isasymsigma}{\isacharprime},\ t{\isacharprime}{\isacharparenright}{\isasymin}T}}{\mbox{{\isacharparenleft}{\isacharparenleft}{\isasymsigma},\ \subnets{s}{t}{\isacharparenright},\ i{\isacharcolon}deliver{\isacharparenleft}d{\isacharparenright},\ {\isacharparenleft}{\isasymsigma}{\isacharprime},\ \subnets{s}{t{\isacharprime}}{\isacharparenright}{\isacharparenright}{\isasymin}\opnetsos\ S\ T}}}}
\snip{opnet_tau1}{0}{0}{\isa{\mbox{}\inferrule{\mbox{{\isacharparenleft}{\isacharparenleft}{\isasymsigma},\ s{\isacharparenright},\ {\isasymtau},\ {\isasymsigma}{\isacharprime},\ s{\isacharprime}{\isacharparenright}{\isasymin}S}}{\mbox{{\isacharparenleft}{\isacharparenleft}{\isasymsigma},\ \subnets{s}{t}{\isacharparenright},\ {\isasymtau},\ {\isacharparenleft}{\isasymsigma}{\isacharprime},\ \subnets{s{\isacharprime}}{t}{\isacharparenright}{\isacharparenright}{\isasymin}\opnetsos\ S\ T}}}}
\snip{opnet_tau2}{0}{0}{\isa{\mbox{}\inferrule{\mbox{{\isacharparenleft}{\isacharparenleft}{\isasymsigma},\ t{\isacharparenright},\ {\isasymtau},\ {\isasymsigma}{\isacharprime},\ t{\isacharprime}{\isacharparenright}{\isasymin}T}}{\mbox{{\isacharparenleft}{\isacharparenleft}{\isasymsigma},\ \subnets{s}{t}{\isacharparenright},\ {\isasymtau},\ {\isacharparenleft}{\isasymsigma}{\isacharprime},\ \subnets{s}{t{\isacharprime}}{\isacharparenright}{\isacharparenright}{\isasymin}\opnetsos\ S\ T}}}}
\snip{opnet_connect}{0}{0}{\isa{\mbox{}\inferrule{\mbox{{\isacharparenleft}{\isacharparenleft}{\isasymsigma},\ s{\isacharparenright},\ connect{\isacharparenleft}i,\ i{\isacharprime}{\isacharparenright},\ {\isasymsigma}{\isacharprime},\ s{\isacharprime}{\isacharparenright}{\isasymin}S}\\\ \mbox{{\isacharparenleft}{\isacharparenleft}{\isasymsigma},\ t{\isacharparenright},\ connect{\isacharparenleft}i,\ i{\isacharprime}{\isacharparenright},\ {\isasymsigma}{\isacharprime},\ t{\isacharprime}{\isacharparenright}{\isasymin}T}}{\mbox{{\isacharparenleft}{\isacharparenleft}{\isasymsigma},\ \subnets{s}{t}{\isacharparenright},\ connect{\isacharparenleft}i,\ i{\isacharprime}{\isacharparenright},\ {\isacharparenleft}{\isasymsigma}{\isacharprime},\ \subnets{s{\isacharprime}}{t{\isacharprime}}{\isacharparenright}{\isacharparenright}{\isasymin}\opnetsos\ S\ T}}}}
\snip{opnet_disconnect}{0}{0}{\isa{\mbox{}\inferrule{\mbox{{\isacharparenleft}{\isacharparenleft}{\isasymsigma},\ s{\isacharparenright},\ disconnect{\isacharparenleft}i,\ i{\isacharprime}{\isacharparenright},\ {\isasymsigma}{\isacharprime},\ s{\isacharprime}{\isacharparenright}{\isasymin}S}\\\ \mbox{{\isacharparenleft}{\isacharparenleft}{\isasymsigma},\ t{\isacharparenright},\ disconnect{\isacharparenleft}i,\ i{\isacharprime}{\isacharparenright},\ {\isasymsigma}{\isacharprime},\ t{\isacharprime}{\isacharparenright}{\isasymin}T}}{\mbox{{\isacharparenleft}{\isacharparenleft}{\isasymsigma},\ \subnets{s}{t}{\isacharparenright},\ disconnect{\isacharparenleft}i,\ i{\isacharprime}{\isacharparenright},\ {\isacharparenleft}{\isasymsigma}{\isacharprime},\ \subnets{s{\isacharprime}}{t{\isacharprime}}{\isacharparenright}{\isacharparenright}{\isasymin}\opnetsos\ S\ T}}}}

\snip{opnet}{0}{0}{\isa{opnet\ onp\ {\isasymlangle}i{\isacharsemicolon}\ \Ri{\isasymrangle}\ {\isacharequal}\ {\isasymlangle}i\ {\isacharcolon}\ onp\ i\ {\isacharcolon}\ \Ri{\isasymrangle}\isactrlsub o\isasep\isanewline%
opnet\ onp\ {\isacharparenleft}p\isactrlsub {\isadigit{1}}\parallelcomp{}p\isactrlsub {\isadigit{2}}{\isacharparenright}\ {\isacharequal}\ {\isasymlparr}init\ {\isacharequal}\ {\isacharbraceleft}{\isacharparenleft}{\isasymsigma},\ \subnets{s\isactrlsub {\isadigit{1}}}{s\isactrlsub {\isadigit{2}}}{\isacharparenright}\ {\isacharbar}\ {\isacharparenleft}{\isasymsigma},\ s\isactrlsub {\isadigit{1}}{\isacharparenright}{\isasymin}init\ {\isacharparenleft}opnet\ onp\ p\isactrlsub {\isadigit{1}}{\isacharparenright}\ {\isasymand}\ {\isacharparenleft}{\isasymsigma},\ s\isactrlsub {\isadigit{2}}{\isacharparenright}{\isasymin}init\ {\isacharparenleft}opnet\ onp\ p\isactrlsub {\isadigit{2}}{\isacharparenright}\ {\isasymand}\ net{\isacharunderscore}ips\ s\isactrlsub {\isadigit{1}}\ {\isasyminter}\ net{\isacharunderscore}ips\ s\isactrlsub {\isadigit{2}}\ {\isacharequal}\ {\isasymemptyset}{\isacharbraceright},\ trans\ {\isacharequal}\ \opnetsos\ {\isacharparenleft}trans\ {\isacharparenleft}opnet\ onp\ p\isactrlsub {\isadigit{1}}{\isacharparenright}{\isacharparenright}\ {\isacharparenleft}trans\ {\isacharparenleft}opnet\ onp\ p\isactrlsub {\isadigit{2}}{\isacharparenright}{\isacharparenright}{\isasymrparr}}}

\snip{ocnet_connect}{0}{0}{\isa{\mbox{}\inferrule{\mbox{{\isacharparenleft}{\isacharparenleft}{\isasymsigma},\ s{\isacharparenright},\ connect{\isacharparenleft}i,\ i{\isacharprime}{\isacharparenright},\ {\isasymsigma}{\isacharprime},\ s{\isacharprime}{\isacharparenright}{\isasymin}S}\\\ \mbox{{\isasymforall}j{\isachardot}\ j\ {\isasymnotin}\ net{\isacharunderscore}ips\ s\ {\isasymlongrightarrow}\ {\isasymsigma}{\isacharprime}\ j\ {\isacharequal}\ {\isasymsigma}\ j}}{\mbox{{\isacharparenleft}{\isacharparenleft}{\isasymsigma},\ s{\isacharparenright},\ connect{\isacharparenleft}i,\ i{\isacharprime}{\isacharparenright},\ {\isasymsigma}{\isacharprime},\ s{\isacharprime}{\isacharparenright}{\isasymin}\ocnetsos\ S}}}}
\snip{ocnet_disconnect}{0}{0}{\isa{\mbox{}\inferrule{\mbox{{\isacharparenleft}{\isacharparenleft}{\isasymsigma},\ s{\isacharparenright},\ disconnect{\isacharparenleft}i,\ i{\isacharprime}{\isacharparenright},\ {\isasymsigma}{\isacharprime},\ s{\isacharprime}{\isacharparenright}{\isasymin}S}\\\ \mbox{{\isasymforall}j{\isachardot}\ j\ {\isasymnotin}\ net{\isacharunderscore}ips\ s\ {\isasymlongrightarrow}\ {\isasymsigma}{\isacharprime}\ j\ {\isacharequal}\ {\isasymsigma}\ j}}{\mbox{{\isacharparenleft}{\isacharparenleft}{\isasymsigma},\ s{\isacharparenright},\ disconnect{\isacharparenleft}i,\ i{\isacharprime}{\isacharparenright},\ {\isasymsigma}{\isacharprime},\ s{\isacharprime}{\isacharparenright}{\isasymin}\ocnetsos\ S}}}}
\snip{ocnet_cast}{0}{0}{\isa{\mbox{}\inferrule{\mbox{{\isacharparenleft}{\isacharparenleft}{\isasymsigma},\ s{\isacharparenright},\ R{\isacharcolon}{\isacharasterisk}cast{\isacharparenleft}m{\isacharparenright},\ {\isasymsigma}{\isacharprime},\ s{\isacharprime}{\isacharparenright}{\isasymin}S}\\\ \mbox{{\isasymforall}j{\isachardot}\ j\ {\isasymnotin}\ net{\isacharunderscore}ips\ s\ {\isasymlongrightarrow}\ {\isasymsigma}{\isacharprime}\ j\ {\isacharequal}\ {\isasymsigma}\ j}}{\mbox{{\isacharparenleft}{\isacharparenleft}{\isasymsigma},\ s{\isacharparenright},\ {\isasymtau},\ {\isasymsigma}{\isacharprime},\ s{\isacharprime}{\isacharparenright}{\isasymin}\ocnetsos\ S}}}}
\snip{ocnet_tau}{0}{0}{\isa{\mbox{}\inferrule{\mbox{{\isacharparenleft}{\isacharparenleft}{\isasymsigma},\ s{\isacharparenright},\ {\isasymtau},\ {\isasymsigma}{\isacharprime},\ s{\isacharprime}{\isacharparenright}{\isasymin}S}\\\ \mbox{{\isasymforall}j{\isachardot}\ j\ {\isasymnotin}\ net{\isacharunderscore}ips\ s\ {\isasymlongrightarrow}\ {\isasymsigma}{\isacharprime}\ j\ {\isacharequal}\ {\isasymsigma}\ j}}{\mbox{{\isacharparenleft}{\isacharparenleft}{\isasymsigma},\ s{\isacharparenright},\ {\isasymtau},\ {\isasymsigma}{\isacharprime},\ s{\isacharprime}{\isacharparenright}{\isasymin}\ocnetsos\ S}}}}
\snip{ocnet_deliver}{0}{0}{\isa{\mbox{}\inferrule{\mbox{{\isacharparenleft}{\isacharparenleft}{\isasymsigma},\ s{\isacharparenright},\ i{\isacharcolon}deliver{\isacharparenleft}d{\isacharparenright},\ {\isasymsigma}{\isacharprime},\ s{\isacharprime}{\isacharparenright}{\isasymin}S}\\\ \mbox{{\isasymforall}j{\isachardot}\ j\ {\isasymnotin}\ net{\isacharunderscore}ips\ s\ {\isasymlongrightarrow}\ {\isasymsigma}{\isacharprime}\ j\ {\isacharequal}\ {\isasymsigma}\ j}}{\mbox{{\isacharparenleft}{\isacharparenleft}{\isasymsigma},\ s{\isacharparenright},\ i{\isacharcolon}deliver{\isacharparenleft}d{\isacharparenright},\ {\isasymsigma}{\isacharprime},\ s{\isacharprime}{\isacharparenright}{\isasymin}\ocnetsos\ S}}}}
\snip{ocnet_newpkt}{0}{0}{\isa{\mbox{}\inferrule{\mbox{{\isacharparenleft}{\isacharparenleft}{\isasymsigma},\ s{\isacharparenright},\ {\isacharbraceleft}i{\isacharbraceright}{\isasymnot}K{\isacharcolon}arrive{\isacharparenleft}newpkt\ {\isacharparenleft}d,\ di{\isacharparenright}{\isacharparenright},\ {\isasymsigma}{\isacharprime},\ s{\isacharprime}{\isacharparenright}{\isasymin}S}\\\ \mbox{{\isasymforall}j{\isachardot}\ j\ {\isasymnotin}\ net{\isacharunderscore}ips\ s\ {\isasymlongrightarrow}\ {\isasymsigma}{\isacharprime}\ j\ {\isacharequal}\ {\isasymsigma}\ j}}{\mbox{{\isacharparenleft}{\isacharparenleft}{\isasymsigma},\ s{\isacharparenright},\ i{\isacharcolon}newpkt{\isacharparenleft}d,\ di{\isacharparenright},\ {\isasymsigma}{\isacharprime},\ s{\isacharprime}{\isacharparenright}{\isasymin}\ocnetsos\ S}}}}

\snip{oclosed}{0}{0}{\isa{oclosed\ {\isacharparenleft}opnet\ {\isacharparenleft}{\isasymlambda}i{\isachardot}\ opaodv\ i\ {\isasymlangle}{\isasymlangle}\isactrlbsub i\isactrlesub \ qmsg{\isacharparenright}\ n{\isacharparenright}}}
\snip{oclosed_withtype}{0}{0}{\isa{oclosed\ {\isacharparenleft}opnet\ {\isacharparenleft}{\isasymlambda}i{\isachardot}\ opaodv\ i\ {\isasymlangle}{\isasymlangle}\isactrlbsub i\isactrlesub \ qmsg{\isacharparenright}\ n{\isacharparenright}{\isasymColon}{\isacharparenleft}{\isacharparenleft}nat\ {\isasymRightarrow}\ state{\isacharparenright}\ {\isasymtimes}\ {\isacharparenleft}{\isacharparenleft}state,\ msg,\ pseqp,\ pseqp\ label{\isacharparenright}\ seqp\ {\isasymtimes}\ msg\ list\ {\isasymtimes}\ {\isacharparenleft}msg\ list,\ msg,\ unit,\ unit\ label{\isacharparenright}\ seqp{\isacharparenright}\ net{\isacharunderscore}state,\ msg\ node{\isacharunderscore}action{\isacharparenright}\ automaton}}
\snip{microstep_choiceI1}{0}{0}{\isa{{\isacharparenleft}p\isactrlsub {\isadigit{1}}\ {\isasymoplus}\ p\isactrlsub {\isadigit{2}}{\isacharparenright}\ {\isasymleadsto}\isactrlbsub {\isasymGamma}\isactrlesub \ p\isactrlsub {\isadigit{1}}}}
\snip{microstep_choiceI2}{0}{0}{\isa{{\isacharparenleft}p\isactrlsub {\isadigit{1}}\ {\isasymoplus}\ p\isactrlsub {\isadigit{2}}{\isacharparenright}\ {\isasymleadsto}\isactrlbsub {\isasymGamma}\isactrlesub \ p\isactrlsub {\isadigit{2}}}}
\snip{microstep_callI}{0}{0}{\isa{{\isacharparenleft}call{\isacharparenleft}pn{\isacharparenright}{\isacharparenright}\ {\isasymleadsto}\isactrlbsub {\isasymGamma}\isactrlesub \ {\isasymGamma}\ pn}}

\snip{wellformed}{0}{0}{\isa{wellformed\ {\isasymGamma}\ {\isacharequal}\ wf\ {\isacharbraceleft}{\isacharparenleft}q,\ p{\isacharparenright}\ {\isacharbar}\ p\ {\isasymleadsto}\isactrlbsub {\isasymGamma}\isactrlesub \ q{\isacharbraceright}}}

\snip{sterms_termination}{0}{0}{\isa{wellformed\ {\isasymGamma}\ {\isasymLongrightarrow}\ sterms{\isacharunderscore}dom\ {\isacharparenleft}{\isasymGamma},\ p{\isacharparenright}}}
\snip{sterms_termination_concl}{0}{0}{\isa{sterms{\isacharunderscore}dom\ {\isacharparenleft}{\isasymGamma},\ p{\isacharparenright}}}

\snip{sterms_choice}{0}{0}{\isa{\mbox{}\inferrule{\mbox{wellformed\ {\isasymGamma}}}{\mbox{sterms\ {\isasymGamma}\ {\isacharparenleft}p\isactrlsub {\isadigit{1}}\ {\isasymoplus}\ p\isactrlsub {\isadigit{2}}{\isacharparenright}\ {\isacharequal}\ sterms\ {\isasymGamma}\ p\isactrlsub {\isadigit{1}}\ {\isasymunion}\ sterms\ {\isasymGamma}\ p\isactrlsub {\isadigit{2}}}}}}

\snip{sterms_call}{0}{0}{\isa{\mbox{}\inferrule{\mbox{wellformed\ {\isasymGamma}}}{\mbox{sterms\ {\isasymGamma}\ {\isacharparenleft}call{\isacharparenleft}pn{\isacharparenright}{\isacharparenright}\ {\isacharequal}\ sterms\ {\isasymGamma}\ {\isacharparenleft}{\isasymGamma}\ pn{\isacharparenright}}}}}
\snip{sterms_other1}{0}{0}{\isa{\mbox{}\inferrule{\mbox{wellformed\ {\isasymGamma}}}{\mbox{sterms\ {\isasymGamma}\ {\isacharparenleft}{\isacharbraceleft}v{\isacharbraceright}{\isasymlangle}va{\isasymrangle}\ vb{\isacharparenright}\ {\isacharequal}\ {\isacharbraceleft}{\isacharbraceleft}v{\isacharbraceright}{\isasymlangle}va{\isasymrangle}\ vb{\isacharbraceright}}}}}
\snip{sterms_other2}{0}{0}{\isa{\mbox{}\inferrule{\mbox{wellformed\ {\isasymGamma}}}{\mbox{sterms\ {\isasymGamma}\ {\isacharparenleft}{\isacharbraceleft}v{\isacharbraceright}{\isasymlbrakk}va{\isasymrbrakk}\ vb{\isacharparenright}\ {\isacharequal}\ {\isacharbraceleft}{\isacharbraceleft}v{\isacharbraceright}{\isasymlbrakk}va{\isasymrbrakk}\ vb{\isacharbraceright}}}}}

\snip{sterms_choice_concl}{0}{0}{\isa{sterms\ {\isasymGamma}\ {\isacharparenleft}p\isactrlsub {\isadigit{1}}\ {\isasymoplus}\ p\isactrlsub {\isadigit{2}}{\isacharparenright}\ {\isacharequal}\ sterms\ {\isasymGamma}\ p\isactrlsub {\isadigit{1}}\ {\isasymunion}\ sterms\ {\isasymGamma}\ p\isactrlsub {\isadigit{2}}}}
\snip{sterms_choice_concl_lhs}{0}{0}{\isa{sterms\ {\isasymGamma}\ {\isacharparenleft}p\isactrlsub {\isadigit{1}}\ {\isasymoplus}\ p\isactrlsub {\isadigit{2}}{\isacharparenright}}}
\snip{sterms_choice_concl_rhs}{0}{0}{\isa{sterms\ {\isasymGamma}\ p\isactrlsub {\isadigit{1}}\ {\isasymunion}\ sterms\ {\isasymGamma}\ p\isactrlsub {\isadigit{2}}}}

\snip{sterms_other}{0}{0}{\isa{sterms\ {\isasymGamma}\ p\ {\isacharequal}\ {\isacharbraceleft}p{\isacharbraceright}}}
\snip{sterms_other_lhs}{0}{0}{\isa{sterms\ {\isasymGamma}\ p}}
\snip{sterms_other_rhs}{0}{0}{\isa{{\isacharbraceleft}p{\isacharbraceright}}}

\snip{sterms_call_concl}{0}{0}{\isa{sterms\ {\isasymGamma}\ {\isacharparenleft}call{\isacharparenleft}pn{\isacharparenright}{\isacharparenright}\ {\isacharequal}\ sterms\ {\isasymGamma}\ {\isacharparenleft}{\isasymGamma}\ pn{\isacharparenright}}}
\snip{sterms_call_concl_lhs}{0}{0}{\isa{sterms\ {\isasymGamma}\ {\isacharparenleft}call{\isacharparenleft}pn{\isacharparenright}{\isacharparenright}}}
\snip{sterms_call_concl_rhs}{0}{0}{\isa{sterms\ {\isasymGamma}\ {\isacharparenleft}{\isasymGamma}\ pn{\isacharparenright}}}

\snip{sterms_other1_concl}{0}{0}{\isa{sterms\ {\isasymGamma}\ {\isacharparenleft}{\isacharbraceleft}v{\isacharbraceright}{\isasymlangle}va{\isasymrangle}\ vb{\isacharparenright}\ {\isacharequal}\ {\isacharbraceleft}{\isacharbraceleft}v{\isacharbraceright}{\isasymlangle}va{\isasymrangle}\ vb{\isacharbraceright}}}
\snip{sterms_other2_concl}{0}{0}{\isa{sterms\ {\isasymGamma}\ {\isacharparenleft}{\isacharbraceleft}v{\isacharbraceright}{\isasymlbrakk}va{\isasymrbrakk}\ vb{\isacharparenright}\ {\isacharequal}\ {\isacharbraceleft}{\isacharbraceleft}v{\isacharbraceright}{\isasymlbrakk}va{\isasymrbrakk}\ vb{\isacharbraceright}}}

\snip{sterms_maximal_microstep}{0}{0}{\isa{\mbox{}\inferrule{\mbox{wellformed\ {\isasymGamma}}}{\mbox{sterms\ {\isasymGamma}\ p\ {\isacharequal}\ {\isacharbraceleft}q\ {\isacharbar}\ p\ {\isasymleadsto}\isactrlbsub {\isasymGamma}\isactrlesub \isactrlsup {\isacharasterisk}\ q\ {\isasymand}\ {\isacharparenleft}{\isasymnexists}q{\isacharprime}{\isachardot}\ q\ {\isasymleadsto}\isactrlbsub {\isasymGamma}\isactrlesub \ q{\isacharprime}{\isacharparenright}{\isacharbraceright}}}}}
\snip{sterms_maximal_microstep_rhs}{0}{0}{\isa{{\isacharbraceleft}q\ {\isacharbar}\ p\ {\isasymleadsto}\isactrlbsub {\isasymGamma}\isactrlesub \isactrlsup {\isacharasterisk}\ q\ {\isasymand}\ {\isacharparenleft}{\isasymnexists}q{\isacharprime}{\isachardot}\ q\ {\isasymleadsto}\isactrlbsub {\isasymGamma}\isactrlesub \ q{\isacharprime}{\isacharparenright}{\isacharbraceright}}}

\snip{dterms_choice}{0}{0}{\isa{\mbox{}\inferrule{\mbox{wellformed\ {\isasymGamma}}}{\mbox{dterms\ {\isasymGamma}\ {\isacharparenleft}p{\isadigit{1}}{\isachardot}{\isadigit{0}}\ {\isasymoplus}\ p{\isadigit{2}}{\isachardot}{\isadigit{0}}{\isacharparenright}\ {\isacharequal}\ dterms\ {\isasymGamma}\ p{\isadigit{1}}{\isachardot}{\isadigit{0}}\ {\isasymunion}\ dterms\ {\isasymGamma}\ p{\isadigit{2}}{\isachardot}{\isadigit{0}}}}}}
\snip{dterms_call}{0}{0}{\isa{\mbox{}\inferrule{\mbox{wellformed\ {\isasymGamma}}}{\mbox{dterms\ {\isasymGamma}\ {\isacharparenleft}call{\isacharparenleft}pn{\isacharparenright}{\isacharparenright}\ {\isacharequal}\ dterms\ {\isasymGamma}\ {\isacharparenleft}{\isasymGamma}\ pn{\isacharparenright}}}}}
\snip{dterms_other1}{0}{0}{\isa{\mbox{}\inferrule{\mbox{wellformed\ {\isasymGamma}}}{\mbox{dterms\ {\isasymGamma}\ {\isacharparenleft}\gray{{\isacharbraceleft}l{\isacharbraceright}}{\isasymlangle}g{\isasymrangle}\ p{\isacharparenright}\ {\isacharequal}\ sterms\ {\isasymGamma}\ p}}}}
\snip{dterms_other2}{0}{0}{\isa{\mbox{}\inferrule{\mbox{wellformed\ {\isasymGamma}}}{\mbox{dterms\ {\isasymGamma}\ {\isacharparenleft}\gray{{\isacharbraceleft}l{\isacharbraceright}}{\isasymlbrakk}u{\isasymrbrakk}\ p{\isacharparenright}\ {\isacharequal}\ sterms\ {\isasymGamma}\ p}}}}

\snip{dterms_choice_concl}{0}{0}{\isa{dterms\ {\isasymGamma}\ {\isacharparenleft}p\isactrlsub {\isadigit{1}}\ {\isasymoplus}\ p\isactrlsub {\isadigit{2}}{\isacharparenright}\ {\isacharequal}\ dterms\ {\isasymGamma}\ p\isactrlsub {\isadigit{1}}\ {\isasymunion}\ dterms\ {\isasymGamma}\ p\isactrlsub {\isadigit{2}}}}
\snip{dterms_choice_concl_lhs}{0}{0}{\isa{dterms\ {\isasymGamma}\ {\isacharparenleft}p\isactrlsub {\isadigit{1}}\ {\isasymoplus}\ p\isactrlsub {\isadigit{2}}{\isacharparenright}}}
\snip{dterms_choice_concl_rhs}{0}{0}{\isa{dterms\ {\isasymGamma}\ p\isactrlsub {\isadigit{1}}\ {\isasymunion}\ dterms\ {\isasymGamma}\ p\isactrlsub {\isadigit{2}}}}

\snip{dterms_call_concl}{0}{0}{\isa{dterms\ {\isasymGamma}\ {\isacharparenleft}call{\isacharparenleft}pn{\isacharparenright}{\isacharparenright}\ {\isacharequal}\ dterms\ {\isasymGamma}\ {\isacharparenleft}{\isasymGamma}\ pn{\isacharparenright}}}
\snip{dterms_call_concl_lhs}{0}{0}{\isa{dterms\ {\isasymGamma}\ {\isacharparenleft}call{\isacharparenleft}pn{\isacharparenright}{\isacharparenright}}}
\snip{dterms_call_concl_rhs}{0}{0}{\isa{dterms\ {\isasymGamma}\ {\isacharparenleft}{\isasymGamma}\ pn{\isacharparenright}}}

\snip{dterms_other1_concl}{0}{0}{\isa{dterms\ {\isasymGamma}\ {\isacharparenleft}\gray{{\isacharbraceleft}l{\isacharbraceright}}{\isasymlangle}g{\isasymrangle}\ p{\isacharparenright}\ {\isacharequal}\ sterms\ {\isasymGamma}\ p}}

\snip{dterms_other2_concl}{0}{0}{\isa{dterms\ {\isasymGamma}\ {\isacharparenleft}\gray{{\isacharbraceleft}l{\isacharbraceright}}{\isasymlbrakk}u{\isasymrbrakk}\ p{\isacharparenright}\ {\isacharequal}\ sterms\ {\isasymGamma}\ p}}
\snip{dterms_other2_concl_lhs}{0}{0}{\isa{dterms\ {\isasymGamma}\ {\isacharparenleft}\gray{{\isacharbraceleft}l{\isacharbraceright}}{\isasymlbrakk}u{\isasymrbrakk}\ p{\isacharparenright}}}
\snip{dterms_other2_concl_rhs}{0}{0}{\isa{sterms\ {\isasymGamma}\ p}}

\snip{dterms_unicast_concl}{0}{0}{\isa{dterms\ {\isasymGamma}\ {\isacharparenleft}\gray{{\isacharbraceleft}l{\isacharbraceright}}unicast{\isacharparenleft}\selip,\ \selmsg{\isacharparenright}\ {\isachardot}\ p\ {\isasymtriangleright}\ q{\isacharparenright}\ {\isacharequal}\ sterms\ {\isasymGamma}\ p\ {\isasymunion}\ sterms\ {\isasymGamma}\ q}}
\snip{dterms_unicast_concl_lhs}{0}{0}{\isa{dterms\ {\isasymGamma}\ {\isacharparenleft}\gray{{\isacharbraceleft}l{\isacharbraceright}}unicast{\isacharparenleft}\selip,\ \selmsg{\isacharparenright}\ {\isachardot}\ p\ {\isasymtriangleright}\ q{\isacharparenright}}}
\snip{dterms_unicast_concl_rhs}{0}{0}{\isa{sterms\ {\isasymGamma}\ p\ {\isasymunion}\ sterms\ {\isasymGamma}\ q}}

\snip{dterms}{0}{0}{\isa{wellformed\ {\isasymGamma}\ {\isasymLongrightarrow}\ dterms\ {\isasymGamma}\ {\isacharparenleft}\gray{{\isacharbraceleft}l{\isacharbraceright}}{\isasymlangle}g{\isasymrangle}\ p{\isacharparenright}\ {\isacharequal}\ sterms\ {\isasymGamma}\ p\isasep\isanewline%
wellformed\ {\isasymGamma}\ {\isasymLongrightarrow}\ dterms\ {\isasymGamma}\ {\isacharparenleft}\gray{{\isacharbraceleft}l{\isacharbraceright}}{\isasymlbrakk}u{\isasymrbrakk}\ p{\isacharparenright}\ {\isacharequal}\ sterms\ {\isasymGamma}\ p\isasep\isanewline%
wellformed\ {\isasymGamma}\ {\isasymLongrightarrow}\ dterms\ {\isasymGamma}\ {\isacharparenleft}p{\isadigit{1}}{\isachardot}{\isadigit{0}}\ {\isasymoplus}\ p{\isadigit{2}}{\isachardot}{\isadigit{0}}{\isacharparenright}\ {\isacharequal}\ dterms\ {\isasymGamma}\ p{\isadigit{1}}{\isachardot}{\isadigit{0}}\ {\isasymunion}\ dterms\ {\isasymGamma}\ p{\isadigit{2}}{\isachardot}{\isadigit{0}}\isasep\isanewline%
wellformed\ {\isasymGamma}\ {\isasymLongrightarrow}\ dterms\ {\isasymGamma}\ {\isacharparenleft}\gray{{\isacharbraceleft}l{\isacharbraceright}}unicast{\isacharparenleft}\selip,\ \selmsg{\isacharparenright}\ {\isachardot}\ p\ {\isasymtriangleright}\ q{\isacharparenright}\ {\isacharequal}\ sterms\ {\isasymGamma}\ p\ {\isasymunion}\ sterms\ {\isasymGamma}\ q\isasep\isanewline%
wellformed\ {\isasymGamma}\ {\isasymLongrightarrow}\ dterms\ {\isasymGamma}\ {\isacharparenleft}\gray{{\isacharbraceleft}l{\isacharbraceright}}broadcast{\isacharparenleft}\selmsg{\isacharparenright}\ {\isachardot}\ p{\isacharparenright}\ {\isacharequal}\ sterms\ {\isasymGamma}\ p\isasep\isanewline%
wellformed\ {\isasymGamma}\ {\isasymLongrightarrow}\ dterms\ {\isasymGamma}\ {\isacharparenleft}\gray{{\isacharbraceleft}l{\isacharbraceright}}groupcast{\isacharparenleft}\selips,\ \selmsg{\isacharparenright}\ {\isachardot}\ p{\isacharparenright}\ {\isacharequal}\ sterms\ {\isasymGamma}\ p\isasep\isanewline%
wellformed\ {\isasymGamma}\ {\isasymLongrightarrow}\ dterms\ {\isasymGamma}\ {\isacharparenleft}\gray{{\isacharbraceleft}l{\isacharbraceright}}send{\isacharparenleft}\selmsg{\isacharparenright}\ {\isachardot}\ p{\isacharparenright}\ {\isacharequal}\ sterms\ {\isasymGamma}\ p\isasep\isanewline%
wellformed\ {\isasymGamma}\ {\isasymLongrightarrow}\ dterms\ {\isasymGamma}\ {\isacharparenleft}\gray{{\isacharbraceleft}l{\isacharbraceright}}deliver{\isacharparenleft}\seldata{\isacharparenright}\ {\isachardot}\ p{\isacharparenright}\ {\isacharequal}\ sterms\ {\isasymGamma}\ p\isasep\isanewline%
wellformed\ {\isasymGamma}\ {\isasymLongrightarrow}\ dterms\ {\isasymGamma}\ {\isacharparenleft}\gray{{\isacharbraceleft}l{\isacharbraceright}}receive{\isacharparenleft}\updmsg{\isacharparenright}\ {\isachardot}\ p{\isacharparenright}\ {\isacharequal}\ sterms\ {\isasymGamma}\ p\isasep\isanewline%
wellformed\ {\isasymGamma}\ {\isasymLongrightarrow}\ dterms\ {\isasymGamma}\ {\isacharparenleft}call{\isacharparenleft}pn{\isacharparenright}{\isacharparenright}\ {\isacharequal}\ dterms\ {\isasymGamma}\ {\isacharparenleft}{\isasymGamma}\ pn{\isacharparenright}}}

\snip{wf_no_direct_calls}{0}{0}{\isa{\mbox{}\inferrule{\mbox{{\isasymAnd}pn{\isachardot}\ {\isasymforall}pn{\isacharprime}{\isachardot}\ call{\isacharparenleft}pn{\isacharprime}{\isacharparenright}\ {\isasymnotin}\ stermsl\ {\isacharparenleft}{\isasymGamma}\ pn{\isacharparenright}}}{\mbox{wellformed\ {\isasymGamma}}}}}
\snip{wf_no_direct_calls_prem_1}{0}{0}{\isa{call{\isacharparenleft}pn{\isacharprime}{\isacharparenright}\ {\isasymnotin}\ stermsl\ {\isacharparenleft}{\isasymGamma}\ pn{\isacharparenright}}}

\snip{cterms_SI}{0}{0}{\isa{\mbox{}\inferrule{\mbox{p{\isasymin}sterms\ {\isasymGamma}\ {\isacharparenleft}{\isasymGamma}\ pn{\isacharparenright}}}{\mbox{p{\isasymin}cterms\ {\isasymGamma}}}}}
\snip{cterms_DI}{0}{0}{\isa{\mbox{}\inferrule{\mbox{pp{\isasymin}cterms\ {\isasymGamma}}\\\ \mbox{p{\isasymin}dterms\ {\isasymGamma}\ pp}}{\mbox{p{\isasymin}cterms\ {\isasymGamma}}}}}

\snip{cterms_def'}{0}{0}{\isa{\mbox{}\inferrule{\mbox{wellformed\ {\isasymGamma}}}{\mbox{cterms\ {\isasymGamma}\ {\isacharequal}\ {\isacharbraceleft}p\ {\isacharbar}\ {\isasymexists}pn{\isachardot}\ p{\isasymin}ctermsl\ {\isacharparenleft}{\isasymGamma}\ pn{\isacharparenright}\ {\isasymand}\ not{\isacharunderscore}call\ p{\isacharbraceright}}}}}
\snip{cterms_def'_concl}{0}{0}{\isa{cterms\ {\isasymGamma}\ {\isacharequal}\ {\isacharbraceleft}p\ {\isacharbar}\ {\isasymexists}pn{\isachardot}\ p{\isasymin}ctermsl\ {\isacharparenleft}{\isasymGamma}\ pn{\isacharparenright}\ {\isasymand}\ not{\isacharunderscore}call\ p{\isacharbraceright}}}

\snip{cterms_subterms}{0}{0}{\isa{\mbox{}\inferrule{\mbox{wellformed\ {\isasymGamma}}}{\mbox{cterms\ {\isasymGamma}\ {\isacharequal}\ {\isacharbraceleft}p\ {\isacharbar}\ {\isasymexists}pn{\isachardot}\ p{\isasymin}subterms\ {\isacharparenleft}{\isasymGamma}\ pn{\isacharparenright}\ {\isasymand}\ not{\isacharunderscore}call\ p\ {\isasymand}\ not{\isacharunderscore}choice\ p{\isacharbraceright}}}}}  
\snip{cterms_subterms_concl}{0}{0}{\isa{cterms\ {\isasymGamma}\ {\isacharequal}\ {\isacharbraceleft}p\ {\isacharbar}\ {\isasymexists}pn{\isachardot}\ p{\isasymin}subterms\ {\isacharparenleft}{\isasymGamma}\ pn{\isacharparenright}\ {\isasymand}\ not{\isacharunderscore}call\ p\ {\isasymand}\ not{\isacharunderscore}choice\ p{\isacharbraceright}}}

\snip{ctermsl_choice}{0}{0}{\isa{ctermsl\ {\isacharparenleft}p\isactrlsub {\isadigit{1}}\ {\isasymoplus}\ p\isactrlsub {\isadigit{2}}{\isacharparenright}\ {\isacharequal}\ ctermsl\ p\isactrlsub {\isadigit{1}}\ {\isasymunion}\ ctermsl\ p\isactrlsub {\isadigit{2}}}}
\snip{ctermsl_choice_lhs}{0}{0}{\isa{ctermsl\ {\isacharparenleft}p\isactrlsub {\isadigit{1}}\ {\isasymoplus}\ p\isactrlsub {\isadigit{2}}{\isacharparenright}}}
\snip{ctermsl_choice_rhs}{0}{0}{\isa{ctermsl\ p\isactrlsub {\isadigit{1}}\ {\isasymunion}\ ctermsl\ p\isactrlsub {\isadigit{2}}}}

\snip{ctermsl_call}{0}{0}{\isa{ctermsl\ {\isacharparenleft}call{\isacharparenleft}pn{\isacharparenright}{\isacharparenright}\ {\isacharequal}\ {\isacharbraceleft}call{\isacharparenleft}pn{\isacharparenright}{\isacharbraceright}}}
\snip{ctermsl_call_lhs}{0}{0}{\isa{ctermsl\ {\isacharparenleft}call{\isacharparenleft}pn{\isacharparenright}{\isacharparenright}}}
\snip{ctermsl_call_rhs}{0}{0}{\isa{{\isacharbraceleft}call{\isacharparenleft}pn{\isacharparenright}{\isacharbraceright}}}

\snip{ctermsl_other2}{0}{0}{\isa{ctermsl\ {\isacharparenleft}\gray{{\isacharbraceleft}l{\isacharbraceright}}{\isasymlbrakk}u{\isasymrbrakk}\ p{\isacharparenright}\ {\isacharequal}\ {\isacharbraceleft}{\isacharbraceleft}l{\isacharbraceright}{\isasymlbrakk}u{\isasymrbrakk}\ p{\isacharbraceright}\ {\isasymunion}\ ctermsl\ p}}
\snip{ctermsl_other2_lhs}{0}{0}{\isa{ctermsl\ {\isacharparenleft}\gray{{\isacharbraceleft}l{\isacharbraceright}}{\isasymlbrakk}u{\isasymrbrakk}\ p{\isacharparenright}}}
\snip{ctermsl_other2_rhs}{0}{0}{\isa{{\isacharbraceleft}\gray{{\isacharbraceleft}l{\isacharbraceright}}{\isasymlbrakk}u{\isasymrbrakk}\ p{\isacharbraceright}\ {\isasymunion}\ ctermsl\ p}}

\snip{ctermsl_unicast}{0}{0}{\isa{ctermsl\ {\isacharparenleft}\gray{{\isacharbraceleft}l{\isacharbraceright}}broadcast{\isacharparenleft}\selmsg{\isacharparenright}\ {\isachardot}\ p{\isacharparenright}\ {\isacharequal}\ {\isacharbraceleft}{\isacharbraceleft}l{\isacharbraceright}broadcast{\isacharparenleft}\selmsg{\isacharparenright}\ {\isachardot}\ p{\isacharbraceright}\ {\isasymunion}\ ctermsl\ p}}

\snip{stermsl_ctermsl}{0}{0}{\isa{\mbox{}\inferrule{\mbox{q{\isasymin}stermsl\ p}}{\mbox{q{\isasymin}ctermsl\ p}}}}
\snip{stermsl_ctermsl_prem}{0}{0}{\isa{q{\isasymin}stermsl\ p}}
\snip{stermsl_ctermsl_concl}{0}{0}{\isa{q{\isasymin}ctermsl\ p}}

\snip{stermsl}{0}{0}{\isa{stermsl}}
\snip{stermsl_choice}{0}{0}{\isa{stermsl\ {\isacharparenleft}p\isactrlsub {\isadigit{1}}\ {\isasymoplus}\ p\isactrlsub {\isadigit{2}}{\isacharparenright}\ {\isacharequal}\ stermsl\ p\isactrlsub {\isadigit{1}}\ {\isasymunion}\ stermsl\ p\isactrlsub {\isadigit{2}}}}
\snip{stermsl_other}{0}{0}{\isa{stermsl\ p\ {\isacharequal}\ {\isacharbraceleft}p{\isacharbraceright}}}

\snip{seq_reachable_in_cterms_prem_1}{0}{0}{\isa{wellformed\ {\isasymGamma}}}
\snip{seq_reachable_in_cterms_prem_2}{0}{0}{\isa{control{\isacharunderscore}within\ {\isasymGamma}\ {\isacharparenleft}init\ A{\isacharparenright}}}
\snip{seq_reachable_in_cterms_prem_3}{0}{0}{\isa{trans\ A\ {\isacharequal}\ \seqpsos\ {\isasymGamma}}}
\snip{seq_reachable_in_cterms_prem_4}{0}{0}{\isa{{\isacharparenleft}{\isasymxi},\ p{\isacharparenright}{\isasymin}reachable\ A\ I}}
\snip{seq_reachable_in_cterms_prem_5}{0}{0}{\isa{q{\isasymin}sterms\ {\isasymGamma}\ p}}
\snip{seq_reachable_in_cterms_concl}{0}{0}{\isa{q{\isasymin}cterms\ {\isasymGamma}}}

\snip{control_within}{0}{0}{\isa{control{\isacharunderscore}within\ {\isasymGamma}\ {\isasymsigma}\ {\isacharequal}\ {\isasymforall}{\isacharparenleft}{\isasymxi},\ p{\isacharparenright}{\isasymin}{\isasymsigma}{\isachardot}\ {\isasymexists}pn{\isachardot}\ p{\isasymin}subterms\ {\isacharparenleft}{\isasymGamma}\ pn{\isacharparenright}}}
\snip{simple_labels}{0}{0}{\isa{simple{\isacharunderscore}labels\ {\isasymGamma}\ {\isacharequal}\ {\isasymforall}pn{\isachardot}\ {\isasymforall}p{\isasymin}subterms\ {\isacharparenleft}{\isasymGamma}\ pn{\isacharparenright}{\isachardot}\ {\isasymexists}\hspace{-.2em}{\isacharbang}\hspace{.15em}l{\isachardot}\ labels\ {\isasymGamma}\ p\ {\isacharequal}\ \gray{{\isacharbraceleft}l{\isacharbraceright}}}}

\snip{seq_invariant_ctermsI_prem_1}{0}{0}{\isa{wellformed\ {\isasymGamma}}}
\snip{seq_invariant_ctermsI_prem_2}{0}{0}{\isa{control{\isacharunderscore}within\ {\isasymGamma}\ {\isacharparenleft}init\ A{\isacharparenright}}}
\snip{seq_invariant_ctermsI_prem_3}{0}{0}{\isa{simple{\isacharunderscore}labels\ {\isasymGamma}}}
\snip{seq_invariant_ctermsI_prem_4}{0}{0}{\isa{trans\ A\ {\isacharequal}\ \seqpsos\ {\isasymGamma}}}

\snip{seq_invariant_ctermsI_init}{0}{0}{\isa{{\isasymAnd}{\isasymxi}\ p\ l{\isachardot}\ \mbox{}\inferrule{\mbox{{\isacharparenleft}{\isasymxi},\ p{\isacharparenright}{\isasymin}init\ A}\\\ \mbox{l{\isasymin}labels\ {\isasymGamma}\ p}}{\mbox{P\ {\isacharparenleft}{\isasymxi},\ l{\isacharparenright}}}}}
\snip{seq_invariant_ctermsI_init_1}{0}{0}{\isa{{\isacharparenleft}{\isasymxi},\ p{\isacharparenright}{\isasymin}init\ A}}
\snip{seq_invariant_ctermsI_init_2}{0}{0}{\isa{l{\isasymin}labels\ {\isasymGamma}\ p}}
\snip{seq_invariant_ctermsI_init_3}{0}{0}{\isa{P\ {\isacharparenleft}{\isasymxi},\ l{\isacharparenright}}}

\snip{seq_invariant_ctermsI_trans_1}{0}{0}{\isa{p{\isasymin}ctermsl\ {\isacharparenleft}{\isasymGamma}\ pn{\isacharparenright}}}
\snip{seq_invariant_ctermsI_trans_2}{0}{0}{\isa{not{\isacharunderscore}call\ p}}
\snip{seq_invariant_ctermsI_trans_3}{0}{0}{\isa{l{\isasymin}labels\ {\isasymGamma}\ p}}
\snip{seq_invariant_ctermsI_trans_4}{0}{0}{\isa{P\ {\isacharparenleft}{\isasymxi},\ l{\isacharparenright}}}
\snip{seq_invariant_ctermsI_trans_5}{0}{0}{\isa{{\isacharparenleft}{\isacharparenleft}{\isasymxi},\ p{\isacharparenright},\ a,\ {\isacharparenleft}{\isasymxi}{\isacharprime},\ q{\isacharparenright}{\isacharparenright}{\isasymin}\seqpsos\ {\isasymGamma}}}
\snip{seq_invariant_ctermsI_trans_6}{0}{0}{\isa{l{\isacharprime}{\isasymin}labels\ {\isasymGamma}\ q}}
\snip{seq_invariant_ctermsI_trans_7}{0}{0}{\isa{{\isacharparenleft}{\isasymxi},\ pp{\isacharparenright}{\isasymin}reachable\ A\ I}}
\snip{seq_invariant_ctermsI_trans_8}{0}{0}{\isa{p{\isasymin}sterms\ {\isasymGamma}\ pp}}
\snip{seq_invariant_ctermsI_trans_9}{0}{0}{\isa{I\ a}}
\snip{seq_invariant_ctermsI_trans_10}{0}{0}{\isa{P\ {\isacharparenleft}{\isasymxi}{\isacharprime},\ l{\isacharprime}{\isacharparenright}}}

\snip{seq_invariant_ctermsI_concl}{0}{0}{\isa{A\ {\isasymTTurnstile}\ {\isacharparenleft}I\ {\isasymrightarrow}{\isacharparenright}\ onl\ {\isasymGamma}\ P}}

\snip{oseq_invariant_ctermsI_trans_1}{0}{0}{\isa{{\isacharparenleft}{\isasymsigma},\ p{\isacharparenright}{\isasymin}oreachable\ A\ S\ U}}
\snip{oseq_invariant_ctermsI_trans_2}{0}{0}{\isa{l{\isasymin}labels\ {\isasymGamma}\ p}}
\snip{oseq_invariant_ctermsI_trans_3}{0}{0}{\isa{P\ {\isacharparenleft}{\isasymsigma},\ l{\isacharparenright}}}
\snip{oseq_invariant_ctermsI_trans_4}{0}{0}{\isa{U\ {\isasymsigma}\ {\isasymsigma}{\isacharprime}}}
\snip{oseq_invariant_ctermsI_trans_5}{0}{0}{\isa{P\ {\isacharparenleft}{\isasymsigma}{\isacharprime},\ l{\isacharparenright}}}

\snip{oseq_invariant_ctermsI_concl}{0}{0}{\isa{A\ {\isasymTurnstile}\ {\isacharparenleft}S,\ U\ {\isasymrightarrow}{\isacharparenright}\ onl\ {\isasymGamma}\ P}}
\snip{seq_invariant_ctermI}{0}{0}{\begin{isabelle}%
{\isasymlbrakk}wellformed\ {\isasymGamma}{\isacharsemicolon}\ control{\isacharunderscore}within\ {\isasymGamma}\ {\isacharparenleft}init\ A{\isacharparenright}{\isacharsemicolon}\ simple{\isacharunderscore}labels\ {\isasymGamma}{\isacharsemicolon}\isanewline
\isaindent{\ }trans\ A\ {\isacharequal}\ \seqpsos\ {\isasymGamma}{\isacharsemicolon}\isanewline
\isaindent{\ }{\isasymAnd}{\isasymxi}\ p\ l{\isachardot}\ {\isasymlbrakk}{\isacharparenleft}{\isasymxi},\ p{\isacharparenright}{\isasymin}init\ A{\isacharsemicolon}\ l{\isasymin}labels\ {\isasymGamma}\ p{\isasymrbrakk}\ {\isasymLongrightarrow}\ P\ {\isacharparenleft}{\isasymxi},\ l{\isacharparenright}{\isacharsemicolon}\isanewline
\isaindent{\ }{\isasymAnd}p\ l\ {\isasymxi}\ a\ q\ l{\isacharprime}\ {\isasymxi}{\isacharprime}\ pp{\isachardot}\isanewline
\isaindent{\ \ \ \ }{\isasymlbrakk}p{\isasymin}cterms\ {\isasymGamma}{\isacharsemicolon}\ l{\isasymin}labels\ {\isasymGamma}\ p{\isacharsemicolon}\ P\ {\isacharparenleft}{\isasymxi},\ l{\isacharparenright}{\isacharsemicolon}\isanewline
\isaindent{\ \ \ \ \ }{\isacharparenleft}{\isacharparenleft}{\isasymxi},\ p{\isacharparenright},\ a,\ {\isacharparenleft}{\isasymxi}{\isacharprime},\ q{\isacharparenright}{\isacharparenright}{\isasymin}\seqpsos\ {\isasymGamma}{\isacharsemicolon}\isanewline
\isaindent{\ \ \ \ \ }{\isacharparenleft}{\isacharparenleft}{\isasymxi},\ p{\isacharparenright},\ a,\ {\isasymxi}{\isacharprime},\ q{\isacharparenright}{\isasymin}trans\ A{\isacharsemicolon}\ l{\isacharprime}{\isasymin}labels\ {\isasymGamma}\ q{\isacharsemicolon}\isanewline
\isaindent{\ \ \ \ \ }{\isacharparenleft}{\isasymxi},\ pp{\isacharparenright}{\isasymin}reachable\ A\ I{\isacharsemicolon}\ p{\isasymin}sterms\ {\isasymGamma}\ pp{\isacharsemicolon}\ {\isacharparenleft}{\isasymxi}{\isacharprime},\ q{\isacharparenright}{\isasymin}reachable\ A\ I{\isacharsemicolon}\ I\ a{\isasymrbrakk}\isanewline
\isaindent{\ \ \ \ }{\isasymLongrightarrow}\ P\ {\isacharparenleft}{\isasymxi}{\isacharprime},\ l{\isacharprime}{\isacharparenright}{\isasymrbrakk}\isanewline
{\isasymLongrightarrow}\ A\ {\isasymTTurnstile}\ {\isacharparenleft}I\ {\isasymrightarrow}{\isacharparenright}\ onl\ {\isasymGamma}\ P%
\end{isabelle}}
\snip{seq_invariant_ctermI_wf}{0}{0}{\isa{wellformed\ {\isasymGamma}}}
\snip{seq_invariant_ctermI_cw}{0}{0}{\isa{control{\isacharunderscore}within\ {\isasymGamma}\ {\isacharparenleft}init\ A{\isacharparenright}}}
\snip{seq_invariant_ctermI_sl}{0}{0}{\isa{simple{\isacharunderscore}labels\ {\isasymGamma}}}
\snip{seq_invariant_ctermI_init}{0}{0}{\begin{isabelle}%
trans\ A\ {\isacharequal}\ \seqpsos\ {\isasymGamma}%
\end{isabelle}}
\snip{seq_invariant_ctermI_step}{0}{0}{\begin{isabelle}%
{\isasymAnd}{\isasymxi}\ p\ l{\isachardot}\ {\isasymlbrakk}{\isacharparenleft}{\isasymxi},\ p{\isacharparenright}{\isasymin}init\ A{\isacharsemicolon}\ l{\isasymin}labels\ {\isasymGamma}\ p{\isasymrbrakk}\ {\isasymLongrightarrow}\ P\ {\isacharparenleft}{\isasymxi},\ l{\isacharparenright}%
\end{isabelle}}
\snip{seq_invariant_ctermI_concl}{0}{0}{\isa{A\ {\isasymTTurnstile}\ {\isacharparenleft}I\ {\isasymrightarrow}{\isacharparenright}\ onl\ {\isasymGamma}\ P}}

\snip{seq_reachable_in_cterms}{0}{0}{\isa{\mbox{}\inferrule{\mbox{wellformed\ {\isasymGamma}}\\\ \mbox{control{\isacharunderscore}within\ {\isasymGamma}\ {\isacharparenleft}init\ A{\isacharparenright}}\\\ \mbox{trans\ A\ {\isacharequal}\ \seqpsos\ {\isasymGamma}}\\\ \mbox{{\isacharparenleft}{\isasymxi},\ p{\isacharparenright}{\isasymin}reachable\ A\ I}\\\ \mbox{p{\isacharprime}{\isasymin}sterms\ {\isasymGamma}\ p}}{\mbox{p{\isacharprime}{\isasymin}cterms\ {\isasymGamma}}}}}
\snip{open_seq_step_invariant}{0}{0}{\isa{\mbox{}\inferrule{\mbox{A\ \stepinv\ {\isacharparenleft}I\ {\isasymrightarrow}{\isacharparenright}\ P}\\\ \mbox{initiali\ i\ {\isacharparenleft}init\ OA{\isacharparenright}\ {\isacharparenleft}init\ A{\isacharparenright}}\\\ \mbox{trans\ OA\ {\isacharequal}\ \oseqpsos\ {\isasymGamma}\ i}\\\ \mbox{trans\ A\ {\isacharequal}\ \seqpsos\ {\isasymGamma}}}{\mbox{OA\ \ostepinv\ {\isacharparenleft}act\ I,\ other\ ANY\ {\isacharbraceleft}i{\isacharbraceright}\ {\isasymrightarrow}{\isacharparenright}\ seqll\ i\ P}}}}

\snip{open_seq_invariant}{0}{0}{\isa{\mbox{}\inferrule{\mbox{A\ {\isasymTTurnstile}\ {\isacharparenleft}I\ {\isasymrightarrow}{\isacharparenright}\ P}\\\ \mbox{initiali\ i\ {\isacharparenleft}init\ OA{\isacharparenright}\ {\isacharparenleft}init\ A{\isacharparenright}}\\\ \mbox{trans\ OA\ {\isacharequal}\ \oseqpsos\ {\isasymGamma}\ i}\\\ \mbox{trans\ A\ {\isacharequal}\ \seqpsos\ {\isasymGamma}}}{\mbox{OA\ {\isasymTurnstile}\ {\isacharparenleft}act\ I,\ other\ ANY\ {\isacharbraceleft}i{\isacharbraceright}\ {\isasymrightarrow}{\isacharparenright}\ seql\ i\ P}}}}
\snip{open_seq_invariant_prem_1}{0}{0}{\isa{A\ {\isasymTTurnstile}\ {\isacharparenleft}I\ {\isasymrightarrow}{\isacharparenright}\ P}}
\snip{open_seq_invariant_prem_2}{0}{0}{\isa{initiali\ i\ {\isacharparenleft}init\ A{\isacharprime}{\isacharparenright}\ {\isacharparenleft}init\ A{\isacharparenright}}}
\snip{open_seq_invariant_prem_2'}{0}{0}{\isa{init\ A\ {\isacharequal}\ {\isacharbraceleft}{\isacharparenleft}{\isasymsigma}\ i,\ p{\isacharparenright}\ {\isacharbar}\ {\isacharparenleft}{\isasymsigma},\ p{\isacharparenright}{\isasymin}init\ A{\isacharprime}{\isacharbraceright}}}
\snip{open_seq_invariant_prem_3}{0}{0}{\isa{trans\ A{\isacharprime}\ {\isacharequal}\ \oseqpsos\ {\isasymGamma}\ i}}
\snip{open_seq_invariant_prem_4}{0}{0}{\isa{trans\ A\ {\isacharequal}\ \seqpsos\ {\isasymGamma}}}
\snip{open_seq_invariant_concl}{0}{0}{\isa{A{\isacharprime}\ {\isasymTurnstile}\ {\isacharparenleft}act\ I,\ other\ U\ {\isacharbraceleft}i{\isacharbraceright}\ {\isasymrightarrow}{\isacharparenright}\ seql\ i\ P}}
\snip{open_seq_invariant_concl'}{0}{0}{\isa{A{\isacharprime}\ {\isasymTurnstile}\ {\isacharparenleft}{\isasymlambda}{\isacharunderscore}\ {\isacharunderscore}{\isachardot}\ I,\ other\ U\ {\isacharbraceleft}i{\isacharbraceright}\ {\isasymrightarrow}{\isacharparenright}\ {\isacharparenleft}{\isasymlambda}{\isacharparenleft}{\isasymsigma},\ p{\isacharparenright}{\isachardot}\ P\ {\isacharparenleft}{\isasymsigma}\ i,\ p{\isacharparenright}{\isacharparenright}}}

\snip{oinvariant}{0}{0}{\begin{isabelle}%
{\isacharparenleft}A\ {\isasymTurnstile}\ {\isacharparenleft}S,\ U\ {\isasymrightarrow}{\isacharparenright}\ P{\isacharparenright}\ {\isacharequal}\ {\isacharparenleft}{\isasymforall}s{\isasymin}oreachable\ A\ S\ U{\isachardot}\ P\ s{\isacharparenright}%
\end{isabelle}}
\snip{oinvariant_lhs}{0}{0}{\isa{A\ {\isasymTurnstile}\ {\isacharparenleft}S,\ U\ {\isasymrightarrow}{\isacharparenright}\ P}}
\snip{oinvariant_rhs}{0}{0}{\isa{{\isasymforall}s{\isasymin}oreachable\ A\ S\ U{\isachardot}\ P\ s}}
\snip{ostep_invariant}{0}{0}{\begin{isabelle}%
{\isacharparenleft}A\ \ostepinv\ {\isacharparenleft}S,\ U\ {\isasymrightarrow}{\isacharparenright}\ P{\isacharparenright}\ {\isacharequal}\isanewline
{\isacharparenleft}{\isasymforall}s{\isasymin}oreachable\ A\ S\ U{\isachardot}\isanewline
\isaindent{{\isacharparenleft}\ \ \ }{\isasymforall}a\ s{\isacharprime}{\isachardot}\isanewline
\isaindent{{\isacharparenleft}\ \ \ \ \ \ }{\isacharparenleft}s,\ a,\ s{\isacharprime}{\isacharparenright}{\isasymin}trans\ A\ {\isasymand}\ S\ {\isacharparenleft}fst\ s{\isacharparenright}\ {\isacharparenleft}fst\ s{\isacharprime}{\isacharparenright}\ a\ {\isasymlongrightarrow}\isanewline
\isaindent{{\isacharparenleft}\ \ \ \ \ \ }P\ {\isacharparenleft}s,\ a,\ s{\isacharprime}{\isacharparenright}{\isacharparenright}%
\end{isabelle}}
\snip{ostep_invariant_lhs}{0}{0}{\isa{A\ \ostepinv\ {\isacharparenleft}S,\ U\ {\isasymrightarrow}{\isacharparenright}\ P}}
\snip{ostep_invariant_rhs}{0}{0}{\isa{{\isasymforall}s{\isasymin}oreachable\ A\ S\ U{\isachardot}\ {\isasymforall}a\ s{\isacharprime}{\isachardot}\ {\isacharparenleft}s,\ a,\ s{\isacharprime}{\isacharparenright}{\isasymin}trans\ A\ {\isasymand}\ S\ {\isacharparenleft}fst\ s{\isacharparenright}\ {\isacharparenleft}fst\ s{\isacharprime}{\isacharparenright}\ a\ {\isasymlongrightarrow}\ P\ {\isacharparenleft}s,\ a,\ s{\isacharprime}{\isacharparenright}}}

\snip{invariant}{0}{0}{\begin{isabelle}%
{\isacharparenleft}A\ {\isasymTTurnstile}\ {\isacharparenleft}I\ {\isasymrightarrow}{\isacharparenright}\ P{\isacharparenright}\ {\isacharequal}\ {\isacharparenleft}{\isasymforall}s{\isasymin}reachable\ A\ I{\isachardot}\ P\ s{\isacharparenright}%
\end{isabelle}}
\snip{invariant_lhs}{0}{0}{\isa{A\ {\isasymTTurnstile}\ {\isacharparenleft}I\ {\isasymrightarrow}{\isacharparenright}\ P}}
\snip{invariant_rhs}{0}{0}{\isa{{\isasymforall}s{\isasymin}reachable\ A\ I{\isachardot}\ P\ s}}
\snip{invariant_TT_lhs}{0}{0}{\isa{A\ {\isasymTTurnstile}\ P}}

\snip{step_invariant}{0}{0}{\begin{isabelle}%
{\isacharparenleft}A\ \stepinv\ {\isacharparenleft}I\ {\isasymrightarrow}{\isacharparenright}\ P{\isacharparenright}\ {\isacharequal}\isanewline
{\isacharparenleft}{\isasymforall}a{\isachardot}\ I\ a\ {\isasymlongrightarrow}\isanewline
\isaindent{{\isacharparenleft}{\isasymforall}a{\isachardot}\ }{\isacharparenleft}{\isasymforall}s{\isasymin}reachable\ A\ I{\isachardot}\isanewline
\isaindent{{\isacharparenleft}{\isasymforall}a{\isachardot}\ {\isacharparenleft}\ \ \ }{\isasymforall}s{\isacharprime}{\isachardot}\ {\isacharparenleft}s,\ a,\ s{\isacharprime}{\isacharparenright}{\isasymin}trans\ A\ {\isasymlongrightarrow}\ P\ {\isacharparenleft}s,\ a,\ s{\isacharprime}{\isacharparenright}{\isacharparenright}{\isacharparenright}%
\end{isabelle}}
\snip{step_invariant_lhs}{0}{0}{\isa{A\ \stepinv\ {\isacharparenleft}I\ {\isasymrightarrow}{\isacharparenright}\ P}}
\snip{step_invariant_rhs}{0}{0}{\isa{{\isasymforall}a{\isachardot}\ I\ a\ {\isasymlongrightarrow}\ {\isacharparenleft}{\isasymforall}s{\isasymin}reachable\ A\ I{\isachardot}\ {\isasymforall}s{\isacharprime}{\isachardot}\ {\isacharparenleft}s,\ a,\ s{\isacharprime}{\isacharparenright}{\isasymin}trans\ A\ {\isasymlongrightarrow}\ P\ {\isacharparenleft}s,\ a,\ s{\isacharprime}{\isacharparenright}{\isacharparenright}}}

\snip{TT}{0}{0}{\isa{TT\ {\isacharequal}\ {\isacharparenleft}{\isasymlambda}{\isacharunderscore}{\isachardot}\ True{\isacharparenright}}}
\snip{TT_rhs}{0}{0}{\isa{{\isasymlambda}{\isacharunderscore}{\isachardot}\ True}}

\snip{open_closed_invariant}{0}{0}{\isa{\mbox{}\inferrule{\mbox{A\ {\isasymTTurnstile}\ {\isacharparenleft}I\ {\isasymrightarrow}{\isacharparenright}\ P}\\\ \mbox{subreachable\ A\ U\ J}\\\ \mbox{{\isasymAnd}{\isasymsigma}\ {\isasymsigma}{\isacharprime}\ s{\isachardot}\ \mbox{}\inferrule{\mbox{{\isasymforall}j{\isasymin}J{\isachardot}\ {\isasymsigma}{\isacharprime}\ j\ {\isacharequal}\ {\isasymsigma}\ j}\\\ \mbox{P\ {\isacharparenleft}{\isasymsigma}{\isacharprime},\ s{\isacharparenright}}}{\mbox{P\ {\isacharparenleft}{\isasymsigma},\ s{\isacharparenright}}}}}{\mbox{A\ {\isasymTurnstile}\ {\isacharparenleft}act\ I,\ U\ {\isasymrightarrow}{\isacharparenright}\ P}}}}  

\snip{oseq_invariant_ctermI}{0}{0}{\begin{isabelle}%
{\isasymlbrakk}wellformed\ {\isasymGamma}{\isacharsemicolon}\ control{\isacharunderscore}within\ {\isasymGamma}\ {\isacharparenleft}init\ A{\isacharparenright}{\isacharsemicolon}\ simple{\isacharunderscore}labels\ {\isasymGamma}{\isacharsemicolon}\isanewline
\isaindent{\ }trans\ A\ {\isacharequal}\ \oseqpsos\ {\isasymGamma}\ i{\isacharsemicolon}\isanewline
\isaindent{\ }{\isasymAnd}{\isasymsigma}\ p\ l{\isachardot}\ {\isasymlbrakk}{\isacharparenleft}{\isasymsigma},\ p{\isacharparenright}{\isasymin}init\ A{\isacharsemicolon}\ l{\isasymin}labels\ {\isasymGamma}\ p{\isasymrbrakk}\ {\isasymLongrightarrow}\ P\ {\isacharparenleft}{\isasymsigma},\ l{\isacharparenright}{\isacharsemicolon}\isanewline
\isaindent{\ }{\isasymAnd}{\isasymsigma}\ {\isasymsigma}{\isacharprime}\ p\ l{\isachardot}\isanewline
\isaindent{\ \ \ \ }{\isasymlbrakk}{\isacharparenleft}{\isasymsigma},\ p{\isacharparenright}{\isasymin}oreachable\ A\ S\ U{\isacharsemicolon}\ l{\isasymin}labels\ {\isasymGamma}\ p{\isacharsemicolon}\ P\ {\isacharparenleft}{\isasymsigma},\ l{\isacharparenright}{\isacharsemicolon}\ U\ {\isasymsigma}\ {\isasymsigma}{\isacharprime}{\isasymrbrakk}\isanewline
\isaindent{\ \ \ \ }{\isasymLongrightarrow}\ P\ {\isacharparenleft}{\isasymsigma}{\isacharprime},\ l{\isacharparenright}{\isacharsemicolon}\isanewline
\isaindent{\ }{\isasymAnd}p\ l\ {\isasymsigma}\ a\ q\ l{\isacharprime}\ {\isasymsigma}{\isacharprime}\ pp{\isachardot}\isanewline
\isaindent{\ \ \ \ }{\isasymlbrakk}p{\isasymin}cterms\ {\isasymGamma}{\isacharsemicolon}\ l{\isasymin}labels\ {\isasymGamma}\ p{\isacharsemicolon}\ P\ {\isacharparenleft}{\isasymsigma},\ l{\isacharparenright}{\isacharsemicolon}\isanewline
\isaindent{\ \ \ \ \ }{\isacharparenleft}{\isacharparenleft}{\isasymsigma},\ p{\isacharparenright},\ a,\ {\isacharparenleft}{\isasymsigma}{\isacharprime},\ q{\isacharparenright}{\isacharparenright}{\isasymin}\oseqpsos\ {\isasymGamma}\ i{\isacharsemicolon}\isanewline
\isaindent{\ \ \ \ \ }{\isacharparenleft}{\isacharparenleft}{\isasymsigma},\ p{\isacharparenright},\ a,\ {\isasymsigma}{\isacharprime},\ q{\isacharparenright}{\isasymin}trans\ A{\isacharsemicolon}\ l{\isacharprime}{\isasymin}labels\ {\isasymGamma}\ q{\isacharsemicolon}\isanewline
\isaindent{\ \ \ \ \ }{\isacharparenleft}{\isasymsigma},\ pp{\isacharparenright}{\isasymin}oreachable\ A\ S\ U{\isacharsemicolon}\ p{\isasymin}sterms\ {\isasymGamma}\ pp{\isacharsemicolon}\isanewline
\isaindent{\ \ \ \ \ }{\isacharparenleft}{\isasymsigma}{\isacharprime},\ q{\isacharparenright}{\isasymin}oreachable\ A\ S\ U{\isacharsemicolon}\ S\ {\isasymsigma}\ {\isasymsigma}{\isacharprime}\ a{\isasymrbrakk}\isanewline
\isaindent{\ \ \ \ }{\isasymLongrightarrow}\ P\ {\isacharparenleft}{\isasymsigma}{\isacharprime},\ l{\isacharprime}{\isacharparenright}{\isasymrbrakk}\isanewline
{\isasymLongrightarrow}\ A\ {\isasymTurnstile}\ {\isacharparenleft}S,\ U\ {\isasymrightarrow}{\isacharparenright}\ onl\ {\isasymGamma}\ P%
\end{isabelle}}
\snip{oseq_invariant_ctermI_wf}{0}{0}{\isa{wellformed\ {\isasymGamma}}}
\snip{oseq_invariant_ctermI_cw}{0}{0}{\isa{control{\isacharunderscore}within\ {\isasymGamma}\ {\isacharparenleft}init\ A{\isacharparenright}}}
\snip{oseq_invariant_ctermI_sl}{0}{0}{\isa{simple{\isacharunderscore}labels\ {\isasymGamma}}}
\snip{oseq_invariant_ctermI_init}{0}{0}{\begin{isabelle}%
trans\ A\ {\isacharequal}\ \oseqpsos\ {\isasymGamma}\ i%
\end{isabelle}}
\snip{oseq_invariant_ctermI_other}{0}{0}{\isa{{\isasymAnd}{\isasymsigma}\ p\ l{\isachardot}\ {\isasymlbrakk}{\isacharparenleft}{\isasymsigma},\ p{\isacharparenright}{\isasymin}init\ A{\isacharsemicolon}\ l{\isasymin}labels\ {\isasymGamma}\ p{\isasymrbrakk}\ {\isasymLongrightarrow}\ P\ {\isacharparenleft}{\isasymsigma},\ l{\isacharparenright}}}
\snip{oseq_invariant_ctermI_step}{0}{0}{\begin{isabelle}%
{\isasymAnd}{\isasymsigma}\ {\isasymsigma}{\isacharprime}\ p\ l{\isachardot}\isanewline
\isaindent{\ \ \ }{\isasymlbrakk}{\isacharparenleft}{\isasymsigma},\ p{\isacharparenright}{\isasymin}oreachable\ A\ S\ U{\isacharsemicolon}\ l{\isasymin}labels\ {\isasymGamma}\ p{\isacharsemicolon}\ P\ {\isacharparenleft}{\isasymsigma},\ l{\isacharparenright}{\isacharsemicolon}\ U\ {\isasymsigma}\ {\isasymsigma}{\isacharprime}{\isasymrbrakk}\isanewline
\isaindent{\ \ \ }{\isasymLongrightarrow}\ P\ {\isacharparenleft}{\isasymsigma}{\isacharprime},\ l{\isacharparenright}%
\end{isabelle}}
\snip{oseq_invariant_ctermI_concl}{0}{0}{\isa{A\ {\isasymTurnstile}\ {\isacharparenleft}S,\ U\ {\isasymrightarrow}{\isacharparenright}\ onl\ {\isasymGamma}\ P}}

\snip{oseq_step_invariant_ctermI}{0}{0}{\begin{isabelle}%
{\isasymlbrakk}wellformed\ {\isasymGamma}{\isacharsemicolon}\ control{\isacharunderscore}within\ {\isasymGamma}\ {\isacharparenleft}init\ A{\isacharparenright}{\isacharsemicolon}\ simple{\isacharunderscore}labels\ {\isasymGamma}{\isacharsemicolon}\isanewline
\isaindent{\ }trans\ A\ {\isacharequal}\ \oseqpsos\ {\isasymGamma}\ i{\isacharsemicolon}\isanewline
\isaindent{\ }{\isasymAnd}p\ l\ {\isasymsigma}\ a\ q\ l{\isacharprime}\ {\isasymsigma}{\isacharprime}\ pp{\isachardot}\isanewline
\isaindent{\ \ \ \ }{\isasymlbrakk}p{\isasymin}cterms\ {\isasymGamma}{\isacharsemicolon}\ l{\isasymin}labels\ {\isasymGamma}\ p{\isacharsemicolon}\ {\isacharparenleft}{\isacharparenleft}{\isasymsigma},\ p{\isacharparenright},\ a,\ {\isacharparenleft}{\isasymsigma}{\isacharprime},\ q{\isacharparenright}{\isacharparenright}{\isasymin}\oseqpsos\ {\isasymGamma}\ i{\isacharsemicolon}\isanewline
\isaindent{\ \ \ \ \ }{\isacharparenleft}{\isacharparenleft}{\isasymsigma},\ p{\isacharparenright},\ a,\ {\isasymsigma}{\isacharprime},\ q{\isacharparenright}{\isasymin}trans\ A{\isacharsemicolon}\ l{\isacharprime}{\isasymin}labels\ {\isasymGamma}\ q{\isacharsemicolon}\isanewline
\isaindent{\ \ \ \ \ }{\isacharparenleft}{\isasymsigma},\ pp{\isacharparenright}{\isasymin}oreachable\ A\ S\ U{\isacharsemicolon}\ p{\isasymin}sterms\ {\isasymGamma}\ pp{\isacharsemicolon}\isanewline
\isaindent{\ \ \ \ \ }{\isacharparenleft}{\isasymsigma}{\isacharprime},\ q{\isacharparenright}{\isasymin}oreachable\ A\ S\ U{\isacharsemicolon}\ S\ {\isasymsigma}\ {\isasymsigma}{\isacharprime}\ a{\isasymrbrakk}\isanewline
\isaindent{\ \ \ \ }{\isasymLongrightarrow}\ P\ {\isacharparenleft}{\isacharparenleft}{\isasymsigma},\ l{\isacharparenright},\ a,\ {\isasymsigma}{\isacharprime},\ l{\isacharprime}{\isacharparenright}{\isasymrbrakk}\isanewline
{\isasymLongrightarrow}\ A\ \ostepinv\ {\isacharparenleft}S,\ U\ {\isasymrightarrow}{\isacharparenright}\ onll\ {\isasymGamma}\ P%
\end{isabelle}}

\snip{reachable_init}{0}{0}{\isa{\mbox{}\inferrule{\mbox{s{\isasymin}init\ A}}{\mbox{s{\isasymin}reachable\ A\ I}}}}
\snip{reachable_step}{0}{0}{\isa{\mbox{}\inferrule{\mbox{s{\isasymin}reachable\ A\ I}\\\ \mbox{{\isacharparenleft}s,\ a,\ s{\isacharprime}{\isacharparenright}{\isasymin}trans\ A}\\\ \mbox{I\ a}}{\mbox{s{\isacharprime}{\isasymin}reachable\ A\ I}}}}

\snip{oreachable_init}{0}{0}{\isa{\mbox{}\inferrule{\mbox{{\isacharparenleft}{\isasymsigma},\ p{\isacharparenright}{\isasymin}init\ A}}{\mbox{{\isacharparenleft}{\isasymsigma},\ p{\isacharparenright}{\isasymin}oreachable\ A\ S\ U}}}}
\snip{oreachable_local}{0}{0}{\isa{\mbox{}\inferrule{\mbox{s{\isasymin}oreachable\ A\ S\ U}\\\ \mbox{{\isacharparenleft}s,\ {\isacharparenleft}a,\ s{\isacharprime}{\isacharparenright}{\isacharparenright}{\isasymin}trans\ A}\\\ \mbox{S\ {\isacharparenleft}fst\ s{\isacharparenright}\ {\isacharparenleft}fst\ s{\isacharprime}{\isacharparenright}\ a}}{\mbox{s{\isacharprime}{\isasymin}oreachable\ A\ S\ U}}}}
\snip{oreachable_other}{0}{0}{\isa{\mbox{}\inferrule{\mbox{s{\isasymin}oreachable\ A\ S\ U}\\\ \mbox{U\ {\isacharparenleft}fst\ s{\isacharparenright}\ {\isasymsigma}{\isacharprime}}}{\mbox{{\isacharparenleft}{\isasymsigma}{\isacharprime},\ snd\ s{\isacharparenright}{\isasymin}oreachable\ A\ S\ U}}}}
\snip{oreachable_local'}{0}{0}{\isa{\mbox{}\inferrule{\mbox{{\isacharparenleft}{\isasymsigma},\ p{\isacharparenright}{\isasymin}oreachable\ A\ S\ U}\\\ \mbox{{\isacharparenleft}{\isacharparenleft}{\isasymsigma},\ p{\isacharparenright},\ a,\ {\isacharparenleft}{\isasymsigma}{\isacharprime},\ p{\isacharprime}{\isacharparenright}{\isacharparenright}{\isasymin}trans\ A}\\\ \mbox{S\ {\isasymsigma}\ {\isasymsigma}{\isacharprime}\ a}}{\mbox{{\isacharparenleft}{\isasymsigma}{\isacharprime},\ p{\isacharprime}{\isacharparenright}{\isasymin}oreachable\ A\ S\ U}}}}
\snip{oreachable_other'}{0}{0}{\isa{\mbox{}\inferrule{\mbox{{\isacharparenleft}{\isasymsigma},\ p{\isacharparenright}{\isasymin}oreachable\ A\ S\ U}\\\ \mbox{U\ {\isasymsigma}\ {\isasymsigma}{\isacharprime}}}{\mbox{{\isacharparenleft}{\isasymsigma}{\isacharprime},\ p{\isacharparenright}{\isasymin}oreachable\ A\ S\ U}}}}

\snip{otherwith}{0}{0}{\isa{otherwith\ {\isacharparenleft}op\ {\isacharequal}{\isacharparenright}\ {\isacharbraceleft}i{\isacharbraceright}\ {\isacharparenleft}orecvmsg\ P{\isacharparenright}}}
\snip{other}{0}{0}{\isa{other\ quality{\isacharunderscore}increases\ {\isacharbraceleft}i{\isacharbraceright}}}
\snip{otherwith_def}{0}{0}{\isa{otherwith\ E\ N\ I\ {\isasymsigma}\ {\isasymsigma}{\isacharprime}\ a\ {\isacharequal}\ {\isacharparenleft}{\isasymforall}i{\isachardot}\ i\ {\isasymnotin}\ N\ {\isasymlongrightarrow}\ E\ {\isacharparenleft}{\isasymsigma}\ i{\isacharparenright}\ {\isacharparenleft}{\isasymsigma}{\isacharprime}\ i{\isacharparenright}{\isacharparenright}\ {\isasymand}\ I\ {\isasymsigma}\ a}}
\snip{other_def}{0}{0}{\isa{other\ P\ I\ {\isasymsigma}\ {\isasymsigma}{\isacharprime}\ {\isacharequal}\ {\isasymforall}i{\isachardot}\ \textsf{if}\ i{\isasymin}I\ \textsf{then}\ {\isasymsigma}{\isacharprime}\ i\ {\isacharequal}\ {\isasymsigma}\ i\ \textsf{else}\ P\ {\isacharparenleft}{\isasymsigma}\ i{\isacharparenright}\ {\isacharparenleft}{\isasymsigma}{\isacharprime}\ i{\isacharparenright}}}
\snip{qmsg_msgs_not_empty}{0}{0}{\begin{isabelle}%
qmsg\ {\isasymTTurnstile}\ onl\ \gammaqmsg\ {\isacharparenleft}{\isasymlambda}{\isacharparenleft}msgs,\ l{\isacharparenright}{\isachardot}\ l\ {\isacharequal}\ {\isacharparenleft}{\isacharparenright}{\isacharminus}{\isacharcolon}{\isadigit{1}}\ {\isasymlongrightarrow}\ msgs\ {\isasymnoteq}\ {\isacharbrackleft}{\isacharbrackright}{\isacharparenright}%
\end{isabelle}}
\snip{qmsg_send_from_queue}{0}{0}{\begin{isabelle}%
qmsg\ \stepinv\ {\isacharparenleft}{\isasymlambda}{\isacharparenleft}{\isacharparenleft}msg,\ q{\isacharparenright},\ a,\ {\isasymzeta}{\isacharprime}{\isacharparenright}{\isachardot}\ sendmsg\ {\isacharparenleft}{\isasymlambda}m{\isachardot}\ m{\isasymin}set\ msg{\isacharparenright}\ a{\isacharparenright}%
\end{isabelle}}
\snip{qmsg_send_receive_or_tau}{0}{0}{\begin{isabelle}%
qmsg\ \stepinv\ {\isacharparenleft}{\isasymlambda}{\isacharparenleft}uu{\isacharunderscore},\ a,\ {\isacharunderscore}{\isacharparenright}{\isachardot}\ {\isasymexists}m{\isachardot}\ a\ {\isacharequal}\ send\ m\ {\isasymor}\ a\ {\isacharequal}\ receive\ m\ {\isasymor}\ a\ {\isacharequal}\ {\isasymtau}{\isacharparenright}%
\end{isabelle}}

\snip{par_qmsg_oreachable}{0}{0}{\begin{isabelle}%
{\isasymlbrakk}{\isacharparenleft}{\isasymsigma},\ {\isasymzeta}{\isacharparenright}\isanewline
\isaindent{{\isasymlbrakk}}{\isasymin}\ oreachable\ {\isacharparenleft}A\ {\isasymlangle}{\isasymlangle}\isactrlbsub i\isactrlesub \ qmsg{\isacharparenright}\ {\isacharparenleft}otherwith\ S\ {\isacharbraceleft}i{\isacharbraceright}\ {\isacharparenleft}orecvmsg\ R{\isacharparenright}{\isacharparenright}\ {\isacharparenleft}other\ U\ {\isacharbraceleft}i{\isacharbraceright}{\isacharparenright}{\isacharsemicolon}\isanewline
\isaindent{\ }A\ \ostepinv\ {\isacharparenleft}otherwith\ S\ {\isacharbraceleft}i{\isacharbraceright}\ {\isacharparenleft}orecvmsg\ R{\isacharparenright},\ other\ U\ {\isacharbraceleft}i{\isacharbraceright}\ {\isasymrightarrow}{\isacharparenright}\isanewline
\isaindent{\ A\ \ostepinv\ \ }globala\ {\isacharparenleft}{\isasymlambda}{\isacharparenleft}{\isasymsigma},\ {\isacharunderscore},\ {\isasymsigma}{\isacharprime}{\isacharparenright}{\isachardot}\ U\ {\isacharparenleft}{\isasymsigma}\ i{\isacharparenright}\ {\isacharparenleft}{\isasymsigma}{\isacharprime}\ i{\isacharparenright}{\isacharparenright}{\isacharsemicolon}\isanewline
\isaindent{\ }{\isasymAnd}{\isasymxi}{\isachardot}\ U\ {\isasymxi}\ {\isasymxi}{\isacharsemicolon}\ {\isasymAnd}{\isasymxi}\ {\isasymxi}{\isacharprime}{\isachardot}\ S\ {\isasymxi}\ {\isasymxi}{\isacharprime}\ {\isasymLongrightarrow}\ U\ {\isasymxi}\ {\isasymxi}{\isacharprime}{\isacharsemicolon}\isanewline
\isaindent{\ }{\isasymAnd}{\isasymsigma}\ {\isasymsigma}{\isacharprime}\ m{\isachardot}\ {\isasymlbrakk}{\isasymforall}j{\isachardot}\ U\ {\isacharparenleft}{\isasymsigma}\ j{\isacharparenright}\ {\isacharparenleft}{\isasymsigma}{\isacharprime}\ j{\isacharparenright}{\isacharsemicolon}\ R\ {\isasymsigma}\ m{\isasymrbrakk}\ {\isasymLongrightarrow}\ R\ {\isasymsigma}{\isacharprime}\ m{\isasymrbrakk}\isanewline
{\isasymLongrightarrow}\ {\isacharparenleft}{\isasymsigma},\ fst\ {\isasymzeta}{\isacharparenright}{\isasymin}oreachable\ A\ {\isacharparenleft}otherwith\ S\ {\isacharbraceleft}i{\isacharbraceright}\ {\isacharparenleft}orecvmsg\ R{\isacharparenright}{\isacharparenright}\ {\isacharparenleft}other\ U\ {\isacharbraceleft}i{\isacharbraceright}{\isacharparenright}\ {\isasymand}\isanewline
\isaindent{{\isasymLongrightarrow}\ }snd\ {\isasymzeta}{\isasymin}reachable\ qmsg\ {\isacharparenleft}recvmsg\ {\isacharparenleft}R\ {\isasymsigma}{\isacharparenright}{\isacharparenright}\ {\isasymand}\ {\isacharparenleft}{\isasymforall}m{\isasymin}set\ {\isacharparenleft}fst\ {\isacharparenleft}snd\ {\isasymzeta}{\isacharparenright}{\isacharparenright}{\isachardot}\ R\ {\isasymsigma}\ m{\isacharparenright}%
\end{isabelle}}
\snip{par_qmsg_oreachable_upreservesq}{0}{0}{\begin{isabelle}%
{\isasymAnd}{\isasymsigma}\ {\isasymsigma}{\isacharprime}\ m{\isachardot}\ {\isasymlbrakk}{\isasymforall}j{\isachardot}\ U\ {\isacharparenleft}{\isasymsigma}\ j{\isacharparenright}\ {\isacharparenleft}{\isasymsigma}{\isacharprime}\ j{\isacharparenright}{\isacharsemicolon}\ R\ {\isasymsigma}\ m{\isasymrbrakk}\ {\isasymLongrightarrow}\ R\ {\isasymsigma}{\isacharprime}\ m%
\end{isabelle}}
\snip{par_qmsg_oreachable_concl3}{0}{0}{\begin{isabelle}%
{\isasymforall}m{\isasymin}set\ {\isacharparenleft}fst\ {\isacharparenleft}snd\ {\isasymzeta}{\isacharparenright}{\isacharparenright}{\isachardot}\ R\ {\isasymsigma}\ m%
\end{isabelle}}

\snip{lift_into_qmsg}{0}{0}{\begin{isabelle}%
{\isasymlbrakk}A\ {\isasymTurnstile}\ {\isacharparenleft}otherwith\ S\ {\isacharbraceleft}i{\isacharbraceright}\ {\isacharparenleft}orecvmsg\ R{\isacharparenright},\ other\ U\ {\isacharbraceleft}i{\isacharbraceright}\ {\isasymrightarrow}{\isacharparenright}\ global\ P{\isacharsemicolon}\ {\isasymAnd}{\isasymxi}{\isachardot}\ U\ {\isasymxi}\ {\isasymxi}{\isacharsemicolon}\isanewline
\isaindent{\ }{\isasymAnd}{\isasymxi}\ {\isasymxi}{\isacharprime}{\isachardot}\ S\ {\isasymxi}\ {\isasymxi}{\isacharprime}\ {\isasymLongrightarrow}\ U\ {\isasymxi}\ {\isasymxi}{\isacharprime}{\isacharsemicolon}\ {\isasymAnd}{\isasymsigma}\ {\isasymsigma}{\isacharprime}\ m{\isachardot}\ {\isasymlbrakk}{\isasymforall}j{\isachardot}\ U\ {\isacharparenleft}{\isasymsigma}\ j{\isacharparenright}\ {\isacharparenleft}{\isasymsigma}{\isacharprime}\ j{\isacharparenright}{\isacharsemicolon}\ R\ {\isasymsigma}\ m{\isasymrbrakk}\ {\isasymLongrightarrow}\ R\ {\isasymsigma}{\isacharprime}\ m{\isacharsemicolon}\isanewline
\isaindent{\ }A\ \ostepinv\ {\isacharparenleft}otherwith\ S\ {\isacharbraceleft}i{\isacharbraceright}\ {\isacharparenleft}orecvmsg\ R{\isacharparenright},\ other\ U\ {\isacharbraceleft}i{\isacharbraceright}\ {\isasymrightarrow}{\isacharparenright}\isanewline
\isaindent{\ A\ \ostepinv\ \ }globala\ {\isacharparenleft}{\isasymlambda}{\isacharparenleft}{\isasymsigma},\ {\isacharunderscore},\ {\isasymsigma}{\isacharprime}{\isacharparenright}{\isachardot}\ U\ {\isacharparenleft}{\isasymsigma}\ i{\isacharparenright}\ {\isacharparenleft}{\isasymsigma}{\isacharprime}\ i{\isacharparenright}{\isacharparenright}{\isasymrbrakk}\isanewline
{\isasymLongrightarrow}\ A\ {\isasymlangle}{\isasymlangle}\isactrlbsub i\isactrlesub \ qmsg\ {\isasymTurnstile}\ {\isacharparenleft}otherwith\ S\ {\isacharbraceleft}i{\isacharbraceright}\ {\isacharparenleft}orecvmsg\ R{\isacharparenright},\ other\ U\ {\isacharbraceleft}i{\isacharbraceright}\ {\isasymrightarrow}{\isacharparenright}\ global\ P%
\end{isabelle}}
\snip{lift_step_into_qmsg}{0}{0}{\begin{isabelle}%
{\isasymlbrakk}A\ \ostepinv\ {\isacharparenleft}otherwith\ S\ {\isacharbraceleft}i{\isacharbraceright}\ {\isacharparenleft}orecvmsg\ R{\isacharparenright},\ other\ U\ {\isacharbraceleft}i{\isacharbraceright}\ {\isasymrightarrow}{\isacharparenright}\ globala\ P{\isacharsemicolon}\ {\isasymAnd}{\isasymxi}{\isachardot}\ U\ {\isasymxi}\ {\isasymxi}{\isacharsemicolon}\isanewline
\isaindent{\ }{\isasymAnd}{\isasymxi}\ {\isasymxi}{\isacharprime}{\isachardot}\ S\ {\isasymxi}\ {\isasymxi}{\isacharprime}\ {\isasymLongrightarrow}\ U\ {\isasymxi}\ {\isasymxi}{\isacharprime}{\isacharsemicolon}\ {\isasymAnd}{\isasymsigma}\ {\isasymsigma}{\isacharprime}\ m{\isachardot}\ {\isasymlbrakk}{\isasymforall}j{\isachardot}\ U\ {\isacharparenleft}{\isasymsigma}\ j{\isacharparenright}\ {\isacharparenleft}{\isasymsigma}{\isacharprime}\ j{\isacharparenright}{\isacharsemicolon}\ R\ {\isasymsigma}\ m{\isasymrbrakk}\ {\isasymLongrightarrow}\ R\ {\isasymsigma}{\isacharprime}\ m{\isacharsemicolon}\isanewline
\isaindent{\ }A\ \ostepinv\ {\isacharparenleft}otherwith\ S\ {\isacharbraceleft}i{\isacharbraceright}\ {\isacharparenleft}orecvmsg\ R{\isacharparenright},\ other\ U\ {\isacharbraceleft}i{\isacharbraceright}\ {\isasymrightarrow}{\isacharparenright}\isanewline
\isaindent{\ A\ \ostepinv\ \ }globala\ {\isacharparenleft}{\isasymlambda}{\isacharparenleft}{\isasymsigma},\ {\isacharunderscore},\ {\isasymsigma}{\isacharprime}{\isacharparenright}{\isachardot}\ U\ {\isacharparenleft}{\isasymsigma}\ i{\isacharparenright}\ {\isacharparenleft}{\isasymsigma}{\isacharprime}\ i{\isacharparenright}{\isacharparenright}{\isacharsemicolon}\isanewline
\isaindent{\ }{\isasymAnd}{\isasymsigma}\ {\isasymsigma}{\isacharprime}\ m{\isachardot}\ {\isasymlbrakk}{\isasymforall}j{\isachardot}\ U\ {\isacharparenleft}{\isasymsigma}\ j{\isacharparenright}\ {\isacharparenleft}{\isasymsigma}{\isacharprime}\ j{\isacharparenright}{\isacharsemicolon}\ {\isasymsigma}{\isacharprime}\ i\ {\isacharequal}\ {\isasymsigma}\ i{\isasymrbrakk}\ {\isasymLongrightarrow}\ P\ {\isacharparenleft}{\isasymsigma},\ receive\ m,\ {\isasymsigma}{\isacharprime}{\isacharparenright}{\isacharsemicolon}\isanewline
\isaindent{\ }{\isasymAnd}{\isasymsigma}\ {\isasymsigma}{\isacharprime}\ m{\isachardot}\ P\ {\isacharparenleft}{\isasymsigma},\ receive\ m,\ {\isasymsigma}{\isacharprime}{\isacharparenright}\ {\isasymLongrightarrow}\ P\ {\isacharparenleft}{\isasymsigma},\ {\isasymtau},\ {\isasymsigma}{\isacharprime}{\isacharparenright}{\isasymrbrakk}\isanewline
{\isasymLongrightarrow}\ A\ {\isasymlangle}{\isasymlangle}\isactrlbsub i\isactrlesub \ qmsg\ \ostepinv\ {\isacharparenleft}otherwith\ S\ {\isacharbraceleft}i{\isacharbraceright}\ {\isacharparenleft}orecvmsg\ R{\isacharparenright},\ other\ U\ {\isacharbraceleft}i{\isacharbraceright}\ {\isasymrightarrow}{\isacharparenright}\ globala\ P%
\end{isabelle}}

\snip{node_lift}{0}{0}{\begin{isabelle}%
{\isasymlbrakk}T\ {\isasymTurnstile}\ {\isacharparenleft}otherwith\ S\ {\isacharbraceleft}ii{\isacharbraceright}\ {\isacharparenleft}orecvmsg\ I{\isacharparenright},\ other\ U\ {\isacharbraceleft}ii{\isacharbraceright}\ {\isasymrightarrow}{\isacharparenright}\ global\ P{\isacharsemicolon}\isanewline
\isaindent{\ }{\isasymAnd}{\isasymxi}\ {\isasymxi}{\isacharprime}{\isachardot}\ S\ {\isasymxi}\ {\isasymxi}{\isacharprime}\ {\isasymLongrightarrow}\ U\ {\isasymxi}\ {\isasymxi}{\isacharprime}{\isasymrbrakk}\isanewline
{\isasymLongrightarrow}\ {\isasymlangle}ii\ {\isacharcolon}\ T\ {\isacharcolon}\ \Ri{\isasymrangle}\isactrlsub o\ {\isasymTurnstile}\ {\isacharparenleft}otherwith\ S\ {\isacharbraceleft}ii{\isacharbraceright}\ {\isacharparenleft}oarrivemsg\ I{\isacharparenright},\ other\ U\ {\isacharbraceleft}ii{\isacharbraceright}\ {\isasymrightarrow}{\isacharparenright}\isanewline
\isaindent{{\isasymLongrightarrow}\ {\isasymlangle}ii\ {\isacharcolon}\ T\ {\isacharcolon}\ \Ri{\isasymrangle}\isactrlsub o\ {\isasymTurnstile}\ \ }global\ P%
\end{isabelle}}
\snip{node_lift_step}{0}{0}{\begin{isabelle}%
{\isasymlbrakk}T\ \ostepinv\ {\isacharparenleft}otherwith\ S\ {\isacharbraceleft}i{\isacharbraceright}\ {\isacharparenleft}orecvmsg\ I{\isacharparenright},\ other\ U\ {\isacharbraceleft}i{\isacharbraceright}\ {\isasymrightarrow}{\isacharparenright}\isanewline
\isaindent{{\isasymlbrakk}T\ \ostepinv\ \ }globala\ {\isacharparenleft}{\isasymlambda}{\isacharparenleft}{\isasymsigma},\ {\isacharunderscore},\ {\isasymsigma}{\isacharprime}{\isacharparenright}{\isachardot}\ Q\ {\isasymsigma}\ {\isasymsigma}{\isacharprime}{\isacharparenright}{\isacharsemicolon}\isanewline
\isaindent{\ }{\isasymAnd}{\isasymsigma}\ {\isasymsigma}{\isacharprime}{\isachardot}\ other\ U\ {\isacharbraceleft}i{\isacharbraceright}\ {\isasymsigma}\ {\isasymsigma}{\isacharprime}\ {\isasymLongrightarrow}\ Q\ {\isasymsigma}\ {\isasymsigma}{\isacharprime}{\isacharsemicolon}\ {\isasymAnd}{\isasymxi}\ {\isasymxi}{\isacharprime}{\isachardot}\ S\ {\isasymxi}\ {\isasymxi}{\isacharprime}\ {\isasymLongrightarrow}\ U\ {\isasymxi}\ {\isasymxi}{\isacharprime}{\isasymrbrakk}\isanewline
{\isasymLongrightarrow}\ {\isasymlangle}i\ {\isacharcolon}\ T\ {\isacharcolon}\ \Ri{\isasymrangle}\isactrlsub o\ \ostepinv\ {\isacharparenleft}otherwith\ S\ {\isacharbraceleft}i{\isacharbraceright}\ {\isacharparenleft}oarrivemsg\ I{\isacharparenright},\ other\ U\ {\isacharbraceleft}i{\isacharbraceright}\ {\isasymrightarrow}{\isacharparenright}\isanewline
\isaindent{{\isasymLongrightarrow}\ {\isasymlangle}i\ {\isacharcolon}\ T\ {\isacharcolon}\ \Ri{\isasymrangle}\isactrlsub o\ \ostepinv\ \ }globala\ {\isacharparenleft}{\isasymlambda}{\isacharparenleft}{\isasymsigma},\ {\isacharunderscore},\ {\isasymsigma}{\isacharprime}{\isacharparenright}{\isachardot}\ Q\ {\isasymsigma}\ {\isasymsigma}{\isacharprime}{\isacharparenright}%
\end{isabelle}}

\snip{opaodv_qmsg_term}{0}{0}{\isa{opaodv\ i\ {\isasymlangle}{\isasymlangle}\isactrlbsub i\isactrlesub \ qmsg}}

\snip{otherwith_r_term}{0}{0}{\isa{otherwith\ S\ {\isacharbraceleft}i{\isacharbraceright}\ {\isacharparenleft}orecvmsg\ R{\isacharparenright}}}
\snip{paodv_prreq_extract}{0}{0}{\begin{isabelle}%
{\isacharbraceleft}PRreq{\isacharminus}{\isacharcolon}{\isadigit{1}}{\isadigit{1}}{\isacharbraceright}{\isasymlbrakk}{\isasymlambda}{\isasymxi}{\isachardot}\ {\isasymxi}{\isasymlparr}pre\ {\isacharcolon}{\isacharequal}\ {\isasymUnion}{\isacharbraceleft}the\ {\isacharparenleft}precs\ {\isacharparenleft}rt\ {\isasymxi}{\isacharparenright}\ rip{\isacharparenright}\ {\isacharbar}\ rip{\isasymin}dom\ {\isacharparenleft}dests\ {\isasymxi}{\isacharparenright}{\isacharbraceright}{\isasymrparr}{\isasymrbrakk}\isanewline
{\isacharbraceleft}PRreq{\isacharminus}{\isacharcolon}{\isadigit{1}}{\isadigit{2}}{\isacharbraceright}{\isasymlbrakk}{\isasymlambda}{\isasymxi}{\isachardot}\ {\isasymxi}{\isasymlparr}dests\ {\isacharcolon}{\isacharequal}\ {\isasymlambda}rip{\isachardot}\ \textsf{if}\ dests\ {\isasymxi}\ rip\ {\isasymnoteq}\ None\ {\isasymand}\ the\ {\isacharparenleft}precs\ {\isacharparenleft}rt\ {\isasymxi}{\isacharparenright}\ rip{\isacharparenright}\ {\isasymnoteq}\ {\isasymemptyset}\ \textsf{then}\ dests\ {\isasymxi}\ rip\ \textsf{else}\ None{\isasymrparr}{\isasymrbrakk}\isanewline
{\isacharbraceleft}PRreq{\isacharminus}{\isacharcolon}{\isadigit{1}}{\isadigit{3}}{\isacharbraceright}groupcast{\isacharparenleft}pre,\ {\isasymlambda}{\isasymxi}{\isachardot}\ rerr\ {\isacharparenleft}dests\ {\isasymxi},\ ip\ {\isasymxi}{\isacharparenright}{\isacharparenright}\ {\isachardot}\isanewline
{\isacharbraceleft}PRreq{\isacharminus}{\isacharcolon}{\isadigit{1}}{\isadigit{4}}{\isacharbraceright}{\isasymlbrakk}clear{\isacharunderscore}locals{\isasymrbrakk}\isanewline
call{\isacharparenleft}PAodv{\isacharparenright}%
\end{isabelle}}

\snip{stmt_guard}{0}{0}{\isa{{\isasymlangle}{\isasymlambda}{\isasymxi}{\isachardot}\ \textsf{if}\ g\ {\isasymxi}\ \textsf{then}\ {\isacharbraceleft}{\isasymxi}{\isacharbraceright}\ \textsf{else}\ {\isasymemptyset}{\isasymrangle}\ p}}
\snip{stmt_bind}{0}{0}{\isa{{\isasymlangle}{\isasymlambda}{\isasymxi}{\isachardot}\ {\isacharbraceleft}{\isasymxi}{\isasymlparr}dip\ {\isacharcolon}{\isacharequal}\ d{\isasymrparr}\ {\isacharbar}\ d{\isasymin}qD\ {\isacharparenleft}store\ {\isasymxi}{\isacharparenright}\ {\isasyminter}\ vD\ {\isacharparenleft}rt\ {\isasymxi}{\isacharparenright}{\isacharbraceright}{\isasymrangle}\ p}}

\snip{act_unicast}{0}{0}{\isa{unicast\ i\ m}}
\snip{act_not_unicast}{0}{0}{\isa{{\isasymnot}unicast\ i}}

\snip{eg1_all}{0}{0}{\isa{closed\ {\isacharparenleft}pnet\ {\isacharparenleft}{\isasymlambda}i{\isachardot}\ paodv\ i\ {\isasymlangle}{\isasymlangle}\ qmsg{\isacharparenright}\ {\isacharparenleft}{\isasymlangle}A{\isacharsemicolon}\ {\isacharbraceleft}B,\ D{\isacharbraceright}{\isasymrangle}\parallelcomp{}{\isacharparenleft}{\isasymlangle}B{\isacharsemicolon}\ {\isacharbraceleft}A,\ C{\isacharbraceright}{\isasymrangle}\parallelcomp{}{\isacharparenleft}{\isasymlangle}C{\isacharsemicolon}\ {\isacharbraceleft}B{\isacharbraceright}{\isasymrangle}\parallelcomp{}{\isasymlangle}D{\isacharsemicolon}\ {\isacharbraceleft}A{\isacharbraceright}{\isasymrangle}{\isacharparenright}{\isacharparenright}{\isacharparenright}{\isacharparenright}}}

\snip{eg1_node_a'}{0}{0}{\isa{{\isasymlangle}A\ {\isacharcolon}\ paodv\ A\ {\isasymlangle}{\isasymlangle}\ qmsg\ {\isacharcolon}\ {\isacharbraceleft}B,\ D{\isacharbraceright}{\isasymrangle}}}
\snip{eg1_node_b'}{0}{0}{\isa{{\isasymlangle}B\ {\isacharcolon}\ paodv\ B\ {\isasymlangle}{\isasymlangle}\ qmsg\ {\isacharcolon}\ {\isacharbraceleft}A,\ C{\isacharbraceright}{\isasymrangle}}}
\snip{eg1_node_c'}{0}{0}{\isa{{\isasymlangle}C\ {\isacharcolon}\ paodv\ C\ {\isasymlangle}{\isasymlangle}\ qmsg\ {\isacharcolon}\ {\isacharbraceleft}B{\isacharbraceright}{\isasymrangle}}}
\snip{eg1_node_d'}{0}{0}{\isa{{\isasymlangle}D\ {\isacharcolon}\ paodv\ D\ {\isasymlangle}{\isasymlangle}\ qmsg\ {\isacharcolon}\ {\isacharbraceleft}A{\isacharbraceright}{\isasymrangle}}}

\snip{eg1_node_a}{0}{0}{\isa{{\isasymlangle}A{\isacharsemicolon}\ {\isacharbraceleft}B,\ D{\isacharbraceright}{\isasymrangle}}}
\snip{eg1_node_b}{0}{0}{\isa{{\isasymlangle}B{\isacharsemicolon}\ {\isacharbraceleft}A,\ C{\isacharbraceright}{\isasymrangle}}}
\snip{eg1_node_c}{0}{0}{\isa{{\isasymlangle}C{\isacharsemicolon}\ {\isacharbraceleft}B{\isacharbraceright}{\isasymrangle}}}
\snip{eg1_node_d}{0}{0}{\isa{{\isasymlangle}D{\isacharsemicolon}\ {\isacharbraceleft}A{\isacharbraceright}{\isasymrangle}}}

\snip{dummy_par}{0}{0}{\isa{\_\parallelcomp{}\_}}

\snip{eg_sterms_foo}{0}{0}{\isa{sterms\ {\isasymGamma}\ {\isacharparenleft}{\isasymGamma}\ foo{\isacharparenright}}}
\snip{eg_sterms_groupcast}{0}{0}{\isa{dterms\ {\isasymGamma}\ {\isacharparenleft}groupcast{\isacharparenleft}\_,\ \_{\isacharparenright}\ {\isachardot}\ p{\isacharparenright}}}
\snip{eg_sterms_p}{0}{0}{\isa{sterms\ {\isasymGamma}\ p}}
\snip{gammap}{0}{0}{\isa{\gammaaodv}}

\snip{gammap_paodv}{0}{0}{\begin{isabelle}%
\gammaaodv\ PAodv\ {\isacharequal}\isanewline
{\isacharbraceleft}PAodv{\isacharminus}{\isacharcolon}{\isadigit{0}}{\isacharbraceright}receive{\isacharparenleft}{\isasymlambda}msg{\isacharprime}{\isachardot}\ msg{\isacharunderscore}update\ {\isacharparenleft}{\isasymlambda}{\isacharunderscore}{\isachardot}\ msg{\isacharprime}{\isacharparenright}{\isacharparenright}\ {\isachardot}\isanewline
{\isacharparenleft}{\isacharbraceleft}PAodv{\isacharminus}{\isacharcolon}{\isadigit{1}}{\isacharbraceright}{\isasymlangle}is{\isacharunderscore}newpkt{\isasymrangle}\isanewline
\isaindent{{\isacharparenleft}}{\isacharbraceleft}PAodv{\isacharminus}{\isacharcolon}{\isadigit{2}}{\isacharbraceright}{\isasymlbrakk}{\isasymlambda}{\isasymxi}{\isachardot}\ clear{\isacharunderscore}locals\ {\isasymxi}{\isasymlparr}data\ {\isacharcolon}{\isacharequal}\ data\ {\isasymxi},\ dip\ {\isacharcolon}{\isacharequal}\ ip\ {\isasymxi}{\isasymrparr}{\isasymrbrakk}\isanewline
\isaindent{{\isacharparenleft}}call{\isacharparenleft}PNewPkt{\isacharparenright}\isanewline
\isaindent{{\isacharparenleft}}{\isasymoplus}\isanewline
\isaindent{{\isacharparenleft}}{\isacharbraceleft}PAodv{\isacharminus}{\isacharcolon}{\isadigit{1}}{\isacharbraceright}{\isasymlangle}is{\isacharunderscore}pkt{\isasymrangle}\isanewline
\isaindent{{\isacharparenleft}}{\isacharbraceleft}PAodv{\isacharminus}{\isacharcolon}{\isadigit{3}}{\isacharbraceright}{\isasymlbrakk}{\isasymlambda}{\isasymxi}{\isachardot}\ clear{\isacharunderscore}locals\ {\isasymxi}{\isasymlparr}data\ {\isacharcolon}{\isacharequal}\ data\ {\isasymxi},\ dip\ {\isacharcolon}{\isacharequal}\ dip\ {\isasymxi},\ oip\ {\isacharcolon}{\isacharequal}\ oip\ {\isasymxi}{\isasymrparr}{\isasymrbrakk}\isanewline
\isaindent{{\isacharparenleft}}call{\isacharparenleft}PPkt{\isacharparenright}\isanewline
\isaindent{{\isacharparenleft}}{\isasymoplus}\isanewline
\isaindent{{\isacharparenleft}}{\isacharbraceleft}PAodv{\isacharminus}{\isacharcolon}{\isadigit{1}}{\isacharbraceright}{\isasymlangle}is{\isacharunderscore}rreq{\isasymrangle}\isanewline
\isaindent{{\isacharparenleft}}{\isacharbraceleft}PAodv{\isacharminus}{\isacharcolon}{\isadigit{4}}{\isacharbraceright}{\isasymlbrakk}{\isasymlambda}{\isasymxi}{\isachardot}\ {\isasymxi}{\isasymlparr}rt\ {\isacharcolon}{\isacharequal}\ update\ {\isacharparenleft}rt\ {\isasymxi}{\isacharparenright}\ {\isacharparenleft}sip\ {\isasymxi}{\isacharparenright}\ {\isacharparenleft}{\isadigit{0}},\ unk,\ val,\ Suc\ {\isadigit{0}},\ sip\ {\isasymxi},\ {\isasymemptyset}{\isacharparenright}{\isasymrparr}{\isasymrbrakk}\isanewline
\isaindent{{\isacharparenleft}}{\isacharbraceleft}PAodv{\isacharminus}{\isacharcolon}{\isadigit{5}}{\isacharbraceright}{\isasymlbrakk}{\isasymlambda}{\isasymxi}{\isachardot}\ clear{\isacharunderscore}locals\ {\isasymxi}{\isasymlparr}hops\ {\isacharcolon}{\isacharequal}\ hops\ {\isasymxi},\ rreqid\ {\isacharcolon}{\isacharequal}\ rreqid\ {\isasymxi},\ dip\ {\isacharcolon}{\isacharequal}\ dip\ {\isasymxi},\ dsn\ {\isacharcolon}{\isacharequal}\ dsn\ {\isasymxi},\ dsk\ {\isacharcolon}{\isacharequal}\ dsk\ {\isasymxi},\ oip\ {\isacharcolon}{\isacharequal}\ oip\ {\isasymxi},\ osn\ {\isacharcolon}{\isacharequal}\ osn\ {\isasymxi},\ sip\ {\isacharcolon}{\isacharequal}\ sip\ {\isasymxi}{\isasymrparr}{\isasymrbrakk}\isanewline
\isaindent{{\isacharparenleft}}call{\isacharparenleft}PRreq{\isacharparenright}\isanewline
\isaindent{{\isacharparenleft}}{\isasymoplus}\isanewline
\isaindent{{\isacharparenleft}}{\isacharbraceleft}PAodv{\isacharminus}{\isacharcolon}{\isadigit{1}}{\isacharbraceright}{\isasymlangle}is{\isacharunderscore}rrep{\isasymrangle}\isanewline
\isaindent{{\isacharparenleft}}{\isacharbraceleft}PAodv{\isacharminus}{\isacharcolon}{\isadigit{6}}{\isacharbraceright}{\isasymlbrakk}{\isasymlambda}{\isasymxi}{\isachardot}\ {\isasymxi}{\isasymlparr}rt\ {\isacharcolon}{\isacharequal}\ update\ {\isacharparenleft}rt\ {\isasymxi}{\isacharparenright}\ {\isacharparenleft}sip\ {\isasymxi}{\isacharparenright}\ {\isacharparenleft}{\isadigit{0}},\ unk,\ val,\ Suc\ {\isadigit{0}},\ sip\ {\isasymxi},\ {\isasymemptyset}{\isacharparenright}{\isasymrparr}{\isasymrbrakk}\isanewline
\isaindent{{\isacharparenleft}}{\isacharbraceleft}PAodv{\isacharminus}{\isacharcolon}{\isadigit{7}}{\isacharbraceright}{\isasymlbrakk}{\isasymlambda}{\isasymxi}{\isachardot}\ clear{\isacharunderscore}locals\ {\isasymxi}{\isasymlparr}hops\ {\isacharcolon}{\isacharequal}\ hops\ {\isasymxi},\ dip\ {\isacharcolon}{\isacharequal}\ dip\ {\isasymxi},\ dsn\ {\isacharcolon}{\isacharequal}\ dsn\ {\isasymxi},\ oip\ {\isacharcolon}{\isacharequal}\ oip\ {\isasymxi},\ sip\ {\isacharcolon}{\isacharequal}\ sip\ {\isasymxi}{\isasymrparr}{\isasymrbrakk}\isanewline
\isaindent{{\isacharparenleft}}call{\isacharparenleft}PRrep{\isacharparenright}\isanewline
\isaindent{{\isacharparenleft}}{\isasymoplus}\isanewline
\isaindent{{\isacharparenleft}}{\isacharbraceleft}PAodv{\isacharminus}{\isacharcolon}{\isadigit{1}}{\isacharbraceright}{\isasymlangle}is{\isacharunderscore}rerr{\isasymrangle}\isanewline
\isaindent{{\isacharparenleft}}{\isacharbraceleft}PAodv{\isacharminus}{\isacharcolon}{\isadigit{8}}{\isacharbraceright}{\isasymlbrakk}{\isasymlambda}{\isasymxi}{\isachardot}\ {\isasymxi}{\isasymlparr}rt\ {\isacharcolon}{\isacharequal}\ update\ {\isacharparenleft}rt\ {\isasymxi}{\isacharparenright}\ {\isacharparenleft}sip\ {\isasymxi}{\isacharparenright}\ {\isacharparenleft}{\isadigit{0}},\ unk,\ val,\ Suc\ {\isadigit{0}},\ sip\ {\isasymxi},\ {\isasymemptyset}{\isacharparenright}{\isasymrparr}{\isasymrbrakk}\isanewline
\isaindent{{\isacharparenleft}}{\isacharbraceleft}PAodv{\isacharminus}{\isacharcolon}{\isadigit{9}}{\isacharbraceright}{\isasymlbrakk}{\isasymlambda}{\isasymxi}{\isachardot}\ clear{\isacharunderscore}locals\ {\isasymxi}{\isasymlparr}dests\ {\isacharcolon}{\isacharequal}\ dests\ {\isasymxi},\ sip\ {\isacharcolon}{\isacharequal}\ sip\ {\isasymxi}{\isasymrparr}{\isasymrbrakk}\isanewline
\isaindent{{\isacharparenleft}}call{\isacharparenleft}PRerr{\isacharparenright}{\isacharparenright}\isanewline
{\isasymoplus}\isanewline
{\isacharbraceleft}PAodv{\isacharminus}{\isacharcolon}{\isadigit{0}}{\isacharbraceright}{\isasymlangle}{\isasymlambda}{\isasymxi}{\isachardot}\ {\isacharbraceleft}{\isasymxi}{\isasymlparr}dip\ {\isacharcolon}{\isacharequal}\ dip{\isasymrparr}\ {\isacharbar}\ dip{\isasymin}qD\ {\isacharparenleft}store\ {\isasymxi}{\isacharparenright}\ {\isasymand}\ dip{\isasymin}vD\ {\isacharparenleft}rt\ {\isasymxi}{\isacharparenright}{\isacharbraceright}{\isasymrangle}\isanewline
{\isacharbraceleft}PAodv{\isacharminus}{\isacharcolon}{\isadigit{1}}{\isadigit{0}}{\isacharbraceright}{\isasymlbrakk}{\isasymlambda}{\isasymxi}{\isachardot}\ {\isasymxi}{\isasymlparr}data\ {\isacharcolon}{\isacharequal}\ hd\ {\isasymsigma}\isactrlbsub queue\isactrlesub {\isacharparenleft}store\ {\isasymxi},\ dip\ {\isasymxi}{\isacharparenright}{\isasymrparr}{\isasymrbrakk}\isanewline
{\isacharbraceleft}PAodv{\isacharminus}{\isacharcolon}{\isadigit{1}}{\isadigit{1}}{\isacharbraceright}unicast{\isacharparenleft}{\isasymlambda}{\isasymxi}{\isachardot}\ the\ {\isacharparenleft}nh{\isacharparenleft}op\ {\isacharequal}{\isacharparenright}rt\ {\isasymxi}{\isacharparenright}\ {\isacharparenleft}dip\ {\isasymxi}{\isacharparenright}{\isacharparenright},\isanewline
\isaindent{{\isacharbraceleft}PAodv{\isacharminus}{\isacharcolon}{\isadigit{1}}{\isadigit{1}}{\isacharbraceright}unicast{\isacharparenleft}\ }{\isasymlambda}{\isasymxi}{\isachardot}\ Pkt\ {\isacharparenleft}data\ {\isasymxi}{\isacharparenright}\ {\isacharparenleft}dip\ {\isasymxi}{\isacharparenright}\ {\isacharparenleft}ip\ {\isasymxi}{\isacharparenright}{\isacharparenright}\ {\isachardot}\isanewline
\isaindent{\ \ \ }{\isacharbraceleft}PAodv{\isacharminus}{\isacharcolon}{\isadigit{1}}{\isadigit{2}}{\isacharbraceright}{\isasymlbrakk}{\isasymlambda}{\isasymxi}{\isachardot}\ {\isasymxi}{\isasymlparr}store\ {\isacharcolon}{\isacharequal}\ the\ {\isacharparenleft}Aodv{\isacharunderscore}Data{\isachardot}dr{\isacharparenleft}op\ {\isacharequal}{\isacharparenright}dip\ {\isasymxi}{\isacharparenright}\ {\isacharparenleft}store\ {\isasymxi}{\isacharparenright}{\isacharparenright}{\isasymrparr}{\isasymrbrakk}\isanewline
\isaindent{\ \ \ }{\isacharbraceleft}PAodv{\isacharminus}{\isacharcolon}{\isadigit{1}}{\isadigit{3}}{\isacharbraceright}{\isasymlbrakk}clear{\isacharunderscore}locals{\isasymrbrakk}\isanewline
\isaindent{\ \ \ }call{\isacharparenleft}PAodv{\isacharparenright}\isanewline
\isaindent{\ \ \ }{\isasymtriangleright}\ {\isacharbraceleft}PAodv{\isacharminus}{\isacharcolon}{\isadigit{1}}{\isadigit{4}}{\isacharbraceright}{\isasymlbrakk}{\isasymlambda}{\isasymxi}{\isachardot}\ {\isasymxi}{\isasymlparr}dests\ {\isacharcolon}{\isacharequal}\ {\isasymlambda}rip{\isachardot}\ \textsf{if}\ rip{\isasymin}vD\ {\isacharparenleft}rt\ {\isasymxi}{\isacharparenright}\ {\isasymand}\ nh{\isacharparenleft}op\ {\isacharequal}{\isacharparenright}rt\ {\isasymxi}{\isacharparenright}\ rip\ {\isacharequal}\ nh{\isacharparenleft}op\ {\isacharequal}{\isacharparenright}rt\ {\isasymxi}{\isacharparenright}\ {\isacharparenleft}dip\ {\isasymxi}{\isacharparenright}\ \textsf{then}\ Some\ {\isacharparenleft}inc\ {\isacharparenleft}sqn\ {\isacharparenleft}rt\ {\isasymxi}{\isacharparenright}\ rip{\isacharparenright}{\isacharparenright}\ \textsf{else}\ None{\isasymrparr}{\isasymrbrakk}\isanewline
\isaindent{\ \ \ \ \ }{\isacharbraceleft}PAodv{\isacharminus}{\isacharcolon}{\isadigit{1}}{\isadigit{5}}{\isacharbraceright}{\isasymlbrakk}{\isasymlambda}{\isasymxi}{\isachardot}\ {\isasymxi}{\isasymlparr}rt\ {\isacharcolon}{\isacharequal}\ invalidate\ {\isacharparenleft}rt\ {\isasymxi}{\isacharparenright}\ {\isacharparenleft}dests\ {\isasymxi}{\isacharparenright}{\isasymrparr}{\isasymrbrakk}\isanewline
\isaindent{\ \ \ \ \ }{\isacharbraceleft}PAodv{\isacharminus}{\isacharcolon}{\isadigit{1}}{\isadigit{6}}{\isacharbraceright}{\isasymlbrakk}{\isasymlambda}{\isasymxi}{\isachardot}\ {\isasymxi}{\isasymlparr}store\ {\isacharcolon}{\isacharequal}\ setRRF\ {\isacharparenleft}store\ {\isasymxi}{\isacharparenright}\ {\isacharparenleft}dests\ {\isasymxi}{\isacharparenright}{\isasymrparr}{\isasymrbrakk}\isanewline
\isaindent{\ \ \ \ \ }{\isacharbraceleft}PAodv{\isacharminus}{\isacharcolon}{\isadigit{1}}{\isadigit{7}}{\isacharbraceright}{\isasymlbrakk}{\isasymlambda}{\isasymxi}{\isachardot}\ {\isasymxi}{\isasymlparr}pre\ {\isacharcolon}{\isacharequal}\ {\isasymUnion}{\isacharbraceleft}the\ {\isacharparenleft}precs\ {\isacharparenleft}rt\ {\isasymxi}{\isacharparenright}\ rip{\isacharparenright}\ {\isacharbar}\ rip{\isasymin}dom\ {\isacharparenleft}dests\ {\isasymxi}{\isacharparenright}{\isacharbraceright}{\isasymrparr}{\isasymrbrakk}\isanewline
\isaindent{\ \ \ \ \ }{\isacharbraceleft}PAodv{\isacharminus}{\isacharcolon}{\isadigit{1}}{\isadigit{8}}{\isacharbraceright}{\isasymlbrakk}{\isasymlambda}{\isasymxi}{\isachardot}\ {\isasymxi}{\isasymlparr}dests\ {\isacharcolon}{\isacharequal}\ {\isasymlambda}rip{\isachardot}\ \textsf{if}\ {\isacharparenleft}{\isasymexists}y{\isachardot}\ dests\ {\isasymxi}\ rip\ {\isacharequal}\ Some\ y{\isacharparenright}\ {\isasymand}\ the\ {\isacharparenleft}precs\ {\isacharparenleft}rt\ {\isasymxi}{\isacharparenright}\ rip{\isacharparenright}\ {\isasymnoteq}\ {\isasymemptyset}\ \textsf{then}\ dests\ {\isasymxi}\ rip\ \textsf{else}\ None{\isasymrparr}{\isasymrbrakk}\isanewline
\isaindent{\ \ \ \ \ }{\isacharbraceleft}PAodv{\isacharminus}{\isacharcolon}{\isadigit{1}}{\isadigit{9}}{\isacharbraceright}groupcast{\isacharparenleft}pre,\ {\isasymlambda}{\isasymxi}{\isachardot}\ Rerr\ {\isacharparenleft}dests\ {\isasymxi}{\isacharparenright}\ {\isacharparenleft}ip\ {\isasymxi}{\isacharparenright}{\isacharparenright}\ {\isachardot}\isanewline
\isaindent{\ \ \ \ \ }{\isacharbraceleft}PAodv{\isacharminus}{\isacharcolon}{\isadigit{2}}{\isadigit{0}}{\isacharbraceright}{\isasymlbrakk}clear{\isacharunderscore}locals{\isasymrbrakk}\isanewline
\isaindent{\ \ \ \ \ }call{\isacharparenleft}PAodv{\isacharparenright}\isanewline
{\isasymoplus}\isanewline
{\isacharbraceleft}PAodv{\isacharminus}{\isacharcolon}{\isadigit{0}}{\isacharbraceright}{\isasymlangle}{\isasymlambda}{\isasymxi}{\isachardot}\ {\isacharbraceleft}{\isasymxi}{\isasymlparr}dip\ {\isacharcolon}{\isacharequal}\ dip{\isasymrparr}\ {\isacharbar}\ dip{\isasymin}qD\ {\isacharparenleft}store\ {\isasymxi}{\isacharparenright}\ {\isasymand}\ dip\ {\isasymnotin}\ vD\ {\isacharparenleft}rt\ {\isasymxi}{\isacharparenright}\ {\isasymand}\ the\ {\isasymsigma}\isactrlbsub p{\isacharminus}flag\isactrlesub {\isacharparenleft}store\ {\isasymxi},\ dip{\isacharparenright}\ {\isacharequal}\ req{\isacharbraceright}{\isasymrangle}\isanewline
{\isacharbraceleft}PAodv{\isacharminus}{\isacharcolon}{\isadigit{2}}{\isadigit{1}}{\isacharbraceright}{\isasymlbrakk}{\isasymlambda}{\isasymxi}{\isachardot}\ {\isasymxi}{\isasymlparr}store\ {\isacharcolon}{\isacharequal}\ unsetRRF\ {\isacharparenleft}store\ {\isasymxi}{\isacharparenright}\ {\isacharparenleft}dip\ {\isasymxi}{\isacharparenright}{\isasymrparr}{\isasymrbrakk}\isanewline
{\isacharbraceleft}PAodv{\isacharminus}{\isacharcolon}{\isadigit{2}}{\isadigit{2}}{\isacharbraceright}{\isasymlbrakk}{\isasymlambda}{\isasymxi}{\isachardot}\ {\isasymxi}{\isasymlparr}sn\ {\isacharcolon}{\isacharequal}\ inc\ {\isacharparenleft}sn\ {\isasymxi}{\isacharparenright}{\isasymrparr}{\isasymrbrakk}\isanewline
{\isacharbraceleft}PAodv{\isacharminus}{\isacharcolon}{\isadigit{2}}{\isadigit{3}}{\isacharbraceright}{\isasymlbrakk}{\isasymlambda}{\isasymxi}{\isachardot}\ {\isasymxi}{\isasymlparr}rreqid\ {\isacharcolon}{\isacharequal}\ nrreqid\ {\isacharparenleft}rreqs\ {\isasymxi}{\isacharparenright}\ {\isacharparenleft}ip\ {\isasymxi}{\isacharparenright}{\isasymrparr}{\isasymrbrakk}\isanewline
{\isacharbraceleft}PAodv{\isacharminus}{\isacharcolon}{\isadigit{2}}{\isadigit{4}}{\isacharbraceright}{\isasymlbrakk}{\isasymlambda}{\isasymxi}{\isachardot}\ {\isasymxi}{\isasymlparr}rreqs\ {\isacharcolon}{\isacharequal}\ {\isacharbraceleft}{\isacharparenleft}ip\ {\isasymxi},\ rreqid\ {\isasymxi}{\isacharparenright}{\isacharbraceright}\ {\isasymunion}\ rreqs\ {\isasymxi}{\isasymrparr}{\isasymrbrakk}\isanewline
{\isacharbraceleft}PAodv{\isacharminus}{\isacharcolon}{\isadigit{2}}{\isadigit{5}}{\isacharbraceright}broadcast{\isacharparenleft}{\isasymlambda}{\isasymxi}{\isachardot}\ Rreq\ {\isadigit{0}}\ {\isacharparenleft}rreqid\ {\isasymxi}{\isacharparenright}\ {\isacharparenleft}dip\ {\isasymxi}{\isacharparenright}\ {\isacharparenleft}sqn\ {\isacharparenleft}rt\ {\isasymxi}{\isacharparenright}\ {\isacharparenleft}dip\ {\isasymxi}{\isacharparenright}{\isacharparenright}\isanewline
\isaindent{{\isacharbraceleft}PAodv{\isacharminus}{\isacharcolon}{\isadigit{2}}{\isadigit{5}}{\isacharbraceright}broadcast{\isacharparenleft}{\isasymlambda}{\isasymxi}{\isachardot}\ \ }{\isacharparenleft}sqnf\ {\isacharparenleft}rt\ {\isasymxi}{\isacharparenright}\ {\isacharparenleft}dip\ {\isasymxi}{\isacharparenright}{\isacharparenright}\ {\isacharparenleft}ip\ {\isasymxi}{\isacharparenright}\ {\isacharparenleft}sn\ {\isasymxi}{\isacharparenright}\ {\isacharparenleft}ip\ {\isasymxi}{\isacharparenright}{\isacharparenright}\ {\isachardot}\isanewline
{\isacharbraceleft}PAodv{\isacharminus}{\isacharcolon}{\isadigit{2}}{\isadigit{6}}{\isacharbraceright}{\isasymlbrakk}clear{\isacharunderscore}locals{\isasymrbrakk}\isanewline
call{\isacharparenleft}PAodv{\isacharparenright}%
\end{isabelle}}
\snip{gammap_ppkt}{0}{0}{\begin{isabelle}%
\gammaaodv\ PNewPkt\ {\isacharequal}\isanewline
{\isacharbraceleft}PNewPkt{\isacharminus}{\isacharcolon}{\isadigit{0}}{\isacharbraceright}{\isasymlangle}{\isasymlambda}{\isasymxi}{\isachardot}\ \textsf{if}\ dip\ {\isasymxi}\ {\isacharequal}\ ip\ {\isasymxi}\ \textsf{then}\ {\isacharbraceleft}{\isasymxi}{\isacharbraceright}\ \textsf{else}\ {\isasymemptyset}{\isasymrangle}\isanewline
{\isacharbraceleft}PNewPkt{\isacharminus}{\isacharcolon}{\isadigit{1}}{\isacharbraceright}deliver{\isacharparenleft}data{\isacharparenright}\ {\isachardot}\isanewline
{\isacharbraceleft}PNewPkt{\isacharminus}{\isacharcolon}{\isadigit{2}}{\isacharbraceright}{\isasymlbrakk}clear{\isacharunderscore}locals{\isasymrbrakk}\isanewline
call{\isacharparenleft}PAodv{\isacharparenright}\isanewline
{\isasymoplus}\isanewline
{\isacharbraceleft}PNewPkt{\isacharminus}{\isacharcolon}{\isadigit{0}}{\isacharbraceright}{\isasymlangle}{\isasymlambda}{\isasymxi}{\isachardot}\ \textsf{if}\ dip\ {\isasymxi}\ {\isasymnoteq}\ ip\ {\isasymxi}\ \textsf{then}\ {\isacharbraceleft}{\isasymxi}{\isacharbraceright}\ \textsf{else}\ {\isasymemptyset}{\isasymrangle}\isanewline
{\isacharbraceleft}PNewPkt{\isacharminus}{\isacharcolon}{\isadigit{3}}{\isacharbraceright}{\isasymlbrakk}{\isasymlambda}{\isasymxi}{\isachardot}\ {\isasymxi}{\isasymlparr}store\ {\isacharcolon}{\isacharequal}\ add\ {\isacharparenleft}data\ {\isasymxi}{\isacharparenright}\ {\isacharparenleft}dip\ {\isasymxi}{\isacharparenright}\ {\isacharparenleft}store\ {\isasymxi}{\isacharparenright}{\isasymrparr}{\isasymrbrakk}\isanewline
{\isacharbraceleft}PNewPkt{\isacharminus}{\isacharcolon}{\isadigit{4}}{\isacharbraceright}{\isasymlbrakk}clear{\isacharunderscore}locals{\isasymrbrakk}\isanewline
call{\isacharparenleft}PAodv{\isacharparenright}%
\end{isabelle}}
\snip{gammap_prreq}{0}{0}{\begin{isabelle}%
\gammaaodv\ PPkt\ {\isacharequal}\isanewline
{\isacharbraceleft}PPkt{\isacharminus}{\isacharcolon}{\isadigit{0}}{\isacharbraceright}{\isasymlangle}{\isasymlambda}{\isasymxi}{\isachardot}\ \textsf{if}\ dip\ {\isasymxi}\ {\isacharequal}\ ip\ {\isasymxi}\ \textsf{then}\ {\isacharbraceleft}{\isasymxi}{\isacharbraceright}\ \textsf{else}\ {\isasymemptyset}{\isasymrangle}\isanewline
{\isacharbraceleft}PPkt{\isacharminus}{\isacharcolon}{\isadigit{1}}{\isacharbraceright}deliver{\isacharparenleft}data{\isacharparenright}\ {\isachardot}\isanewline
{\isacharbraceleft}PPkt{\isacharminus}{\isacharcolon}{\isadigit{2}}{\isacharbraceright}{\isasymlbrakk}clear{\isacharunderscore}locals{\isasymrbrakk}\isanewline
call{\isacharparenleft}PAodv{\isacharparenright}\isanewline
{\isasymoplus}\isanewline
{\isacharbraceleft}PPkt{\isacharminus}{\isacharcolon}{\isadigit{0}}{\isacharbraceright}{\isasymlangle}{\isasymlambda}{\isasymxi}{\isachardot}\ \textsf{if}\ dip\ {\isasymxi}\ {\isasymnoteq}\ ip\ {\isasymxi}\ \textsf{then}\ {\isacharbraceleft}{\isasymxi}{\isacharbraceright}\ \textsf{else}\ {\isasymemptyset}{\isasymrangle}\isanewline
{\isacharparenleft}{\isacharbraceleft}PPkt{\isacharminus}{\isacharcolon}{\isadigit{3}}{\isacharbraceright}{\isasymlangle}{\isasymlambda}{\isasymxi}{\isachardot}\ \textsf{if}\ dip\ {\isasymxi}{\isasymin}vD\ {\isacharparenleft}rt\ {\isasymxi}{\isacharparenright}\ \textsf{then}\ {\isacharbraceleft}{\isasymxi}{\isacharbraceright}\ \textsf{else}\ {\isasymemptyset}{\isasymrangle}\isanewline
\isaindent{{\isacharparenleft}}{\isacharbraceleft}PPkt{\isacharminus}{\isacharcolon}{\isadigit{4}}{\isacharbraceright}unicast{\isacharparenleft}{\isasymlambda}{\isasymxi}{\isachardot}\ the\ {\isacharparenleft}nh{\isacharparenleft}op\ {\isacharequal}{\isacharparenright}rt\ {\isasymxi}{\isacharparenright}\ {\isacharparenleft}dip\ {\isasymxi}{\isacharparenright}{\isacharparenright},\isanewline
\isaindent{{\isacharparenleft}{\isacharbraceleft}PPkt{\isacharminus}{\isacharcolon}{\isadigit{4}}{\isacharbraceright}unicast{\isacharparenleft}\ }{\isasymlambda}{\isasymxi}{\isachardot}\ Pkt\ {\isacharparenleft}data\ {\isasymxi}{\isacharparenright}\ {\isacharparenleft}dip\ {\isasymxi}{\isacharparenright}\ {\isacharparenleft}oip\ {\isasymxi}{\isacharparenright}{\isacharparenright}\ {\isachardot}\isanewline
\isaindent{{\isacharparenleft}\ \ \ }{\isacharbraceleft}PPkt{\isacharminus}{\isacharcolon}{\isadigit{5}}{\isacharbraceright}{\isasymlbrakk}clear{\isacharunderscore}locals{\isasymrbrakk}\isanewline
\isaindent{{\isacharparenleft}\ \ \ }call{\isacharparenleft}PAodv{\isacharparenright}\isanewline
\isaindent{{\isacharparenleft}\ \ \ }{\isasymtriangleright}\ {\isacharbraceleft}PPkt{\isacharminus}{\isacharcolon}{\isadigit{6}}{\isacharbraceright}{\isasymlbrakk}{\isasymlambda}{\isasymxi}{\isachardot}\ {\isasymxi}{\isasymlparr}dests\ {\isacharcolon}{\isacharequal}\ {\isasymlambda}rip{\isachardot}\ \textsf{if}\ rip{\isasymin}vD\ {\isacharparenleft}rt\ {\isasymxi}{\isacharparenright}\ {\isasymand}\ nh{\isacharparenleft}op\ {\isacharequal}{\isacharparenright}rt\ {\isasymxi}{\isacharparenright}\ rip\ {\isacharequal}\ nh{\isacharparenleft}op\ {\isacharequal}{\isacharparenright}rt\ {\isasymxi}{\isacharparenright}\ {\isacharparenleft}dip\ {\isasymxi}{\isacharparenright}\ \textsf{then}\ Some\ {\isacharparenleft}inc\ {\isacharparenleft}sqn\ {\isacharparenleft}rt\ {\isasymxi}{\isacharparenright}\ rip{\isacharparenright}{\isacharparenright}\ \textsf{else}\ None{\isasymrparr}{\isasymrbrakk}\isanewline
\isaindent{{\isacharparenleft}\ \ \ \ \ }{\isacharbraceleft}PPkt{\isacharminus}{\isacharcolon}{\isadigit{7}}{\isacharbraceright}{\isasymlbrakk}{\isasymlambda}{\isasymxi}{\isachardot}\ {\isasymxi}{\isasymlparr}rt\ {\isacharcolon}{\isacharequal}\ invalidate\ {\isacharparenleft}rt\ {\isasymxi}{\isacharparenright}\ {\isacharparenleft}dests\ {\isasymxi}{\isacharparenright}{\isasymrparr}{\isasymrbrakk}\isanewline
\isaindent{{\isacharparenleft}\ \ \ \ \ }{\isacharbraceleft}PPkt{\isacharminus}{\isacharcolon}{\isadigit{8}}{\isacharbraceright}{\isasymlbrakk}{\isasymlambda}{\isasymxi}{\isachardot}\ {\isasymxi}{\isasymlparr}store\ {\isacharcolon}{\isacharequal}\ setRRF\ {\isacharparenleft}store\ {\isasymxi}{\isacharparenright}\ {\isacharparenleft}dests\ {\isasymxi}{\isacharparenright}{\isasymrparr}{\isasymrbrakk}\isanewline
\isaindent{{\isacharparenleft}\ \ \ \ \ }{\isacharbraceleft}PPkt{\isacharminus}{\isacharcolon}{\isadigit{9}}{\isacharbraceright}{\isasymlbrakk}{\isasymlambda}{\isasymxi}{\isachardot}\ {\isasymxi}{\isasymlparr}pre\ {\isacharcolon}{\isacharequal}\ {\isasymUnion}{\isacharbraceleft}the\ {\isacharparenleft}precs\ {\isacharparenleft}rt\ {\isasymxi}{\isacharparenright}\ rip{\isacharparenright}\ {\isacharbar}\ rip{\isasymin}dom\ {\isacharparenleft}dests\ {\isasymxi}{\isacharparenright}{\isacharbraceright}{\isasymrparr}{\isasymrbrakk}\isanewline
\isaindent{{\isacharparenleft}\ \ \ \ \ }{\isacharbraceleft}PPkt{\isacharminus}{\isacharcolon}{\isadigit{1}}{\isadigit{0}}{\isacharbraceright}{\isasymlbrakk}{\isasymlambda}{\isasymxi}{\isachardot}\ {\isasymxi}{\isasymlparr}dests\ {\isacharcolon}{\isacharequal}\ {\isasymlambda}rip{\isachardot}\ \textsf{if}\ {\isacharparenleft}{\isasymexists}y{\isachardot}\ dests\ {\isasymxi}\ rip\ {\isacharequal}\ Some\ y{\isacharparenright}\ {\isasymand}\ the\ {\isacharparenleft}precs\ {\isacharparenleft}rt\ {\isasymxi}{\isacharparenright}\ rip{\isacharparenright}\ {\isasymnoteq}\ {\isasymemptyset}\ \textsf{then}\ dests\ {\isasymxi}\ rip\ \textsf{else}\ None{\isasymrparr}{\isasymrbrakk}\isanewline
\isaindent{{\isacharparenleft}\ \ \ \ \ }{\isacharbraceleft}PPkt{\isacharminus}{\isacharcolon}{\isadigit{1}}{\isadigit{1}}{\isacharbraceright}groupcast{\isacharparenleft}pre,\ {\isasymlambda}{\isasymxi}{\isachardot}\ Rerr\ {\isacharparenleft}dests\ {\isasymxi}{\isacharparenright}\ {\isacharparenleft}ip\ {\isasymxi}{\isacharparenright}{\isacharparenright}\ {\isachardot}\isanewline
\isaindent{{\isacharparenleft}\ \ \ \ \ }{\isacharbraceleft}PPkt{\isacharminus}{\isacharcolon}{\isadigit{1}}{\isadigit{2}}{\isacharbraceright}{\isasymlbrakk}clear{\isacharunderscore}locals{\isasymrbrakk}\isanewline
\isaindent{{\isacharparenleft}\ \ \ \ \ }call{\isacharparenleft}PAodv{\isacharparenright}\isanewline
\isaindent{{\isacharparenleft}}{\isasymoplus}\isanewline
\isaindent{{\isacharparenleft}}{\isacharbraceleft}PPkt{\isacharminus}{\isacharcolon}{\isadigit{3}}{\isacharbraceright}{\isasymlangle}{\isasymlambda}{\isasymxi}{\isachardot}\ \textsf{if}\ dip\ {\isasymxi}\ {\isasymnotin}\ vD\ {\isacharparenleft}rt\ {\isasymxi}{\isacharparenright}\ \textsf{then}\ {\isacharbraceleft}{\isasymxi}{\isacharbraceright}\ \textsf{else}\ {\isasymemptyset}{\isasymrangle}\isanewline
\isaindent{{\isacharparenleft}}{\isacharparenleft}{\isacharbraceleft}PPkt{\isacharminus}{\isacharcolon}{\isadigit{1}}{\isadigit{3}}{\isacharbraceright}{\isasymlangle}{\isasymlambda}{\isasymxi}{\isachardot}\ \textsf{if}\ dip\ {\isasymxi}{\isasymin}iD\ {\isacharparenleft}rt\ {\isasymxi}{\isacharparenright}\ \textsf{then}\ {\isacharbraceleft}{\isasymxi}{\isacharbraceright}\ \textsf{else}\ {\isasymemptyset}{\isasymrangle}\isanewline
\isaindent{{\isacharparenleft}{\isacharparenleft}}{\isacharbraceleft}PPkt{\isacharminus}{\isacharcolon}{\isadigit{1}}{\isadigit{4}}{\isacharbraceright}groupcast{\isacharparenleft}{\isasymlambda}{\isasymxi}{\isachardot}\ the\ {\isacharparenleft}precs\ {\isacharparenleft}rt\ {\isasymxi}{\isacharparenright}\ {\isacharparenleft}dip\ {\isasymxi}{\isacharparenright}{\isacharparenright},\isanewline
\isaindent{{\isacharparenleft}{\isacharparenleft}{\isacharbraceleft}PPkt{\isacharminus}{\isacharcolon}{\isadigit{1}}{\isadigit{4}}{\isacharbraceright}groupcast{\isacharparenleft}\ }{\isasymlambda}{\isasymxi}{\isachardot}\ Rerr\ {\isacharbrackleft}dip\ {\isasymxi}\ {\isasymmapsto}\ sqn\ {\isacharparenleft}rt\ {\isasymxi}{\isacharparenright}\ {\isacharparenleft}dip\ {\isasymxi}{\isacharparenright}{\isacharbrackright}\ {\isacharparenleft}ip\ {\isasymxi}{\isacharparenright}{\isacharparenright}\ {\isachardot}\isanewline
\isaindent{{\isacharparenleft}{\isacharparenleft}}{\isacharbraceleft}PPkt{\isacharminus}{\isacharcolon}{\isadigit{1}}{\isadigit{5}}{\isacharbraceright}{\isasymlbrakk}clear{\isacharunderscore}locals{\isasymrbrakk}\isanewline
\isaindent{{\isacharparenleft}{\isacharparenleft}}call{\isacharparenleft}PAodv{\isacharparenright}\isanewline
\isaindent{{\isacharparenleft}{\isacharparenleft}}{\isasymoplus}\isanewline
\isaindent{{\isacharparenleft}{\isacharparenleft}}{\isacharbraceleft}PPkt{\isacharminus}{\isacharcolon}{\isadigit{1}}{\isadigit{3}}{\isacharbraceright}{\isasymlangle}{\isasymlambda}{\isasymxi}{\isachardot}\ \textsf{if}\ dip\ {\isasymxi}\ {\isasymnotin}\ iD\ {\isacharparenleft}rt\ {\isasymxi}{\isacharparenright}\ \textsf{then}\ {\isacharbraceleft}{\isasymxi}{\isacharbraceright}\ \textsf{else}\ {\isasymemptyset}{\isasymrangle}\isanewline
\isaindent{{\isacharparenleft}{\isacharparenleft}}{\isacharbraceleft}PPkt{\isacharminus}{\isacharcolon}{\isadigit{1}}{\isadigit{6}}{\isacharbraceright}{\isasymlbrakk}clear{\isacharunderscore}locals{\isasymrbrakk}\isanewline
\isaindent{{\isacharparenleft}{\isacharparenleft}}call{\isacharparenleft}PAodv{\isacharparenright}{\isacharparenright}{\isacharparenright}%
\end{isabelle}}
\snip{gammap_prrep}{0}{0}{\begin{isabelle}%
\gammaaodv\ PRreq\ {\isacharequal}\isanewline
{\isacharbraceleft}PRreq{\isacharminus}{\isacharcolon}{\isadigit{0}}{\isacharbraceright}{\isasymlangle}{\isasymlambda}{\isasymxi}{\isachardot}\ \textsf{if}\ {\isacharparenleft}oip\ {\isasymxi},\ rreqid\ {\isasymxi}{\isacharparenright}{\isasymin}rreqs\ {\isasymxi}\ \textsf{then}\ {\isacharbraceleft}{\isasymxi}{\isacharbraceright}\ \textsf{else}\ {\isasymemptyset}{\isasymrangle}\isanewline
{\isacharbraceleft}PRreq{\isacharminus}{\isacharcolon}{\isadigit{1}}{\isacharbraceright}{\isasymlbrakk}clear{\isacharunderscore}locals{\isasymrbrakk}\isanewline
call{\isacharparenleft}PAodv{\isacharparenright}\isanewline
{\isasymoplus}\isanewline
{\isacharbraceleft}PRreq{\isacharminus}{\isacharcolon}{\isadigit{0}}{\isacharbraceright}{\isasymlangle}{\isasymlambda}{\isasymxi}{\isachardot}\ \textsf{if}\ {\isacharparenleft}oip\ {\isasymxi},\ rreqid\ {\isasymxi}{\isacharparenright}\ {\isasymnotin}\ rreqs\ {\isasymxi}\ \textsf{then}\ {\isacharbraceleft}{\isasymxi}{\isacharbraceright}\ \textsf{else}\ {\isasymemptyset}{\isasymrangle}\isanewline
{\isacharbraceleft}PRreq{\isacharminus}{\isacharcolon}{\isadigit{2}}{\isacharbraceright}{\isasymlbrakk}{\isasymlambda}{\isasymxi}{\isachardot}\ {\isasymxi}{\isasymlparr}rt\ {\isacharcolon}{\isacharequal}\ update\ {\isacharparenleft}rt\ {\isasymxi}{\isacharparenright}\ {\isacharparenleft}oip\ {\isasymxi}{\isacharparenright}\ {\isacharparenleft}osn\ {\isasymxi},\ kno,\ val,\ Suc\ {\isacharparenleft}hops\ {\isasymxi}{\isacharparenright},\ sip\ {\isasymxi},\ {\isasymemptyset}{\isacharparenright}{\isasymrparr}{\isasymrbrakk}\isanewline
{\isacharbraceleft}PRreq{\isacharminus}{\isacharcolon}{\isadigit{3}}{\isacharbraceright}{\isasymlbrakk}{\isasymlambda}{\isasymxi}{\isachardot}\ {\isasymxi}{\isasymlparr}rreqs\ {\isacharcolon}{\isacharequal}\ {\isacharbraceleft}{\isacharparenleft}oip\ {\isasymxi},\ rreqid\ {\isasymxi}{\isacharparenright}{\isacharbraceright}\ {\isasymunion}\ rreqs\ {\isasymxi}{\isasymrparr}{\isasymrbrakk}\isanewline
{\isacharparenleft}{\isacharbraceleft}PRreq{\isacharminus}{\isacharcolon}{\isadigit{4}}{\isacharbraceright}{\isasymlangle}{\isasymlambda}{\isasymxi}{\isachardot}\ \textsf{if}\ dip\ {\isasymxi}\ {\isacharequal}\ ip\ {\isasymxi}\ \textsf{then}\ {\isacharbraceleft}{\isasymxi}{\isacharbraceright}\ \textsf{else}\ {\isasymemptyset}{\isasymrangle}\isanewline
\isaindent{{\isacharparenleft}}{\isacharbraceleft}PRreq{\isacharminus}{\isacharcolon}{\isadigit{5}}{\isacharbraceright}{\isasymlbrakk}{\isasymlambda}{\isasymxi}{\isachardot}\ {\isasymxi}{\isasymlparr}sn\ {\isacharcolon}{\isacharequal}\ max\ {\isacharparenleft}sn\ {\isasymxi}{\isacharparenright}\ {\isacharparenleft}dsn\ {\isasymxi}{\isacharparenright}{\isasymrparr}{\isasymrbrakk}\isanewline
\isaindent{{\isacharparenleft}}{\isacharbraceleft}PRreq{\isacharminus}{\isacharcolon}{\isadigit{6}}{\isacharbraceright}unicast{\isacharparenleft}{\isasymlambda}{\isasymxi}{\isachardot}\ the\ {\isacharparenleft}nh{\isacharparenleft}op\ {\isacharequal}{\isacharparenright}rt\ {\isasymxi}{\isacharparenright}\ {\isacharparenleft}oip\ {\isasymxi}{\isacharparenright}{\isacharparenright},\isanewline
\isaindent{{\isacharparenleft}{\isacharbraceleft}PRreq{\isacharminus}{\isacharcolon}{\isadigit{6}}{\isacharbraceright}unicast{\isacharparenleft}\ }{\isasymlambda}{\isasymxi}{\isachardot}\ Rrep\ {\isadigit{0}}\ {\isacharparenleft}dip\ {\isasymxi}{\isacharparenright}\ {\isacharparenleft}sn\ {\isasymxi}{\isacharparenright}\ {\isacharparenleft}oip\ {\isasymxi}{\isacharparenright}\ {\isacharparenleft}ip\ {\isasymxi}{\isacharparenright}{\isacharparenright}\ {\isachardot}\isanewline
\isaindent{{\isacharparenleft}\ \ \ }{\isacharbraceleft}PRreq{\isacharminus}{\isacharcolon}{\isadigit{7}}{\isacharbraceright}{\isasymlbrakk}clear{\isacharunderscore}locals{\isasymrbrakk}\isanewline
\isaindent{{\isacharparenleft}\ \ \ }call{\isacharparenleft}PAodv{\isacharparenright}\isanewline
\isaindent{{\isacharparenleft}\ \ \ }{\isasymtriangleright}\ {\isacharbraceleft}PRreq{\isacharminus}{\isacharcolon}{\isadigit{8}}{\isacharbraceright}{\isasymlbrakk}{\isasymlambda}{\isasymxi}{\isachardot}\ {\isasymxi}{\isasymlparr}dests\ {\isacharcolon}{\isacharequal}\ {\isasymlambda}rip{\isachardot}\ \textsf{if}\ rip{\isasymin}vD\ {\isacharparenleft}rt\ {\isasymxi}{\isacharparenright}\ {\isasymand}\ nh{\isacharparenleft}op\ {\isacharequal}{\isacharparenright}rt\ {\isasymxi}{\isacharparenright}\ rip\ {\isacharequal}\ nh{\isacharparenleft}op\ {\isacharequal}{\isacharparenright}rt\ {\isasymxi}{\isacharparenright}\ {\isacharparenleft}oip\ {\isasymxi}{\isacharparenright}\ \textsf{then}\ Some\ {\isacharparenleft}inc\ {\isacharparenleft}sqn\ {\isacharparenleft}rt\ {\isasymxi}{\isacharparenright}\ rip{\isacharparenright}{\isacharparenright}\ \textsf{else}\ None{\isasymrparr}{\isasymrbrakk}\isanewline
\isaindent{{\isacharparenleft}\ \ \ \ \ }{\isacharbraceleft}PRreq{\isacharminus}{\isacharcolon}{\isadigit{9}}{\isacharbraceright}{\isasymlbrakk}{\isasymlambda}{\isasymxi}{\isachardot}\ {\isasymxi}{\isasymlparr}rt\ {\isacharcolon}{\isacharequal}\ invalidate\ {\isacharparenleft}rt\ {\isasymxi}{\isacharparenright}\ {\isacharparenleft}dests\ {\isasymxi}{\isacharparenright}{\isasymrparr}{\isasymrbrakk}\isanewline
\isaindent{{\isacharparenleft}\ \ \ \ \ }{\isacharbraceleft}PRreq{\isacharminus}{\isacharcolon}{\isadigit{1}}{\isadigit{0}}{\isacharbraceright}{\isasymlbrakk}{\isasymlambda}{\isasymxi}{\isachardot}\ {\isasymxi}{\isasymlparr}store\ {\isacharcolon}{\isacharequal}\ setRRF\ {\isacharparenleft}store\ {\isasymxi}{\isacharparenright}\ {\isacharparenleft}dests\ {\isasymxi}{\isacharparenright}{\isasymrparr}{\isasymrbrakk}\isanewline
\isaindent{{\isacharparenleft}\ \ \ \ \ }{\isacharbraceleft}PRreq{\isacharminus}{\isacharcolon}{\isadigit{1}}{\isadigit{1}}{\isacharbraceright}{\isasymlbrakk}{\isasymlambda}{\isasymxi}{\isachardot}\ {\isasymxi}{\isasymlparr}pre\ {\isacharcolon}{\isacharequal}\ {\isasymUnion}{\isacharbraceleft}the\ {\isacharparenleft}precs\ {\isacharparenleft}rt\ {\isasymxi}{\isacharparenright}\ rip{\isacharparenright}\ {\isacharbar}\ rip{\isasymin}dom\ {\isacharparenleft}dests\ {\isasymxi}{\isacharparenright}{\isacharbraceright}{\isasymrparr}{\isasymrbrakk}\isanewline
\isaindent{{\isacharparenleft}\ \ \ \ \ }{\isacharbraceleft}PRreq{\isacharminus}{\isacharcolon}{\isadigit{1}}{\isadigit{2}}{\isacharbraceright}{\isasymlbrakk}{\isasymlambda}{\isasymxi}{\isachardot}\ {\isasymxi}{\isasymlparr}dests\ {\isacharcolon}{\isacharequal}\ {\isasymlambda}rip{\isachardot}\ \textsf{if}\ {\isacharparenleft}{\isasymexists}y{\isachardot}\ dests\ {\isasymxi}\ rip\ {\isacharequal}\ Some\ y{\isacharparenright}\ {\isasymand}\ the\ {\isacharparenleft}precs\ {\isacharparenleft}rt\ {\isasymxi}{\isacharparenright}\ rip{\isacharparenright}\ {\isasymnoteq}\ {\isasymemptyset}\ \textsf{then}\ dests\ {\isasymxi}\ rip\ \textsf{else}\ None{\isasymrparr}{\isasymrbrakk}\isanewline
\isaindent{{\isacharparenleft}\ \ \ \ \ }{\isacharbraceleft}PRreq{\isacharminus}{\isacharcolon}{\isadigit{1}}{\isadigit{3}}{\isacharbraceright}groupcast{\isacharparenleft}pre,\ {\isasymlambda}{\isasymxi}{\isachardot}\ Rerr\ {\isacharparenleft}dests\ {\isasymxi}{\isacharparenright}\ {\isacharparenleft}ip\ {\isasymxi}{\isacharparenright}{\isacharparenright}\ {\isachardot}\isanewline
\isaindent{{\isacharparenleft}\ \ \ \ \ }{\isacharbraceleft}PRreq{\isacharminus}{\isacharcolon}{\isadigit{1}}{\isadigit{4}}{\isacharbraceright}{\isasymlbrakk}clear{\isacharunderscore}locals{\isasymrbrakk}\isanewline
\isaindent{{\isacharparenleft}\ \ \ \ \ }call{\isacharparenleft}PAodv{\isacharparenright}\isanewline
\isaindent{{\isacharparenleft}}{\isasymoplus}\isanewline
\isaindent{{\isacharparenleft}}{\isacharbraceleft}PRreq{\isacharminus}{\isacharcolon}{\isadigit{4}}{\isacharbraceright}{\isasymlangle}{\isasymlambda}{\isasymxi}{\isachardot}\ \textsf{if}\ dip\ {\isasymxi}\ {\isasymnoteq}\ ip\ {\isasymxi}\ \textsf{then}\ {\isacharbraceleft}{\isasymxi}{\isacharbraceright}\ \textsf{else}\ {\isasymemptyset}{\isasymrangle}\isanewline
\isaindent{{\isacharparenleft}}{\isacharparenleft}{\isacharbraceleft}PRreq{\isacharminus}{\isacharcolon}{\isadigit{1}}{\isadigit{5}}{\isacharbraceright}{\isasymlangle}{\isasymlambda}{\isasymxi}{\isachardot}\ \textsf{if}\ dip\ {\isasymxi}{\isasymin}vD\ {\isacharparenleft}rt\ {\isasymxi}{\isacharparenright}\ {\isasymand}\ dsn\ {\isasymxi}\ {\isasymle}\ sqn\ {\isacharparenleft}rt\ {\isasymxi}{\isacharparenright}\ {\isacharparenleft}dip\ {\isasymxi}{\isacharparenright}\ {\isasymand}\ sqnf\ {\isacharparenleft}rt\ {\isasymxi}{\isacharparenright}\ {\isacharparenleft}dip\ {\isasymxi}{\isacharparenright}\ {\isacharequal}\ kno\ \textsf{then}\ {\isacharbraceleft}{\isasymxi}{\isacharbraceright}\ \textsf{else}\ {\isasymemptyset}{\isasymrangle}\isanewline
\isaindent{{\isacharparenleft}{\isacharparenleft}}{\isacharbraceleft}PRreq{\isacharminus}{\isacharcolon}{\isadigit{1}}{\isadigit{6}}{\isacharbraceright}{\isasymlbrakk}{\isasymlambda}{\isasymxi}{\isachardot}\ {\isasymxi}{\isasymlparr}rt\ {\isacharcolon}{\isacharequal}\ the\ {\isacharparenleft}addpreRT\ {\isacharparenleft}rt\ {\isasymxi}{\isacharparenright}\ {\isacharparenleft}dip\ {\isasymxi}{\isacharparenright}\ {\isacharbraceleft}sip\ {\isasymxi}{\isacharbraceright}{\isacharparenright}{\isasymrparr}{\isasymrbrakk}\isanewline
\isaindent{{\isacharparenleft}{\isacharparenleft}}{\isacharbraceleft}PRreq{\isacharminus}{\isacharcolon}{\isadigit{1}}{\isadigit{7}}{\isacharbraceright}{\isasymlbrakk}{\isasymlambda}{\isasymxi}{\isachardot}\ {\isasymxi}{\isasymlparr}rt\ {\isacharcolon}{\isacharequal}\ the\ {\isacharparenleft}addpreRT\ {\isacharparenleft}rt\ {\isasymxi}{\isacharparenright}\ {\isacharparenleft}oip\ {\isasymxi}{\isacharparenright}\ {\isacharbraceleft}the\ {\isacharparenleft}nh{\isacharparenleft}op\ {\isacharequal}{\isacharparenright}rt\ {\isasymxi}{\isacharparenright}\ {\isacharparenleft}dip\ {\isasymxi}{\isacharparenright}{\isacharparenright}{\isacharbraceright}{\isacharparenright}{\isasymrparr}{\isasymrbrakk}\isanewline
\isaindent{{\isacharparenleft}{\isacharparenleft}}{\isacharbraceleft}PRreq{\isacharminus}{\isacharcolon}{\isadigit{1}}{\isadigit{8}}{\isacharbraceright}unicast{\isacharparenleft}{\isasymlambda}{\isasymxi}{\isachardot}\ the\ {\isacharparenleft}nh{\isacharparenleft}op\ {\isacharequal}{\isacharparenright}rt\ {\isasymxi}{\isacharparenright}\ {\isacharparenleft}oip\ {\isasymxi}{\isacharparenright}{\isacharparenright},\isanewline
\isaindent{{\isacharparenleft}{\isacharparenleft}{\isacharbraceleft}PRreq{\isacharminus}{\isacharcolon}{\isadigit{1}}{\isadigit{8}}{\isacharbraceright}unicast{\isacharparenleft}\ }{\isasymlambda}{\isasymxi}{\isachardot}\ Rrep\ {\isacharparenleft}the\ {\isacharparenleft}dhops\ {\isacharparenleft}rt\ {\isasymxi}{\isacharparenright}\ {\isacharparenleft}dip\ {\isasymxi}{\isacharparenright}{\isacharparenright}{\isacharparenright}\ {\isacharparenleft}dip\ {\isasymxi}{\isacharparenright}\isanewline
\isaindent{{\isacharparenleft}{\isacharparenleft}{\isacharbraceleft}PRreq{\isacharminus}{\isacharcolon}{\isadigit{1}}{\isadigit{8}}{\isacharbraceright}unicast{\isacharparenleft}\ {\isasymlambda}{\isasymxi}{\isachardot}\ \ }{\isacharparenleft}sqn\ {\isacharparenleft}rt\ {\isasymxi}{\isacharparenright}\ {\isacharparenleft}dip\ {\isasymxi}{\isacharparenright}{\isacharparenright}\ {\isacharparenleft}oip\ {\isasymxi}{\isacharparenright}\ {\isacharparenleft}ip\ {\isasymxi}{\isacharparenright}{\isacharparenright}\ {\isachardot}\isanewline
\isaindent{{\isacharparenleft}{\isacharparenleft}\ \ \ }{\isacharbraceleft}PRreq{\isacharminus}{\isacharcolon}{\isadigit{1}}{\isadigit{9}}{\isacharbraceright}{\isasymlbrakk}clear{\isacharunderscore}locals{\isasymrbrakk}\isanewline
\isaindent{{\isacharparenleft}{\isacharparenleft}\ \ \ }call{\isacharparenleft}PAodv{\isacharparenright}\isanewline
\isaindent{{\isacharparenleft}{\isacharparenleft}\ \ \ }{\isasymtriangleright}\ {\isacharbraceleft}PRreq{\isacharminus}{\isacharcolon}{\isadigit{2}}{\isadigit{0}}{\isacharbraceright}{\isasymlbrakk}{\isasymlambda}{\isasymxi}{\isachardot}\ {\isasymxi}{\isasymlparr}dests\ {\isacharcolon}{\isacharequal}\ {\isasymlambda}rip{\isachardot}\ \textsf{if}\ rip{\isasymin}vD\ {\isacharparenleft}rt\ {\isasymxi}{\isacharparenright}\ {\isasymand}\ nh{\isacharparenleft}op\ {\isacharequal}{\isacharparenright}rt\ {\isasymxi}{\isacharparenright}\ rip\ {\isacharequal}\ nh{\isacharparenleft}op\ {\isacharequal}{\isacharparenright}rt\ {\isasymxi}{\isacharparenright}\ {\isacharparenleft}oip\ {\isasymxi}{\isacharparenright}\ \textsf{then}\ Some\ {\isacharparenleft}inc\ {\isacharparenleft}sqn\ {\isacharparenleft}rt\ {\isasymxi}{\isacharparenright}\ rip{\isacharparenright}{\isacharparenright}\ \textsf{else}\ None{\isasymrparr}{\isasymrbrakk}\isanewline
\isaindent{{\isacharparenleft}{\isacharparenleft}\ \ \ \ \ }{\isacharbraceleft}PRreq{\isacharminus}{\isacharcolon}{\isadigit{2}}{\isadigit{1}}{\isacharbraceright}{\isasymlbrakk}{\isasymlambda}{\isasymxi}{\isachardot}\ {\isasymxi}{\isasymlparr}rt\ {\isacharcolon}{\isacharequal}\ invalidate\ {\isacharparenleft}rt\ {\isasymxi}{\isacharparenright}\ {\isacharparenleft}dests\ {\isasymxi}{\isacharparenright}{\isasymrparr}{\isasymrbrakk}\isanewline
\isaindent{{\isacharparenleft}{\isacharparenleft}\ \ \ \ \ }{\isacharbraceleft}PRreq{\isacharminus}{\isacharcolon}{\isadigit{2}}{\isadigit{2}}{\isacharbraceright}{\isasymlbrakk}{\isasymlambda}{\isasymxi}{\isachardot}\ {\isasymxi}{\isasymlparr}store\ {\isacharcolon}{\isacharequal}\ setRRF\ {\isacharparenleft}store\ {\isasymxi}{\isacharparenright}\ {\isacharparenleft}dests\ {\isasymxi}{\isacharparenright}{\isasymrparr}{\isasymrbrakk}\isanewline
\isaindent{{\isacharparenleft}{\isacharparenleft}\ \ \ \ \ }{\isacharbraceleft}PRreq{\isacharminus}{\isacharcolon}{\isadigit{2}}{\isadigit{3}}{\isacharbraceright}{\isasymlbrakk}{\isasymlambda}{\isasymxi}{\isachardot}\ {\isasymxi}{\isasymlparr}pre\ {\isacharcolon}{\isacharequal}\ {\isasymUnion}{\isacharbraceleft}the\ {\isacharparenleft}precs\ {\isacharparenleft}rt\ {\isasymxi}{\isacharparenright}\ rip{\isacharparenright}\ {\isacharbar}\ rip{\isasymin}dom\ {\isacharparenleft}dests\ {\isasymxi}{\isacharparenright}{\isacharbraceright}{\isasymrparr}{\isasymrbrakk}\isanewline
\isaindent{{\isacharparenleft}{\isacharparenleft}\ \ \ \ \ }{\isacharbraceleft}PRreq{\isacharminus}{\isacharcolon}{\isadigit{2}}{\isadigit{4}}{\isacharbraceright}{\isasymlbrakk}{\isasymlambda}{\isasymxi}{\isachardot}\ {\isasymxi}{\isasymlparr}dests\ {\isacharcolon}{\isacharequal}\ {\isasymlambda}rip{\isachardot}\ \textsf{if}\ {\isacharparenleft}{\isasymexists}y{\isachardot}\ dests\ {\isasymxi}\ rip\ {\isacharequal}\ Some\ y{\isacharparenright}\ {\isasymand}\ the\ {\isacharparenleft}precs\ {\isacharparenleft}rt\ {\isasymxi}{\isacharparenright}\ rip{\isacharparenright}\ {\isasymnoteq}\ {\isasymemptyset}\ \textsf{then}\ dests\ {\isasymxi}\ rip\ \textsf{else}\ None{\isasymrparr}{\isasymrbrakk}\isanewline
\isaindent{{\isacharparenleft}{\isacharparenleft}\ \ \ \ \ }{\isacharbraceleft}PRreq{\isacharminus}{\isacharcolon}{\isadigit{2}}{\isadigit{5}}{\isacharbraceright}groupcast{\isacharparenleft}pre,\ {\isasymlambda}{\isasymxi}{\isachardot}\ Rerr\ {\isacharparenleft}dests\ {\isasymxi}{\isacharparenright}\ {\isacharparenleft}ip\ {\isasymxi}{\isacharparenright}{\isacharparenright}\ {\isachardot}\isanewline
\isaindent{{\isacharparenleft}{\isacharparenleft}\ \ \ \ \ }{\isacharbraceleft}PRreq{\isacharminus}{\isacharcolon}{\isadigit{2}}{\isadigit{6}}{\isacharbraceright}{\isasymlbrakk}clear{\isacharunderscore}locals{\isasymrbrakk}\isanewline
\isaindent{{\isacharparenleft}{\isacharparenleft}\ \ \ \ \ }call{\isacharparenleft}PAodv{\isacharparenright}\isanewline
\isaindent{{\isacharparenleft}{\isacharparenleft}}{\isasymoplus}\isanewline
\isaindent{{\isacharparenleft}{\isacharparenleft}}{\isacharbraceleft}PRreq{\isacharminus}{\isacharcolon}{\isadigit{1}}{\isadigit{5}}{\isacharbraceright}{\isasymlangle}{\isasymlambda}{\isasymxi}{\isachardot}\ \textsf{if}\ dip\ {\isasymxi}{\isasymin}vD\ {\isacharparenleft}rt\ {\isasymxi}{\isacharparenright}\ {\isasymlongrightarrow}\ sqn\ {\isacharparenleft}rt\ {\isasymxi}{\isacharparenright}\ {\isacharparenleft}dip\ {\isasymxi}{\isacharparenright}\ {\isacharless}\ dsn\ {\isasymxi}\ {\isasymor}\ {\isasympi}\isactrlsub {\isadigit{3}}\ {\isacharparenleft}the\ {\isacharparenleft}rt\ {\isasymxi}\ {\isacharparenleft}dip\ {\isasymxi}{\isacharparenright}{\isacharparenright}{\isacharparenright}\ {\isacharequal}\ unk\ \textsf{then}\ {\isacharbraceleft}{\isasymxi}{\isacharbraceright}\ \textsf{else}\ {\isasymemptyset}{\isasymrangle}\isanewline
\isaindent{{\isacharparenleft}{\isacharparenleft}}{\isacharbraceleft}PRreq{\isacharminus}{\isacharcolon}{\isadigit{2}}{\isadigit{7}}{\isacharbraceright}broadcast{\isacharparenleft}{\isasymlambda}{\isasymxi}{\isachardot}\ Rreq\ {\isacharparenleft}Suc\ {\isacharparenleft}hops\ {\isasymxi}{\isacharparenright}{\isacharparenright}\ {\isacharparenleft}rreqid\ {\isasymxi}{\isacharparenright}\ {\isacharparenleft}dip\ {\isasymxi}{\isacharparenright}\isanewline
\isaindent{{\isacharparenleft}{\isacharparenleft}{\isacharbraceleft}PRreq{\isacharminus}{\isacharcolon}{\isadigit{2}}{\isadigit{7}}{\isacharbraceright}broadcast{\isacharparenleft}{\isasymlambda}{\isasymxi}{\isachardot}\ \ }{\isacharparenleft}max\ {\isacharparenleft}sqn\ {\isacharparenleft}rt\ {\isasymxi}{\isacharparenright}\ {\isacharparenleft}dip\ {\isasymxi}{\isacharparenright}{\isacharparenright}\ {\isacharparenleft}dsn\ {\isasymxi}{\isacharparenright}{\isacharparenright}\ {\isacharparenleft}dsk\ {\isasymxi}{\isacharparenright}\isanewline
\isaindent{{\isacharparenleft}{\isacharparenleft}{\isacharbraceleft}PRreq{\isacharminus}{\isacharcolon}{\isadigit{2}}{\isadigit{7}}{\isacharbraceright}broadcast{\isacharparenleft}{\isasymlambda}{\isasymxi}{\isachardot}\ \ }{\isacharparenleft}oip\ {\isasymxi}{\isacharparenright}\ {\isacharparenleft}osn\ {\isasymxi}{\isacharparenright}\ {\isacharparenleft}ip\ {\isasymxi}{\isacharparenright}{\isacharparenright}\ {\isachardot}\isanewline
\isaindent{{\isacharparenleft}{\isacharparenleft}}{\isacharbraceleft}PRreq{\isacharminus}{\isacharcolon}{\isadigit{2}}{\isadigit{8}}{\isacharbraceright}{\isasymlbrakk}clear{\isacharunderscore}locals{\isasymrbrakk}\isanewline
\isaindent{{\isacharparenleft}{\isacharparenleft}}call{\isacharparenleft}PAodv{\isacharparenright}{\isacharparenright}{\isacharparenright}%
\end{isabelle}}
\snip{gammap_prerr}{0}{0}{\begin{isabelle}%
\gammaaodv\ PRrep\ {\isacharequal}\isanewline
{\isacharbraceleft}PRrep{\isacharminus}{\isacharcolon}{\isadigit{0}}{\isacharbraceright}{\isasymlangle}{\isasymlambda}{\isasymxi}{\isachardot}\ \textsf{if}\ rt\ {\isasymxi}\ {\isasymnoteq}\ update\ {\isacharparenleft}rt\ {\isasymxi}{\isacharparenright}\ {\isacharparenleft}dip\ {\isasymxi}{\isacharparenright}\ {\isacharparenleft}dsn\ {\isasymxi},\ kno,\ val,\ Suc\ {\isacharparenleft}hops\ {\isasymxi}{\isacharparenright},\ sip\ {\isasymxi},\ {\isasymemptyset}{\isacharparenright}\ \textsf{then}\ {\isacharbraceleft}{\isasymxi}{\isacharbraceright}\ \textsf{else}\ {\isasymemptyset}{\isasymrangle}\isanewline
{\isacharbraceleft}PRrep{\isacharminus}{\isacharcolon}{\isadigit{1}}{\isacharbraceright}{\isasymlbrakk}{\isasymlambda}{\isasymxi}{\isachardot}\ {\isasymxi}{\isasymlparr}rt\ {\isacharcolon}{\isacharequal}\ update\ {\isacharparenleft}rt\ {\isasymxi}{\isacharparenright}\ {\isacharparenleft}dip\ {\isasymxi}{\isacharparenright}\ {\isacharparenleft}dsn\ {\isasymxi},\ kno,\ val,\ Suc\ {\isacharparenleft}hops\ {\isasymxi}{\isacharparenright},\ sip\ {\isasymxi},\ {\isasymemptyset}{\isacharparenright}{\isasymrparr}{\isasymrbrakk}\isanewline
{\isacharparenleft}{\isacharbraceleft}PRrep{\isacharminus}{\isacharcolon}{\isadigit{2}}{\isacharbraceright}{\isasymlangle}{\isasymlambda}{\isasymxi}{\isachardot}\ \textsf{if}\ oip\ {\isasymxi}\ {\isacharequal}\ ip\ {\isasymxi}\ \textsf{then}\ {\isacharbraceleft}{\isasymxi}{\isacharbraceright}\ \textsf{else}\ {\isasymemptyset}{\isasymrangle}\isanewline
\isaindent{{\isacharparenleft}}{\isacharbraceleft}PRrep{\isacharminus}{\isacharcolon}{\isadigit{3}}{\isacharbraceright}{\isasymlbrakk}clear{\isacharunderscore}locals{\isasymrbrakk}\isanewline
\isaindent{{\isacharparenleft}}call{\isacharparenleft}PAodv{\isacharparenright}\isanewline
\isaindent{{\isacharparenleft}}{\isasymoplus}\isanewline
\isaindent{{\isacharparenleft}}{\isacharbraceleft}PRrep{\isacharminus}{\isacharcolon}{\isadigit{2}}{\isacharbraceright}{\isasymlangle}{\isasymlambda}{\isasymxi}{\isachardot}\ \textsf{if}\ oip\ {\isasymxi}\ {\isasymnoteq}\ ip\ {\isasymxi}\ \textsf{then}\ {\isacharbraceleft}{\isasymxi}{\isacharbraceright}\ \textsf{else}\ {\isasymemptyset}{\isasymrangle}\isanewline
\isaindent{{\isacharparenleft}}{\isacharparenleft}{\isacharbraceleft}PRrep{\isacharminus}{\isacharcolon}{\isadigit{4}}{\isacharbraceright}{\isasymlangle}{\isasymlambda}{\isasymxi}{\isachardot}\ \textsf{if}\ oip\ {\isasymxi}{\isasymin}vD\ {\isacharparenleft}rt\ {\isasymxi}{\isacharparenright}\ \textsf{then}\ {\isacharbraceleft}{\isasymxi}{\isacharbraceright}\ \textsf{else}\ {\isasymemptyset}{\isasymrangle}\isanewline
\isaindent{{\isacharparenleft}{\isacharparenleft}}{\isacharbraceleft}PRrep{\isacharminus}{\isacharcolon}{\isadigit{5}}{\isacharbraceright}{\isasymlbrakk}{\isasymlambda}{\isasymxi}{\isachardot}\ {\isasymxi}{\isasymlparr}rt\ {\isacharcolon}{\isacharequal}\ the\ {\isacharparenleft}addpreRT\ {\isacharparenleft}rt\ {\isasymxi}{\isacharparenright}\ {\isacharparenleft}dip\ {\isasymxi}{\isacharparenright}\ {\isacharbraceleft}the\ {\isacharparenleft}nh{\isacharparenleft}op\ {\isacharequal}{\isacharparenright}rt\ {\isasymxi}{\isacharparenright}\ {\isacharparenleft}oip\ {\isasymxi}{\isacharparenright}{\isacharparenright}{\isacharbraceright}{\isacharparenright}{\isasymrparr}{\isasymrbrakk}\isanewline
\isaindent{{\isacharparenleft}{\isacharparenleft}}{\isacharbraceleft}PRrep{\isacharminus}{\isacharcolon}{\isadigit{6}}{\isacharbraceright}{\isasymlbrakk}{\isasymlambda}{\isasymxi}{\isachardot}\ {\isasymxi}{\isasymlparr}rt\ {\isacharcolon}{\isacharequal}\ the\ {\isacharparenleft}addpreRT\ {\isacharparenleft}rt\ {\isasymxi}{\isacharparenright}\ {\isacharparenleft}the\ {\isacharparenleft}nh{\isacharparenleft}op\ {\isacharequal}{\isacharparenright}rt\ {\isasymxi}{\isacharparenright}\ {\isacharparenleft}dip\ {\isasymxi}{\isacharparenright}{\isacharparenright}{\isacharparenright}\ {\isacharbraceleft}the\ {\isacharparenleft}nh{\isacharparenleft}op\ {\isacharequal}{\isacharparenright}rt\ {\isasymxi}{\isacharparenright}\ {\isacharparenleft}oip\ {\isasymxi}{\isacharparenright}{\isacharparenright}{\isacharbraceright}{\isacharparenright}{\isasymrparr}{\isasymrbrakk}\isanewline
\isaindent{{\isacharparenleft}{\isacharparenleft}}{\isacharbraceleft}PRrep{\isacharminus}{\isacharcolon}{\isadigit{7}}{\isacharbraceright}unicast{\isacharparenleft}{\isasymlambda}{\isasymxi}{\isachardot}\ the\ {\isacharparenleft}nh{\isacharparenleft}op\ {\isacharequal}{\isacharparenright}rt\ {\isasymxi}{\isacharparenright}\ {\isacharparenleft}oip\ {\isasymxi}{\isacharparenright}{\isacharparenright},\isanewline
\isaindent{{\isacharparenleft}{\isacharparenleft}{\isacharbraceleft}PRrep{\isacharminus}{\isacharcolon}{\isadigit{7}}{\isacharbraceright}unicast{\isacharparenleft}\ }{\isasymlambda}{\isasymxi}{\isachardot}\ Rrep\ {\isacharparenleft}Suc\ {\isacharparenleft}hops\ {\isasymxi}{\isacharparenright}{\isacharparenright}\ {\isacharparenleft}dip\ {\isasymxi}{\isacharparenright}\ {\isacharparenleft}dsn\ {\isasymxi}{\isacharparenright}\ {\isacharparenleft}oip\ {\isasymxi}{\isacharparenright}\isanewline
\isaindent{{\isacharparenleft}{\isacharparenleft}{\isacharbraceleft}PRrep{\isacharminus}{\isacharcolon}{\isadigit{7}}{\isacharbraceright}unicast{\isacharparenleft}\ {\isasymlambda}{\isasymxi}{\isachardot}\ \ }{\isacharparenleft}ip\ {\isasymxi}{\isacharparenright}{\isacharparenright}\ {\isachardot}\isanewline
\isaindent{{\isacharparenleft}{\isacharparenleft}\ \ \ }{\isacharbraceleft}PRrep{\isacharminus}{\isacharcolon}{\isadigit{8}}{\isacharbraceright}{\isasymlbrakk}clear{\isacharunderscore}locals{\isasymrbrakk}\isanewline
\isaindent{{\isacharparenleft}{\isacharparenleft}\ \ \ }call{\isacharparenleft}PAodv{\isacharparenright}\isanewline
\isaindent{{\isacharparenleft}{\isacharparenleft}\ \ \ }{\isasymtriangleright}\ {\isacharbraceleft}PRrep{\isacharminus}{\isacharcolon}{\isadigit{9}}{\isacharbraceright}{\isasymlbrakk}{\isasymlambda}{\isasymxi}{\isachardot}\ {\isasymxi}{\isasymlparr}dests\ {\isacharcolon}{\isacharequal}\ {\isasymlambda}rip{\isachardot}\ \textsf{if}\ rip{\isasymin}vD\ {\isacharparenleft}rt\ {\isasymxi}{\isacharparenright}\ {\isasymand}\ nh{\isacharparenleft}op\ {\isacharequal}{\isacharparenright}rt\ {\isasymxi}{\isacharparenright}\ rip\ {\isacharequal}\ nh{\isacharparenleft}op\ {\isacharequal}{\isacharparenright}rt\ {\isasymxi}{\isacharparenright}\ {\isacharparenleft}oip\ {\isasymxi}{\isacharparenright}\ \textsf{then}\ Some\ {\isacharparenleft}inc\ {\isacharparenleft}sqn\ {\isacharparenleft}rt\ {\isasymxi}{\isacharparenright}\ rip{\isacharparenright}{\isacharparenright}\ \textsf{else}\ None{\isasymrparr}{\isasymrbrakk}\isanewline
\isaindent{{\isacharparenleft}{\isacharparenleft}\ \ \ \ \ }{\isacharbraceleft}PRrep{\isacharminus}{\isacharcolon}{\isadigit{1}}{\isadigit{0}}{\isacharbraceright}{\isasymlbrakk}{\isasymlambda}{\isasymxi}{\isachardot}\ {\isasymxi}{\isasymlparr}rt\ {\isacharcolon}{\isacharequal}\ invalidate\ {\isacharparenleft}rt\ {\isasymxi}{\isacharparenright}\ {\isacharparenleft}dests\ {\isasymxi}{\isacharparenright}{\isasymrparr}{\isasymrbrakk}\isanewline
\isaindent{{\isacharparenleft}{\isacharparenleft}\ \ \ \ \ }{\isacharbraceleft}PRrep{\isacharminus}{\isacharcolon}{\isadigit{1}}{\isadigit{1}}{\isacharbraceright}{\isasymlbrakk}{\isasymlambda}{\isasymxi}{\isachardot}\ {\isasymxi}{\isasymlparr}store\ {\isacharcolon}{\isacharequal}\ setRRF\ {\isacharparenleft}store\ {\isasymxi}{\isacharparenright}\ {\isacharparenleft}dests\ {\isasymxi}{\isacharparenright}{\isasymrparr}{\isasymrbrakk}\isanewline
\isaindent{{\isacharparenleft}{\isacharparenleft}\ \ \ \ \ }{\isacharbraceleft}PRrep{\isacharminus}{\isacharcolon}{\isadigit{1}}{\isadigit{2}}{\isacharbraceright}{\isasymlbrakk}{\isasymlambda}{\isasymxi}{\isachardot}\ {\isasymxi}{\isasymlparr}pre\ {\isacharcolon}{\isacharequal}\ {\isasymUnion}{\isacharbraceleft}the\ {\isacharparenleft}precs\ {\isacharparenleft}rt\ {\isasymxi}{\isacharparenright}\ rip{\isacharparenright}\ {\isacharbar}\ rip{\isasymin}dom\ {\isacharparenleft}dests\ {\isasymxi}{\isacharparenright}{\isacharbraceright}{\isasymrparr}{\isasymrbrakk}\isanewline
\isaindent{{\isacharparenleft}{\isacharparenleft}\ \ \ \ \ }{\isacharbraceleft}PRrep{\isacharminus}{\isacharcolon}{\isadigit{1}}{\isadigit{3}}{\isacharbraceright}{\isasymlbrakk}{\isasymlambda}{\isasymxi}{\isachardot}\ {\isasymxi}{\isasymlparr}dests\ {\isacharcolon}{\isacharequal}\ {\isasymlambda}rip{\isachardot}\ \textsf{if}\ {\isacharparenleft}{\isasymexists}y{\isachardot}\ dests\ {\isasymxi}\ rip\ {\isacharequal}\ Some\ y{\isacharparenright}\ {\isasymand}\ the\ {\isacharparenleft}precs\ {\isacharparenleft}rt\ {\isasymxi}{\isacharparenright}\ rip{\isacharparenright}\ {\isasymnoteq}\ {\isasymemptyset}\ \textsf{then}\ dests\ {\isasymxi}\ rip\ \textsf{else}\ None{\isasymrparr}{\isasymrbrakk}\isanewline
\isaindent{{\isacharparenleft}{\isacharparenleft}\ \ \ \ \ }{\isacharbraceleft}PRrep{\isacharminus}{\isacharcolon}{\isadigit{1}}{\isadigit{4}}{\isacharbraceright}groupcast{\isacharparenleft}pre,\ {\isasymlambda}{\isasymxi}{\isachardot}\ Rerr\ {\isacharparenleft}dests\ {\isasymxi}{\isacharparenright}\ {\isacharparenleft}ip\ {\isasymxi}{\isacharparenright}{\isacharparenright}\ {\isachardot}\isanewline
\isaindent{{\isacharparenleft}{\isacharparenleft}\ \ \ \ \ }{\isacharbraceleft}PRrep{\isacharminus}{\isacharcolon}{\isadigit{1}}{\isadigit{5}}{\isacharbraceright}{\isasymlbrakk}clear{\isacharunderscore}locals{\isasymrbrakk}\isanewline
\isaindent{{\isacharparenleft}{\isacharparenleft}\ \ \ \ \ }call{\isacharparenleft}PAodv{\isacharparenright}\isanewline
\isaindent{{\isacharparenleft}{\isacharparenleft}}{\isasymoplus}\isanewline
\isaindent{{\isacharparenleft}{\isacharparenleft}}{\isacharbraceleft}PRrep{\isacharminus}{\isacharcolon}{\isadigit{4}}{\isacharbraceright}{\isasymlangle}{\isasymlambda}{\isasymxi}{\isachardot}\ \textsf{if}\ oip\ {\isasymxi}\ {\isasymnotin}\ vD\ {\isacharparenleft}rt\ {\isasymxi}{\isacharparenright}\ \textsf{then}\ {\isacharbraceleft}{\isasymxi}{\isacharbraceright}\ \textsf{else}\ {\isasymemptyset}{\isasymrangle}\isanewline
\isaindent{{\isacharparenleft}{\isacharparenleft}}{\isacharbraceleft}PRrep{\isacharminus}{\isacharcolon}{\isadigit{1}}{\isadigit{6}}{\isacharbraceright}{\isasymlbrakk}clear{\isacharunderscore}locals{\isasymrbrakk}\isanewline
\isaindent{{\isacharparenleft}{\isacharparenleft}}call{\isacharparenleft}PAodv{\isacharparenright}{\isacharparenright}{\isacharparenright}\isanewline
{\isasymoplus}\isanewline
{\isacharbraceleft}PRrep{\isacharminus}{\isacharcolon}{\isadigit{0}}{\isacharbraceright}{\isasymlangle}{\isasymlambda}{\isasymxi}{\isachardot}\ \textsf{if}\ rt\ {\isasymxi}\ {\isacharequal}\ update\ {\isacharparenleft}rt\ {\isasymxi}{\isacharparenright}\ {\isacharparenleft}dip\ {\isasymxi}{\isacharparenright}\ {\isacharparenleft}dsn\ {\isasymxi},\ kno,\ val,\ Suc\ {\isacharparenleft}hops\ {\isasymxi}{\isacharparenright},\ sip\ {\isasymxi},\ {\isasymemptyset}{\isacharparenright}\ \textsf{then}\ {\isacharbraceleft}{\isasymxi}{\isacharbraceright}\ \textsf{else}\ {\isasymemptyset}{\isasymrangle}\isanewline
{\isacharbraceleft}PRrep{\isacharminus}{\isacharcolon}{\isadigit{1}}{\isadigit{7}}{\isacharbraceright}{\isasymlbrakk}clear{\isacharunderscore}locals{\isasymrbrakk}\isanewline
call{\isacharparenleft}PAodv{\isacharparenright}%
\end{isabelle}}

\snip{sigmap}{0}{0}{\isa{\sigmaaodv\ i\ {\isacharequal}\ {\isacharbraceleft}{\isacharparenleft}aodv{\isacharunderscore}init\ i,\ \gammaaodv\ PAodv{\isacharparenright}{\isacharbraceright}}}
\snip{paodv}{0}{0}{\isa{paodv}}
\snip{paodv_term}{0}{0}{\isa{{\isasymlparr}init\ {\isacharequal}\ {\isacharbraceleft}{\isacharparenleft}aodv{\isacharunderscore}init\ i,\ \gammaaodv\ PAodv{\isacharparenright}{\isacharbraceright},\ trans\ {\isacharequal}\ \seqpsos\ \gammaaodv{\isasymrparr}}}
\snip{paodv_term'}{0}{0}{\isa{{\isasymlparr}init\ {\isacharequal}\ \sigmaaodv\ i,\ trans\ {\isacharequal}\ \seqpsos\ \gammaaodv{\isasymrparr}}}
\snip{paodv_type}{0}{0}{\isa{nat\ {\isasymRightarrow}\ {\isacharparenleft}state\ {\isasymtimes}\ {\isacharparenleft}state,\ msg,\ pseqp,\ pseqp\ label{\isacharparenright}\ seqp,\ msg\ seq{\isacharunderscore}action{\isacharparenright}\ automaton}}

\snip{sigmap'}{0}{0}{\isa{\sigmaaodv{\isacharprime}\ {\isacharequal}\ {\isacharbraceleft}{\isacharparenleft}aodv{\isacharunderscore}init,\ \gammaaodv\ PAodv{\isacharparenright}{\isacharbraceright}}}
\snip{opaodv}{0}{0}{\isa{opaodv}}
\snip{opaodv_term}{0}{0}{\isa{{\isasymlparr}init\ {\isacharequal}\ {\isacharbraceleft}{\isacharparenleft}aodv{\isacharunderscore}init,\ \gammaaodv\ PAodv{\isacharparenright}{\isacharbraceright},\ trans\ {\isacharequal}\ \oseqpsos\ \gammaaodv\ i{\isasymrparr}}}

\snip{gammaq}{0}{0}{\isa{\gammaqmsg\ {\isacharparenleft}{\isacharparenright}}}
\snip{gammaq_unit}{0}{0}{\begin{isabelle}%
\gammaqmsg\ {\isacharparenleft}{\isacharparenright}\ {\isacharequal}\isanewline
{\isacharbraceleft}{\isacharparenleft}{\isacharparenright}{\isacharminus}{\isacharcolon}{\isadigit{0}}{\isacharbraceright}receive{\isacharparenleft}{\isasymlambda}msg\ msgs{\isachardot}\ msgs\ {\isacharat}\ {\isacharbrackleft}msg{\isacharbrackright}{\isacharparenright}\ {\isachardot}\isanewline
call{\isacharparenleft}{\isacharparenleft}{\isacharparenright}{\isacharparenright}\isanewline
{\isasymoplus}\isanewline
{\isacharbraceleft}{\isacharparenleft}{\isacharparenright}{\isacharminus}{\isacharcolon}{\isadigit{0}}{\isacharbraceright}{\isasymlangle}{\isasymlambda}msgs{\isachardot}\ \textsf{if}\ msgs\ {\isasymnoteq}\ {\isacharbrackleft}{\isacharbrackright}\ \textsf{then}\ {\isacharbraceleft}msgs{\isacharbraceright}\ \textsf{else}\ {\isasymemptyset}{\isasymrangle}\isanewline
{\isacharparenleft}{\isacharbraceleft}{\isacharparenleft}{\isacharparenright}{\isacharminus}{\isacharcolon}{\isadigit{1}}{\isacharbraceright}send{\isacharparenleft}hd{\isacharparenright}\ {\isachardot}\isanewline
\isaindent{{\isacharparenleft}}{\isacharbraceleft}{\isacharparenleft}{\isacharparenright}{\isacharminus}{\isacharcolon}{\isadigit{2}}{\isacharbraceright}{\isasymlbrakk}tl{\isasymrbrakk}\isanewline
\isaindent{{\isacharparenleft}}call{\isacharparenleft}{\isacharparenleft}{\isacharparenright}{\isacharparenright}\isanewline
\isaindent{{\isacharparenleft}}{\isasymoplus}\isanewline
\isaindent{{\isacharparenleft}}{\isacharbraceleft}{\isacharparenleft}{\isacharparenright}{\isacharminus}{\isacharcolon}{\isadigit{1}}{\isacharbraceright}receive{\isacharparenleft}{\isasymlambda}msg\ msgs{\isachardot}\ msgs\ {\isacharat}\ {\isacharbrackleft}msg{\isacharbrackright}{\isacharparenright}\ {\isachardot}\isanewline
\isaindent{{\isacharparenleft}}call{\isacharparenleft}{\isacharparenleft}{\isacharparenright}{\isacharparenright}{\isacharparenright}%
\end{isabelle}}
\snip{sigmaq}{0}{0}{\isa{\gammaqmsg\ {\isacharequal}\ {\isacharbraceleft}{\isacharparenleft}{\isacharbrackleft}{\isacharbrackright},\ \gammaqmsg\ {\isacharparenleft}{\isacharparenright}{\isacharparenright}{\isacharbraceright}}}
\snip{qmsg}{0}{0}{\isa{qmsg}}
\snip{qmsg_term}{0}{0}{\isa{{\isasymlparr}init\ {\isacharequal}\ \gammaqmsg,\ trans\ {\isacharequal}\ \seqpsos\ \gammaqmsg{\isasymrparr}}}

\snip{clear_locals_sip}{0}{0}{\isa{{\isasymxi}{\isasymlparr}sip\ {\isacharcolon}{\isacharequal}\ SOME\ x{\isachardot}\ x\ {\isasymnoteq}\ ip\ {\isasymxi}{\isasymrparr}}}

\snip{paodv_qmsg_term}{0}{0}{\isa{paodv\ i\ {\isasymlangle}{\isasymlangle}\ qmsg}}
\snip{opnet_p}{0}{0}{\isa{opnet\ p}}
\snip{opnet_node}{0}{0}{\isa{opnet\ {\isacharparenleft}{\isasymlambda}i{\isachardot}\ opaodv\ i\ {\isasymlangle}{\isasymlangle}\isactrlbsub i\isactrlesub \ qmsg{\isacharparenright}\ {\isasymlangle}i{\isacharsemicolon}\ R{\isasymrangle}}}
\snip{net_tree_ips_node}{0}{0}{\isa{net{\isacharunderscore}tree{\isacharunderscore}ips\ {\isasymlangle}i{\isacharsemicolon}\ R{\isasymrangle}\ {\isacharequal}\ {\isacharbraceleft}i{\isacharbraceright}}}

\snip{opnet_p1_par_p2}{0}{0}{\isa{opnet\ {\isacharparenleft}{\isasymlambda}i{\isachardot}\ opaodv\ i\ {\isasymlangle}{\isasymlangle}\isactrlbsub i\isactrlesub \ qmsg{\isacharparenright}\ {\isacharparenleft}p\isactrlsub {\isadigit{1}}\parallelcomp{}p\isactrlsub {\isadigit{2}}{\isacharparenright}}}
\snip{net_tree_ips_par}{0}{0}{\isa{net{\isacharunderscore}tree{\isacharunderscore}ips\ {\isacharparenleft}p{\isadigit{1}}{\isachardot}{\isadigit{0}}\parallelcomp{}p{\isadigit{2}}{\isachardot}{\isadigit{0}}{\isacharparenright}\ {\isacharequal}\ net{\isacharunderscore}tree{\isacharunderscore}ips\ p{\isadigit{1}}{\isachardot}{\isadigit{0}}\ {\isasymunion}\ net{\isacharunderscore}tree{\isacharunderscore}ips\ p{\isadigit{2}}{\isachardot}{\isadigit{0}}}}

\snip{pnet_lift_prem1}{0}{0}{\isa{opnet\ onp\ {\isasymlangle}i{\isacharsemicolon}\ R{\isasymrangle}\ {\isasymTurnstile}\ {\isacharparenleft}otherwith\ S\ {\isacharbraceleft}i{\isacharbraceright}\ {\isacharparenleft}oarrivemsg\ I{\isacharparenright},\ other\ U\ {\isacharbraceleft}i{\isacharbraceright}\ {\isasymrightarrow}{\isacharparenright}\ global\ {\isacharparenleft}P\ i{\isacharparenright}}}

\snip{pnet_lift_assume_s}{0}{0}{\isa{otherwith\ S\ {\isacharparenleft}net{\isacharunderscore}tree{\isacharunderscore}ips\ p{\isacharparenright}\ {\isacharparenleft}oarrivemsg\ I{\isacharparenright}}}
\snip{pnet_lift_assume_u}{0}{0}{\isa{other\ U\ {\isacharparenleft}net{\isacharunderscore}tree{\isacharunderscore}ips\ p{\isacharparenright}}}
\snip{pnet_lift_shows}{0}{0}{\isa{global\ {\isacharparenleft}{\isasymlambda}{\isasymsigma}{\isachardot}\ {\isasymforall}i{\isasymin}net{\isacharunderscore}tree{\isacharunderscore}ips\ p{\isachardot}\ P\ i\ {\isasymsigma}{\isacharparenright}}}

\snip{pnet_lift_castmsg_pred}{0}{0}{\isa{castmsg\ {\isacharparenleft}I\ {\isasymsigma}{\isacharparenright}}}
\snip{pnet_lift_arrivemsg_pred}{0}{0}{\isa{oarrivemsg\ I\ {\isasymsigma}}}

\snip{pnet_lift_castmsg}{0}{0}{\isa{opnet\ onp\ {\isasymlangle}i{\isacharsemicolon}\ R{\isasymrangle}\ \ostepinv\ {\isacharparenleft}{\isasymlambda}{\isasymsigma}\ {\isacharunderscore}{\isachardot}\ oarrivemsg\ I\ {\isasymsigma},\ other\ U\ {\isacharbraceleft}i{\isacharbraceright}\ {\isasymrightarrow}{\isacharparenright}\ globala\ {\isacharparenleft}{\isasymlambda}{\isacharparenleft}{\isasymsigma},\ a,\ {\isasymsigma}{\isacharprime}{\isacharparenright}{\isachardot}\ castmsg\ {\isacharparenleft}I\ {\isasymsigma}{\isacharparenright}\ a{\isacharparenright}}}
\snip{pnet_lift_s}{0}{0}{\isa{opnet\ onp\ {\isasymlangle}i{\isacharsemicolon}\ R{\isasymrangle}\ \ostepinv\ {\isacharparenleft}{\isasymlambda}{\isasymsigma}\ {\isacharunderscore}{\isachardot}\ oarrivemsg\ I\ {\isasymsigma},\ other\ U\ {\isacharbraceleft}i{\isacharbraceright}\ {\isasymrightarrow}{\isacharparenright}\ globala\ {\isacharparenleft}{\isasymlambda}{\isacharparenleft}{\isasymsigma},\ a,\ {\isasymsigma}{\isacharprime}{\isacharparenright}{\isachardot}\ S\ {\isacharparenleft}{\isasymsigma}\ i{\isacharparenright}\ {\isacharparenleft}{\isasymsigma}{\isacharprime}\ i{\isacharparenright}{\isacharparenright}}}
\snip{pnet_lift_u}{0}{0}{\isa{opnet\ onp\ {\isasymlangle}i{\isacharsemicolon}\ R{\isasymrangle}\ \ostepinv\ {\isacharparenleft}{\isasymlambda}{\isasymsigma}\ {\isacharunderscore}{\isachardot}\ oarrivemsg\ I\ {\isasymsigma},\ other\ U\ {\isacharbraceleft}i{\isacharbraceright}\ {\isasymrightarrow}{\isacharparenright}\ globala\ {\isacharparenleft}{\isasymlambda}{\isacharparenleft}{\isasymsigma},\ a,\ {\isasymsigma}{\isacharprime}{\isacharparenright}{\isachardot}\ U\ {\isacharparenleft}{\isasymsigma}\ i{\isacharparenright}\ {\isacharparenleft}{\isasymsigma}{\isacharprime}\ i{\isacharparenright}{\isacharparenright}}}

\snip{pnet_lift_concl}{0}{0}{\isa{opnet\ onp\ p\ {\isasymTurnstile}\ {\isacharparenleft}otherwith\ S\ {\isacharparenleft}net{\isacharunderscore}tree{\isacharunderscore}ips\ p{\isacharparenright}\ {\isacharparenleft}oarrivemsg\ I{\isacharparenright},\ other\ U\ {\isacharparenleft}net{\isacharunderscore}tree{\isacharunderscore}ips\ p{\isacharparenright}\ {\isasymrightarrow}{\isacharparenright}\ global\ {\isacharparenleft}{\isasymlambda}{\isasymsigma}{\isachardot}\ {\isasymforall}i{\isasymin}net{\isacharunderscore}tree{\isacharunderscore}ips\ p{\isachardot}\ P\ i\ {\isasymsigma}{\isacharparenright}}}
\snip{other_net_tree_ips_par_left}{0}{0}{\isa{\mbox{}\inferrule{\mbox{other\ U\ {\isacharparenleft}net{\isacharunderscore}tree{\isacharunderscore}ips\ {\isacharparenleft}p\isactrlsub {\isadigit{1}}\parallelcomp{}p\isactrlsub {\isadigit{2}}{\isacharparenright}{\isacharparenright}\ {\isasymsigma}\ {\isasymsigma}{\isacharprime}}\\\ \mbox{{\isasymAnd}{\isasymxi}{\isachardot}\ U\ {\isasymxi}\ {\isasymxi}}}{\mbox{other\ U\ {\isacharparenleft}net{\isacharunderscore}tree{\isacharunderscore}ips\ p\isactrlsub {\isadigit{1}}{\isacharparenright}\ {\isasymsigma}\ {\isasymsigma}{\isacharprime}}}}}
\snip{net_tree_ips_p1_par_p2}{0}{0}{\isa{i{\isasymin}net{\isacharunderscore}tree{\isacharunderscore}ips\ {\isacharparenleft}p\isactrlsub {\isadigit{1}}\parallelcomp{}p\isactrlsub {\isadigit{2}}{\isacharparenright}}}
\snip{net_tree_ips_not_p1_par_p2}{0}{0}{\isa{i\ {\isasymnotin}\ net{\isacharunderscore}tree{\isacharunderscore}ips\ {\isacharparenleft}p\isactrlsub {\isadigit{1}}\parallelcomp{}p\isactrlsub {\isadigit{2}}{\isacharparenright}}}
\snip{net_tree_ips_p1}{0}{0}{\isa{i{\isasymin}net{\isacharunderscore}tree{\isacharunderscore}ips\ p\isactrlsub {\isadigit{1}}}}
\snip{net_tree_ips_p2}{0}{0}{\isa{i{\isasymin}net{\isacharunderscore}tree{\isacharunderscore}ips\ p\isactrlsub {\isadigit{2}}}}
\snip{net_tree_ips_not_p1_sigma}{0}{0}{\isa{{\isasymforall}j{\isachardot}\ j\ {\isasymnotin}\ net{\isacharunderscore}tree{\isacharunderscore}ips\ p\isactrlsub {\isadigit{1}}\ {\isasymlongrightarrow}\ {\isasymsigma}{\isacharprime}\ j\ {\isacharequal}\ {\isasymsigma}\ j}}
\snip{node_proc_reachable_prem_1}{0}{0}{\isa{{\isacharparenleft}{\isasymsigma},\ \NodeS\ i\ s\ R{\isacharparenright}{\isasymin}oreachable\ {\isacharparenleft}{\isasymlangle}i\ {\isacharcolon}\ A\ {\isacharcolon}\ \Ri{\isasymrangle}\isactrlsub o{\isacharparenright}\ {\isacharparenleft}otherwith\ E\ {\isacharbraceleft}i{\isacharbraceright}\ {\isacharparenleft}oarrivemsg\ I{\isacharparenright}{\isacharparenright}\ {\isacharparenleft}other\ F\ {\isacharbraceleft}i{\isacharbraceright}{\isacharparenright}}}

\snip{node_proc_reachable_prem_2}{0}{0}{\isa{{\isasymAnd}{\isasymxi}\ {\isasymxi}{\isacharprime}{\isachardot}\ E\ {\isasymxi}\ {\isasymxi}{\isacharprime}\ {\isasymLongrightarrow}\ F\ {\isasymxi}\ {\isasymxi}{\isacharprime}}}
\snip{node_proc_reachable_prem_2_prem}{0}{0}{\isa{E\ {\isasymxi}\ {\isasymxi}{\isacharprime}}}
\snip{node_proc_reachable_prem_2_concl}{0}{0}{\isa{F\ {\isasymxi}\ {\isasymxi}{\isacharprime}}}
\snip{node_proc_reachable_concl}{0}{0}{\isa{{\isacharparenleft}{\isasymsigma},\ s{\isacharparenright}{\isasymin}oreachable\ A\ {\isacharparenleft}otherwith\ E\ {\isacharbraceleft}i{\isacharbraceright}\ {\isacharparenleft}orecvmsg\ I{\isacharparenright}{\isacharparenright}\ {\isacharparenleft}other\ F\ {\isacharbraceleft}i{\isacharbraceright}{\isacharparenright}}}

\snip{subnet_oreachable_prem_1}{0}{0}{\isa{{\isacharparenleft}{\isasymsigma},\ \subnets{s}{t}{\isacharparenright}{\isasymin}oreachable\ {\isacharparenleft}opnet\ onp\ {\isacharparenleft}p\isactrlsub {\isadigit{1}}\parallelcomp{}p\isactrlsub {\isadigit{2}}{\isacharparenright}{\isacharparenright}\ S\ U}}
\snip{subnet_oreachable_prem_1_L}{0}{0}{\isa{S\ {\isacharequal}\ otherwith\ E\ {\isacharparenleft}net{\isacharunderscore}tree{\isacharunderscore}ips\ {\isacharparenleft}p\isactrlsub {\isadigit{1}}\parallelcomp{}p\isactrlsub {\isadigit{2}}{\isacharparenright}{\isacharparenright}\ {\isacharparenleft}oarrivemsg\ I{\isacharparenright}}}
\snip{subnet_oreachable_prem_1_E}{0}{0}{\isa{U\ {\isacharequal}\ other\ F\ {\isacharparenleft}net{\isacharunderscore}tree{\isacharunderscore}ips\ {\isacharparenleft}p\isactrlsub {\isadigit{1}}\parallelcomp{}p\isactrlsub {\isadigit{2}}{\isacharparenright}{\isacharparenright}}}

\snip{subnet_oreachable_prem_2}{0}{0}{\isa{{\isasymAnd}{\isasymxi}{\isachardot}\ E\ {\isasymxi}\ {\isasymxi}}}
\snip{subnet_oreachable_prem_3}{0}{0}{\isa{{\isasymAnd}{\isasymxi}{\isachardot}\ F\ {\isasymxi}\ {\isasymxi}}}
\snip{subnet_oreachable_prem_4}{0}{0}{\isa{{\isasymAnd}i\ R{\isachardot}\ {\isasymlangle}i\ {\isacharcolon}\ onp\ i\ {\isacharcolon}\ R{\isasymrangle}\isactrlsub o\ \ostepinv\ {\isacharparenleft}{\isasymlambda}{\isasymsigma}\ {\isacharunderscore}{\isachardot}\ oarrivemsg\ I\ {\isasymsigma},\ other\ F\ {\isacharbraceleft}i{\isacharbraceright}\ {\isasymrightarrow}{\isacharparenright}\ globala\ {\isacharparenleft}{\isasymlambda}{\isacharparenleft}{\isasymsigma},\ a,\ {\isacharunderscore}{\isacharparenright}{\isachardot}\ castmsg\ {\isacharparenleft}I\ {\isasymsigma}{\isacharparenright}\ a{\isacharparenright}}}

\snip{subnet_oreachable_prem_4_p}{0}{0}{\isa{p\ i\ \ostepinv\ {\isacharparenleft}{\isasymlambda}{\isasymsigma}\ {\isacharunderscore}{\isachardot}\ oarrivemsg\ I\ {\isasymsigma},\ other\ F\ {\isacharbraceleft}i{\isacharbraceright}\ {\isasymrightarrow}{\isacharparenright}\ {\isacharparenleft}{\isasymlambda}{\isacharparenleft}{\isacharparenleft}{\isasymsigma},\ {\isacharunderscore}{\isacharparenright},\ a,\ ({\isasymsigma}{\isacharprime},\ {\isacharunderscore}){\isacharparenright}{\isachardot}\ castmsg\ {\isacharparenleft}I\ {\isasymsigma}{\isacharparenright}\ a{\isacharparenright}}}
\snip{subnet_oreachable_prem_4_p'}{0}{0}{\isa{{\isasymlangle}i\ {\isacharcolon}\ onp\ i\ {\isacharcolon}\ R{\isasymrangle}\isactrlsub o}}

\snip{subnet_oreachable_prem_5}{0}{0}{\isa{{\isasymAnd}i\ R{\isachardot}\ {\isasymlangle}i\ {\isacharcolon}\ onp\ i\ {\isacharcolon}\ R{\isasymrangle}\isactrlsub o\ \ostepinv\ {\isacharparenleft}{\isasymlambda}{\isasymsigma}\ {\isacharunderscore}{\isachardot}\ oarrivemsg\ I\ {\isasymsigma},\ other\ F\ {\isacharbraceleft}i{\isacharbraceright}\ {\isasymrightarrow}{\isacharparenright}\ globala\ {\isacharparenleft}{\isasymlambda}{\isacharparenleft}{\isasymsigma},\ a,\ {\isasymsigma}{\isacharprime}{\isacharparenright}{\isachardot}\ a\ {\isasymnoteq}\ {\isasymtau}\ {\isasymand}\ {\isacharparenleft}{\isasymforall}i\ d{\isachardot}\ a\ {\isasymnoteq}\ i{\isacharcolon}deliver{\isacharparenleft}d{\isacharparenright}{\isacharparenright}\ {\isasymlongrightarrow}\ E\ {\isacharparenleft}{\isasymsigma}\ i{\isacharparenright}\ {\isacharparenleft}{\isasymsigma}{\isacharprime}\ i{\isacharparenright}{\isacharparenright}}}
\snip{subnet_oreachable_prem_5_pred}{0}{0}{\isa{E\ {\isacharparenleft}{\isasymsigma}\ i{\isacharparenright}\ {\isacharparenleft}{\isasymsigma}{\isacharprime}\ i{\isacharparenright}}}

\snip{subnet_oreachable_prem_6}{0}{0}{\isa{{\isasymAnd}i\ R{\isachardot}\ {\isasymlangle}i\ {\isacharcolon}\ onp\ i\ {\isacharcolon}\ R{\isasymrangle}\isactrlsub o\ \ostepinv\ {\isacharparenleft}{\isasymlambda}{\isasymsigma}\ {\isacharunderscore}{\isachardot}\ oarrivemsg\ I\ {\isasymsigma},\ other\ F\ {\isacharbraceleft}i{\isacharbraceright}\ {\isasymrightarrow}{\isacharparenright}\ globala\ {\isacharparenleft}{\isasymlambda}{\isacharparenleft}{\isasymsigma},\ a,\ {\isasymsigma}{\isacharprime}{\isacharparenright}{\isachardot}\ a\ {\isacharequal}\ {\isasymtau}\ {\isasymor}\ {\isacharparenleft}{\isasymexists}d{\isachardot}\ a\ {\isacharequal}\ i{\isacharcolon}deliver{\isacharparenleft}d{\isacharparenright}{\isacharparenright}\ {\isasymlongrightarrow}\ F\ {\isacharparenleft}{\isasymsigma}\ i{\isacharparenright}\ {\isacharparenleft}{\isasymsigma}{\isacharprime}\ i{\isacharparenright}{\isacharparenright}}}
\snip{subnet_oreachable_prem_6_pred}{0}{0}{\isa{F\ {\isacharparenleft}{\isasymsigma}\ i{\isacharparenright}\ {\isacharparenleft}{\isasymsigma}{\isacharprime}\ i{\isacharparenright}}}

\snip{subnet_oreachable_concl_1}{0}{0}{\isa{{\isacharparenleft}{\isasymsigma},\ s{\isacharparenright}{\isasymin}oreachable\ {\isacharparenleft}opnet\ onp\ p\isactrlsub {\isadigit{1}}{\isacharparenright}\ S\isactrlsub {\isadigit{1}}\ U\isactrlsub {\isadigit{1}}}}
\snip{subnet_oreachable_concl_2}{0}{0}{\isa{{\isacharparenleft}{\isasymsigma},\ t{\isacharparenright}{\isasymin}oreachable\ {\isacharparenleft}opnet\ onp\ p\isactrlsub {\isadigit{2}}{\isacharparenright}\ S\isactrlsub {\isadigit{2}}\ U\isactrlsub {\isadigit{2}}}}
\snip{subnet_oreachable_concl_3}{0}{0}{\isa{net{\isacharunderscore}tree{\isacharunderscore}ips\ p\isactrlsub {\isadigit{1}}\ {\isasyminter}\ net{\isacharunderscore}tree{\isacharunderscore}ips\ p\isactrlsub {\isadigit{2}}\ {\isacharequal}\ {\isasymemptyset}}}

\snip{S1}{0}{0}{\isa{S\isactrlsub {\isadigit{1}}}}
\snip{U1}{0}{0}{\isa{U\isactrlsub {\isadigit{1}}}}
\snip{S2}{0}{0}{\isa{S\isactrlsub {\isadigit{2}}}}
\snip{U2}{0}{0}{\isa{U\isactrlsub {\isadigit{2}}}}

\snip{subnet_oreachable_proof_step}{0}{0}{\isa{{\isacharparenleft}{\isacharparenleft}{\isasymsigma},\ \subnets{s}{t}{\isacharparenright},\ a,\ {\isacharparenleft}{\isasymsigma}{\isacharprime},\ \subnets{s{\isacharprime}}{t{\isacharprime}}{\isacharparenright}{\isacharparenright}{\isasymin}trans\ {\isacharparenleft}opnet\ onp\ {\isacharparenleft}p\isactrlsub {\isadigit{1}}\parallelcomp{}p\isactrlsub {\isadigit{2}}{\isacharparenright}{\isacharparenright}}}
\snip{subnet_oreachable_proof_jL}{0}{0}{\isa{j\ {\isasymnotin}\ net{\isacharunderscore}tree{\isacharunderscore}ips\ {\isacharparenleft}p\isactrlsub {\isadigit{1}}\parallelcomp{}p\isactrlsub {\isadigit{2}}{\isacharparenright}}}
\snip{subnet_oreachable_proof_oarrivemsg}{0}{0}{\isa{oarrivemsg\ I\ {\isasymsigma}}}

\snip{closed_reachable_oreachable'_prem_2}{0}{0}{\isa{s{\isasymin}reachable\ {\isacharparenleft}closed\ {\isacharparenleft}pnet\ {\isacharparenleft}{\isasymlambda}i{\isachardot}\ paodv\ i\ {\isasymlangle}{\isasymlangle}\ qmsg{\isacharparenright}\ n{\isacharparenright}{\isacharparenright}\ {\isacharparenleft}{\isasymlambda}{\isacharunderscore}{\isachardot}\ True{\isacharparenright}}}
\snip{closed_reachable_oreachable'_concl}{0}{0}{\isa{netgmap\ {\isacharparenleft}{\isasymlambda}{\isacharparenleft}p,\ q{\isacharparenright}{\isachardot}\ {\isacharparenleft}fst\ {\isacharparenleft}id\ p{\isacharparenright},\ snd\ {\isacharparenleft}id\ p{\isacharparenright},\ q{\isacharparenright}{\isacharparenright}\ s{\isasymin}netmask\ {\isacharparenleft}net{\isacharunderscore}tree{\isacharunderscore}ips\ n{\isacharparenright}\ {\isacharbackquote}\ oreachable\ {\isacharparenleft}oclosed\ {\isacharparenleft}opnet\ {\isacharparenleft}{\isasymlambda}i{\isachardot}\ opaodv\ i\ {\isasymlangle}{\isasymlangle}\isactrlbsub i\isactrlesub \ qmsg{\isacharparenright}\ n{\isacharparenright}{\isacharparenright}\ {\isacharparenleft}{\isasymlambda}{\isacharunderscore}\ {\isacharunderscore}\ {\isacharunderscore}{\isachardot}\ True{\isacharparenright}\ U}}
\snip{pnet_reachable_transfer}{0}{0}{\isa{{\isasymlbrakk}openproc\ np\ onp\ sr{\isacharsemicolon}\ wf{\isacharunderscore}net{\isacharunderscore}tree\ n{\isacharsemicolon}\ s{\isasymin}reachable\ {\isacharparenleft}closed\ {\isacharparenleft}pnet\ np\ n{\isacharparenright}{\isacharparenright}\ TT{\isasymrbrakk}\ {\isasymLongrightarrow}\ {\isacharparenleft}default\ {\isacharparenleft}someinit\ np\ sr{\isacharparenright}\ {\isacharparenleft}netlift\ sr\ s{\isacharparenright},\ netliftl\ sr\ s{\isacharparenright}{\isasymin}oreachable\ {\isacharparenleft}oclosed\ {\isacharparenleft}opnet\ onp\ n{\isacharparenright}{\isacharparenright}\ {\isacharparenleft}{\isasymlambda}{\isacharunderscore}\ {\isacharunderscore}\ {\isacharunderscore}{\isachardot}\ True{\isacharparenright}\ U}}
\snip{pnet_reachable_transfer_prem_1}{0}{0}{\isa{openproc\ np\ onp\ sr}}
\snip{pnet_reachable_transfer_prem_2}{0}{0}{\isa{wf{\isacharunderscore}net{\isacharunderscore}tree\ n}}
\snip{pnet_reachable_transfer_prem_3}{0}{0}{\isa{s{\isasymin}reachable\ {\isacharparenleft}closed\ {\isacharparenleft}pnet\ np\ n{\isacharparenright}{\isacharparenright}\ {\isacharparenleft}{\isasymlambda}{\isacharunderscore}{\isachardot}\ True{\isacharparenright}}}
\snip{pnet_reachable_transfer_concl}{0}{0}{\isa{{\isacharparenleft}default\ {\isacharparenleft}someinit\ np\ sr{\isacharparenright}\ {\isacharparenleft}netlift\ sr\ s{\isacharparenright},\ netliftl\ sr\ s{\isacharparenright}\\\mbox{}\hfill{}{\isasymin}oreachable\ {\isacharparenleft}oclosed\ {\isacharparenleft}opnet\ onp\ n{\isacharparenright}{\isacharparenright}\ {\isacharparenleft}{\isasymlambda}{\isacharunderscore}\ {\isacharunderscore}\ {\isacharunderscore}{\isachardot}\ True{\isacharparenright}\ U}}

\snip{someinit}{0}{0}{\isa{openproc\ np\ onp\ sr\ {\isasymLongrightarrow}\ someinit\ np\ sr\ i\ {\isacharequal}\ SOME\ x{\isachardot}\ x{\isasymin}{\isacharparenleft}fst\ {\isasymcirc}\ sr{\isacharparenright}\ {\isacharbackquote}\ init\ {\isacharparenleft}np\ i{\isacharparenright}}}
\snip{someinit_concl}{0}{0}{\isa{someinit\ np\ sr\ i\ {\isacharequal}\ SOME\ x{\isachardot}\ x{\isasymin}{\isacharparenleft}fst\ {\isasymcirc}\ sr{\isacharparenright}\ {\isacharbackquote}\ init\ {\isacharparenleft}np\ i{\isacharparenright}}}
\snip{someinit_concl_lhs}{0}{0}{\isa{someinit\ np\ sr\ i}}
\snip{someinit_concl_rhs}{0}{0}{\isa{SOME\ x{\isachardot}\ x{\isasymin}{\isacharparenleft}fst\ {\isasymcirc}\ sr{\isacharparenright}\ {\isacharbackquote}\ init\ {\isacharparenleft}np\ i{\isacharparenright}}}

\snip{openproc_init}{0}{0}{\isa{openproc\ np\ onp\ sr\ {\isasymLongrightarrow}\ {\isacharbraceleft}uu{\isacharunderscore}\ {\isacharbar}\ {\isasymexists}{\isasymsigma}\ {\isasymzeta}\ s{\isachardot}\ {\isacharunderscore}\ {\isacharequal}\ {\isacharparenleft}{\isasymsigma},\ {\isasymzeta}{\isacharparenright}\ {\isasymand}\ s{\isasymin}init\ {\isacharparenleft}np\ i{\isacharparenright}\ {\isasymand}\ {\isacharparenleft}{\isasymsigma}\ i,\ {\isasymzeta}{\isacharparenright}\ {\isacharequal}\ sr\ s\ {\isasymand}\ {\isacharparenleft}{\isasymforall}j{\isachardot}\ j\ {\isasymnoteq}\ i\ {\isasymlongrightarrow}\ {\isasymsigma}\ j{\isasymin}{\isacharparenleft}fst\ {\isasymcirc}\ sr{\isacharparenright}\ {\isacharbackquote}\ init\ {\isacharparenleft}np\ j{\isacharparenright}{\isacharparenright}{\isacharbraceright}\ {\isasymsubseteq}\ init\ {\isacharparenleft}onp\ i{\isacharparenright}}}
\snip{openproc_init_notempty}{0}{0}{\isa{openproc\ np\ onp\ sr\ {\isasymLongrightarrow}\ {\isasymforall}j{\isachardot}\ init\ {\isacharparenleft}np\ j{\isacharparenright}\ {\isasymnoteq}\ {\isasymemptyset}}}
\snip{openproc_trans}{0}{0}{\isa{{\isasymlbrakk}openproc\ np\ onp\ sr{\isacharsemicolon}\ {\isasymsigma}\ i\ {\isacharequal}\ fst\ {\isacharparenleft}sr\ s{\isacharparenright}{\isacharsemicolon}\ {\isasymsigma}{\isacharprime}\ i\ {\isacharequal}\ fst\ {\isacharparenleft}sr\ s{\isacharprime}{\isacharparenright}{\isacharsemicolon}\ {\isacharparenleft}s,\ a,\ s{\isacharprime}{\isacharparenright}{\isasymin}trans\ {\isacharparenleft}np\ i{\isacharparenright}{\isasymrbrakk}\ {\isasymLongrightarrow}\ {\isacharparenleft}{\isacharparenleft}{\isasymsigma},\ snd\ {\isacharparenleft}sr\ s{\isacharparenright}{\isacharparenright},\ a,\ ({\isasymsigma}{\isacharprime},\ snd\ {\isacharparenleft}sr\ s{\isacharprime}{\isacharparenright}){\isacharparenright}{\isasymin}trans\ {\isacharparenleft}onp\ i{\isacharparenright}}}

\snip{openproc_trans_prem_1}{0}{0}{\isa{openproc\ np\ onp\ sr}}
\snip{openproc_trans_prem_2}{0}{0}{\isa{{\isasymsigma}\ i\ {\isacharequal}\ fst\ {\isacharparenleft}sr\ s{\isacharparenright}}}
\snip{openproc_trans_prem_3}{0}{0}{\isa{{\isasymsigma}{\isacharprime}\ i\ {\isacharequal}\ fst\ {\isacharparenleft}sr\ s{\isacharprime}{\isacharparenright}}}
\snip{openproc_trans_prem_4}{0}{0}{\isa{{\isacharparenleft}s,\ a,\ s{\isacharprime}{\isacharparenright}{\isasymin}trans\ {\isacharparenleft}np\ i{\isacharparenright}}}
\snip{openproc_trans_concl}{0}{0}{\isa{{\isacharparenleft}{\isacharparenleft}{\isasymsigma},\ snd\ {\isacharparenleft}sr\ s{\isacharparenright}{\isacharparenright},\ a,\ ({\isasymsigma}{\isacharprime},\ snd\ {\isacharparenleft}sr\ s{\isacharprime}{\isacharparenright}){\isacharparenright}{\isasymin}trans\ {\isacharparenleft}onp\ i{\isacharparenright}}}

\snip{openproc_np_type}{0}{0}{\isa{ip\ {\isasymRightarrow}\ {\isacharparenleft}{\isacharprime}s,\ {\isacharprime}m\ seq{\isacharunderscore}action{\isacharparenright}\ automaton}}
\snip{openproc_onp_type}{0}{0}{\isa{ip\ {\isasymRightarrow}\ {\isacharparenleft}{\isacharparenleft}ip\ {\isasymRightarrow}\ {\isacharprime}g{\isacharparenright}\ {\isasymtimes}\ {\isacharprime}l,\ {\isacharprime}m\ seq{\isacharunderscore}action{\isacharparenright}\ automaton}}
\snip{openproc_sr_type}{0}{0}{\isa{{\isacharprime}s\ {\isasymRightarrow}\ {\isacharprime}g\ {\isasymtimes}\ {\isacharprime}l}}
\snip{par_qmsg_oreachable_prem_1}{0}{0}{\isa{{\isacharparenleft}{\isasymsigma},\ (s,\ (q,\ t)){\isacharparenright}{\isasymin}oreachable\ {\isacharparenleft}A\ {\isasymlangle}{\isasymlangle}\isactrlbsub i\isactrlesub \ qmsg{\isacharparenright}\ S\ U}}
\snip{par_qmsg_oreachable_prem_2}{0}{0}{\isa{A\ \ostepinv\ {\isacharparenleft}S,\ U\ {\isasymrightarrow}{\isacharparenright}\ {\isacharparenleft}{\isasymlambda}{\isacharparenleft}{\isacharparenleft}{\isasymsigma},\ {\isacharunderscore}{\isacharparenright},\ {\isacharunderscore},\ ({\isasymsigma}{\isacharprime},\ {\isacharunderscore}){\isacharparenright}{\isachardot}\ F\ {\isacharparenleft}{\isasymsigma}\ i{\isacharparenright}\ {\isacharparenleft}{\isasymsigma}{\isacharprime}\ i{\isacharparenright}{\isacharparenright}}}
\snip{par_qmsg_oreachable_prem_3}{0}{0}{\isa{{\isasymAnd}{\isasymxi}{\isachardot}\ F\ {\isasymxi}\ {\isasymxi}}}
\snip{par_qmsg_oreachable_prem_4}{0}{0}{\isa{{\isasymAnd}{\isasymxi}\ {\isasymxi}{\isacharprime}{\isachardot}\ E\ {\isasymxi}\ {\isasymxi}{\isacharprime}\ {\isasymLongrightarrow}\ F\ {\isasymxi}\ {\isasymxi}{\isacharprime}}}
\snip{par_qmsg_oreachable_prem_5}{0}{0}{\isa{{\isasymAnd}{\isasymsigma}\ {\isasymsigma}{\isacharprime}\ m{\isachardot}\ {\isasymlbrakk}{\isasymforall}j{\isachardot}\ F\ {\isacharparenleft}{\isasymsigma}\ j{\isacharparenright}\ {\isacharparenleft}{\isasymsigma}{\isacharprime}\ j{\isacharparenright}{\isacharsemicolon}\ R\ {\isasymsigma}\ m{\isasymrbrakk}\ {\isasymLongrightarrow}\ R\ {\isasymsigma}{\isacharprime}\ m}}

\snip{par_qmsg_oreachable_L}{0}{0}{\isa{S\ {\isacharequal}\ otherwith\ E\ {\isacharbraceleft}i{\isacharbraceright}\ {\isacharparenleft}orecvmsg\ R{\isacharparenright}}}
\snip{par_qmsg_oreachable_E}{0}{0}{\isa{U\ {\isacharequal}\ other\ F\ {\isacharbraceleft}i{\isacharbraceright}}}

\snip{par_qmsg_oreachable_conj1}{0}{0}{\isa{{\isacharparenleft}{\isasymsigma},\ s{\isacharparenright}{\isasymin}oreachable\ A\ S\ U}}
\snip{par_qmsg_oreachable_conj2}{0}{0}{\isa{{\isacharparenleft}q,\ t{\isacharparenright}{\isasymin}reachable\ qmsg\ {\isacharparenleft}recvmsg\ {\isacharparenleft}R\ {\isasymsigma}{\isacharparenright}{\isacharparenright}}}
\snip{par_qmsg_oreachable_conj3}{0}{0}{\isa{{\isasymforall}m{\isasymin}set\ q{\isachardot}\ R\ {\isasymsigma}\ m}}

\snip{par_qmsg_oreachable_Exixi'}{0}{0}{\isa{E\ {\isasymxi}\ {\isasymxi}{\isacharprime}}}
\snip{par_qmsg_oreachable_Fxixi'}{0}{0}{\isa{F\ {\isasymxi}\ {\isasymxi}{\isacharprime}}}
\snip{par_qmsg_oreachable_Fsigmaj}{0}{0}{\isa{{\isasymforall}j{\isachardot}\ F\ {\isacharparenleft}{\isasymsigma}\ j{\isacharparenright}\ {\isacharparenleft}{\isasymsigma}{\isacharprime}\ j{\isacharparenright}}}
\snip{par_qmsg_oreachable_Rsigmam}{0}{0}{\isa{R\ {\isasymsigma}\ m}}
\snip{par_qmsg_oreachable_Rsigma'm}{0}{0}{\isa{R\ {\isasymsigma}{\isacharprime}\ m}}
\snip{sequence_number_increases}{0}{0}{\begin{isabelle}%
paodv\ i\ \stepinv\ onll\ \gammaaodv\ {\isacharparenleft}{\isasymlambda}{\isacharparenleft}{\isacharparenleft}{\isasymxi},\ l{\isacharparenright},\ a,\ {\isasymxi}{\isacharprime},\ l{\isacharprime}{\isacharparenright}{\isachardot}\ sn\ {\isasymxi}\ {\isasymle}\ sn\ {\isasymxi}{\isacharprime}{\isacharparenright}%
\end{isabelle}}
\snip{sequence_number_one_or_bigger}{0}{0}{\begin{isabelle}%
paodv\ i\ {\isasymTTurnstile}\ onl\ \gammaaodv\ {\isacharparenleft}{\isasymlambda}{\isacharparenleft}{\isasymxi},\ l{\isacharparenright}{\isachardot}\ {\isadigit{1}}\ {\isasymle}\ sn\ {\isasymxi}{\isacharparenright}%
\end{isabelle}}

\snip{sequence_number_increases'}{0}{0}{\begin{isabelle}%
paodv\ i\ \stepinv\ {\isacharparenleft}{\isasymlambda}{\isacharparenleft}{\isacharparenleft}{\isasymxi},\ p{\isacharparenright},\ a,\ {\isasymxi}{\isacharprime},\ p{\isacharprime}{\isacharparenright}{\isachardot}\ sn\ {\isasymxi}\ {\isasymle}\ sn\ {\isasymxi}{\isacharprime}{\isacharparenright}%
\end{isabelle}}
\snip{sequence_number_one_or_bigger'}{0}{0}{\begin{isabelle}%
paodv\ i\ {\isasymTTurnstile}\ {\isacharparenleft}{\isasymlambda}{\isacharparenleft}{\isasymxi},\ p{\isacharparenright}{\isachardot}\ {\isadigit{1}}\ {\isasymle}\ sn\ {\isasymxi}{\isacharparenright}%
\end{isabelle}}

\snip{hop_count_positive}{0}{0}{\isa{paodv\ i\ {\isasymTTurnstile}\ onl\ \gammaaodv\ {\isacharparenleft}{\isasymlambda}{\isacharparenleft}{\isasymxi},\ {\isacharminus}{\isacharparenright}{\isachardot}\ {\isasymforall}ip{\isasymin}kD\ {\isacharparenleft}rt\ {\isasymxi}{\isacharparenright}{\isachardot}\ {\isadigit{1}}\ {\isasymle}\ the\ {\isacharparenleft}dhops\ {\isacharparenleft}rt\ {\isasymxi}{\isacharparenright}\ ip{\isacharparenright}{\isacharparenright}}}
\snip{osn_rreq}{0}{0}{\isa{paodv\ i\ {\isasymTTurnstile}\ {\isacharparenleft}recvmsg\ rreq{\isacharunderscore}rrep{\isacharunderscore}sn\ {\isasymrightarrow}{\isacharparenright}\ onl\ \gammaaodv\ {\isacharparenleft}{\isasymlambda}{\isacharparenleft}{\isasymxi},\ l{\isacharparenright}{\isachardot}\ l{\isasymin}{\isacharbraceleft}PAodv{\isacharminus}{\isacharcolon}{\isadigit{4}},\ PAodv{\isacharminus}{\isacharcolon}{\isadigit{5}}{\isacharbraceright}\ {\isasymunion}\ {\isacharbraceleft}PRreq{\isacharminus}{\isacharcolon}n\ {\isacharbar}\ True{\isacharbraceright}\ {\isasymlongrightarrow}\ {\isadigit{1}}\ {\isasymle}\ osn\ {\isasymxi}{\isacharparenright}}}
\snip{received_msg_inv}{0}{0}{\isa{paodv\ i\ {\isasymTTurnstile}\ {\isacharparenleft}recvmsg\ P\ {\isasymrightarrow}{\isacharparenright}\ onl\ \gammaaodv\ {\isacharparenleft}{\isasymlambda}{\isacharparenleft}{\isasymxi},\ l{\isacharparenright}{\isachardot}\ l{\isasymin}{\isacharbraceleft}PAodv{\isacharminus}{\isacharcolon}{\isadigit{1}}{\isacharbraceright}\ {\isasymlongrightarrow}\ P\ {\isacharparenleft}msg\ {\isasymxi}{\isacharparenright}{\isacharparenright}}}
\snip{rreq_dip_in_vD_dip_eq_ip}{0}{0}{\isa{paodv\ i\ {\isasymTTurnstile}\ onl\ \gammaaodv\ {\isacharparenleft}{\isasymlambda}{\isacharparenleft}{\isasymxi},\ l{\isacharparenright}{\isachardot}\ {\isacharparenleft}l{\isasymin}{\isacharbraceleft}PRreq{\isacharminus}{\isacharcolon}{\isadigit{1}}{\isadigit{6}}{\isachardot}{\isachardot}PRreq{\isacharminus}{\isacharcolon}{\isadigit{1}}{\isadigit{8}}{\isacharbraceright}\ {\isasymlongrightarrow}\ dip\ {\isasymxi}{\isasymin}vD\ {\isacharparenleft}rt\ {\isasymxi}{\isacharparenright}{\isacharparenright}\ {\isasymand}\ {\isacharparenleft}l{\isasymin}{\isacharbraceleft}PRreq{\isacharminus}{\isacharcolon}{\isadigit{5}},\ PRreq{\isacharminus}{\isacharcolon}{\isadigit{6}}{\isacharbraceright}\ {\isasymlongrightarrow}\ dip\ {\isasymxi}\ {\isacharequal}\ ip\ {\isasymxi}{\isacharparenright}\ {\isasymand}\ {\isacharparenleft}l{\isasymin}{\isacharbraceleft}PRreq{\isacharminus}{\isacharcolon}{\isadigit{1}}{\isadigit{5}}{\isachardot}{\isachardot}PRreq{\isacharminus}{\isacharcolon}{\isadigit{1}}{\isadigit{8}}{\isacharbraceright}\ {\isasymlongrightarrow}\ dip\ {\isasymxi}\ {\isasymnoteq}\ ip\ {\isasymxi}{\isacharparenright}{\isacharparenright}}}

\snip{rreq_rrep_sn_any_step_invariant}{0}{0}{\isa{paodv\ i\ \stepinv\ {\isacharparenleft}recvmsg\ rreq{\isacharunderscore}rrep{\isacharunderscore}sn\ {\isasymrightarrow}{\isacharparenright}\ onll\ \gammaaodv\ {\isacharparenleft}{\isasymlambda}{\isacharparenleft}s,\ a,\ s{\isacharprime}{\isacharparenright}{\isachardot}\ anycast\ rreq{\isacharunderscore}rrep{\isacharunderscore}sn\ a{\isacharparenright}}}
\snip{rreq_rrep_sn}{0}{0}{\begin{isabelle}%
rreq{\isacharunderscore}rrep{\isacharunderscore}sn\ m\ {\isacharequal}\isanewline
\textsf{case}\ m\ \textsf{of}\ Rreq\ hops\ rreqid\ dip\ dsn\ dsk\ oip\ osn\ sip\ {\isasymRightarrow}\ {\isadigit{1}}\ {\isasymle}\ osn\isanewline
{\isacharbar}\ Rrep\ hops\ dip\ dsn\ oip\ sip\ {\isasymRightarrow}\ {\isadigit{1}}\ {\isasymle}\ dsn\ {\isacharbar}\ {\isacharunderscore}\ {\isasymRightarrow}\ True%
\end{isabelle}}

\snip{seq_nhop_quality_increases'}{0}{0}{\isa{opaodv\ i\ {\isasymTurnstile}\ {\isacharparenleft}otherwith\ {\isacharparenleft}op\ {\isacharequal}{\isacharparenright}\ {\isacharbraceleft}i{\isacharbraceright}\ {\isacharparenleft}orecvmsg\ {\isacharparenleft}{\isasymlambda}{\isasymsigma}\ m{\isachardot}\ msg{\isacharunderscore}fresh\ {\isasymsigma}\ m\ {\isasymand}\ msg{\isacharunderscore}zhops\ m{\isacharparenright}{\isacharparenright},\ other\ quality{\isacharunderscore}increases\ {\isacharbraceleft}i{\isacharbraceright}\ {\isasymrightarrow}{\isacharparenright}\ onl\ \gammaaodv\ {\isacharparenleft}{\isasymlambda}{\isacharparenleft}{\isasymsigma},\ {\isacharunderscore}{\isacharparenright}{\isachardot}\ {\isasymforall}dip{\isachardot}\ \textsf{let}\ nhip\ {\isacharequal}\ the\ {\isacharparenleft}nh{\isacharparenleft}op\ {\isacharequal}{\isacharparenright}rt\ {\isacharparenleft}{\isasymsigma}\ i{\isacharparenright}{\isacharparenright}\ dip{\isacharparenright}\ \textsf{in}\ dip{\isasymin}vD\ {\isacharparenleft}rt\ {\isacharparenleft}{\isasymsigma}\ i{\isacharparenright}{\isacharparenright}\ {\isasyminter}\ vD\ {\isacharparenleft}rt\ {\isacharparenleft}{\isasymsigma}\ nhip{\isacharparenright}{\isacharparenright}\ {\isasymand}\ nhip\ {\isasymnoteq}\ dip\ {\isasymlongrightarrow}\ rt\ {\isacharparenleft}{\isasymsigma}\ i{\isacharparenright}\ {\isasymsqsubset}\isactrlbsub dip\isactrlesub \ rt\ {\isacharparenleft}{\isasymsigma}\ nhip{\isacharparenright}{\isacharparenright}}}
\snip{closed_reachable_oreachable}{0}{0}{\isa{\mbox{}\inferrule{\mbox{wf{\isacharunderscore}net{\isacharunderscore}tree\ n}\\\ \mbox{s{\isasymin}reachable\ {\isacharparenleft}closed\ {\isacharparenleft}pnet\ {\isacharparenleft}{\isasymlambda}i{\isachardot}\ paodv\ i\ {\isasymlangle}{\isasymlangle}\ qmsg{\isacharparenright}\ n{\isacharparenright}{\isacharparenright}\ TT}}{\mbox{initmissing\ {\isacharparenleft}netgmap\ {\isacharparenleft}{\isasymlambda}{\isacharparenleft}p,\ q{\isacharparenright}{\isachardot}\ {\isacharparenleft}fst\ {\isacharparenleft}id\ p{\isacharparenright},\ snd\ {\isacharparenleft}id\ p{\isacharparenright},\ q{\isacharparenright}{\isacharparenright}\ s{\isacharparenright}{\isasymin}oreachable\ {\isacharparenleft}oclosed\ {\isacharparenleft}opnet\ {\isacharparenleft}{\isasymlambda}i{\isachardot}\ opaodv\ i\ {\isasymlangle}{\isasymlangle}\isactrlbsub i\isactrlesub \ qmsg{\isacharparenright}\ n{\isacharparenright}{\isacharparenright}\ {\isacharparenleft}{\isasymlambda}{\isacharunderscore}\ {\isacharunderscore}\ {\isacharunderscore}{\isachardot}\ True{\isacharparenright}\ U}}}}

\snip{close_opnet}{0}{0}{\isa{\mbox{}\inferrule{\mbox{wf{\isacharunderscore}net{\isacharunderscore}tree\ n}\\\ \mbox{oclosed\ {\isacharparenleft}opnet\ opaodv\ n{\isacharparenright}\ {\isasymTurnstile}\ {\isacharparenleft}{\isasymlambda}{\isacharunderscore}\ {\isacharunderscore}\ {\isacharunderscore}{\isachardot}\ True,\ U\ {\isasymrightarrow}{\isacharparenright}\ global\ P}}{\mbox{closed\ {\isacharparenleft}pnet\ paodv\ n{\isacharparenright}\ {\isasymTTurnstile}\ openproc{\isachardot}netglobal\ paodv\ id\ P}}}}
\snip{close_opnet_prem_1}{0}{0}{\isa{wf{\isacharunderscore}net{\isacharunderscore}tree\ n}}
\snip{close_opnet_prem_2}{0}{0}{\isa{oclosed\ {\isacharparenleft}opnet\ opaodv\ n{\isacharparenright}\ {\isasymTurnstile}\ {\isacharparenleft}{\isasymlambda}{\isacharunderscore}\ {\isacharunderscore}\ {\isacharunderscore}{\isachardot}\ True,\ U\ {\isasymrightarrow}{\isacharparenright}\ {\isacharparenleft}{\isasymlambda}{\isacharparenleft}{\isasymsigma},\ {\isacharunderscore}{\isacharparenright}{\isachardot}\ P\ {\isasymsigma}{\isacharparenright}}}
\snip{close_opnet_concl}{0}{0}{\isa{closed\ {\isacharparenleft}pnet\ paodv\ n{\isacharparenright}\ {\isasymTTurnstile}\ openproc{\isachardot}netglobal\ paodv\ id\ P}}

\snip{rt_graph_lhs}{0}{0}{\isa{rt{\isacharunderscore}graph\ {\isasymsigma}}}
\snip{rt_graph_rhs}{0}{0}{\isa{{\isasymlambda}dip{\isachardot}\ {\isacharbraceleft}edge\ {\isacharbar}\ {\isasymexists}ip\ ip{\isacharprime}\ dsn\ dsk\ hops\ pre{\isachardot}\ edge\ {\isacharequal}\ {\isacharparenleft}ip,\ ip{\isacharprime}{\isacharparenright}\ {\isasymand}\ ip\ {\isasymnoteq}\ dip\ {\isasymand}\ rt\ {\isacharparenleft}{\isasymsigma}\ ip{\isacharparenright}\ dip\ {\isacharequal}\ Some\ {\isacharparenleft}dsn,\ dsk,\ val,\ hops,\ ip{\isacharprime},\ pre{\isacharparenright}{\isacharbraceright}}}

\snip{rt_strictly_fresher_than_term}{0}{0}{\isa{rt\isactrlsub {\isadigit{1}}\ {\isasymsqsubset}\isactrlbsub i\isactrlesub \ rt\isactrlsub {\isadigit{2}}}}

\snip{loop_freedom}{0}{0}{\isa{\mbox{}\inferrule{\mbox{{\isasymforall}i\ dip{\isachardot}\ \textsf{let}\ nhip\ {\isacharequal}\ the\ {\isacharparenleft}nh{\isacharparenleft}op\ {\isacharequal}{\isacharparenright}rt\ {\isacharparenleft}{\isasymsigma}\ i{\isacharparenright}{\isacharparenright}\ dip{\isacharparenright}\ \textsf{in}\ dip{\isasymin}vD\ {\isacharparenleft}rt\ {\isacharparenleft}{\isasymsigma}\ i{\isacharparenright}{\isacharparenright}\ {\isasyminter}\ vD\ {\isacharparenleft}rt\ {\isacharparenleft}{\isasymsigma}\ nhip{\isacharparenright}{\isacharparenright}\ {\isasymand}\ nhip\ {\isasymnoteq}\ dip\ {\isasymlongrightarrow}\ rt\ {\isacharparenleft}{\isasymsigma}\ i{\isacharparenright}\ {\isasymsqsubset}\isactrlbsub dip\isactrlesub \ rt\ {\isacharparenleft}{\isasymsigma}\ nhip{\isacharparenright}}}{\mbox{{\isasymforall}dip{\isachardot}\ irrefl\ {\isacharparenleft}{\isacharparenleft}rt{\isacharunderscore}graph\ {\isasymsigma}\ dip{\isacharparenright}\isactrlsup {\isacharplus}{\isacharparenright}}}}}

\snip{aodv_loop_freedom}{0}{0}{\isa{\mbox{}\inferrule{\mbox{wf{\isacharunderscore}net{\isacharunderscore}tree\ n}}{\mbox{closed\ {\isacharparenleft}pnet\ {\isacharparenleft}{\isasymlambda}i{\isachardot}\ paodv\ i\ {\isasymlangle}{\isasymlangle}\ qmsg{\isacharparenright}\ n{\isacharparenright}\ {\isasymTTurnstile}\ netglobal\ {\isacharparenleft}{\isasymlambda}{\isasymsigma}{\isachardot}\ {\isasymforall}dip{\isachardot}\ irrefl\ {\isacharparenleft}{\isacharparenleft}rt{\isacharunderscore}graph\ {\isasymsigma}\ dip{\isacharparenright}\isactrlsup {\isacharplus}{\isacharparenright}{\isacharparenright}}}}}
\snip{aodv_loop_freedom_concl}{0}{0}{\isa{closed\ {\isacharparenleft}pnet\ {\isacharparenleft}{\isasymlambda}i{\isachardot}\ paodv\ i\ {\isasymlangle}{\isasymlangle}\ qmsg{\isacharparenright}\ n{\isacharparenright}\ {\isasymTTurnstile}\ netglobal\ {\isacharparenleft}{\isasymlambda}{\isasymsigma}{\isachardot}\ {\isasymforall}dip{\isachardot}\ irrefl\ {\isacharparenleft}{\isacharparenleft}rt{\isacharunderscore}graph\ {\isasymsigma}\ dip{\isacharparenright}\isactrlsup {\isacharplus}{\isacharparenright}{\isacharparenright}}}
\snip{rt_strictly_fresher}{0}{0}{\isa{{\isacharparenleft}rt\isactrlsub {\isadigit{1}}\ {\isasymsqsubset}\isactrlbsub i\isactrlesub \ rt\isactrlsub {\isadigit{2}}{\isacharparenright}\ {\isacharequal}\ {\isacharparenleft}rt\isactrlsub {\isadigit{1}}\ {\isasymsqsubseteq}\isactrlbsub i\isactrlesub \ rt\isactrlsub {\isadigit{2}}\ {\isasymand}\ {\isasymnot}\ rt\isactrlsub {\isadigit{2}}\ {\isasymsqsubseteq}\isactrlbsub i\isactrlesub \ rt\isactrlsub {\isadigit{1}}{\isacharparenright}}}
\snip{rt_fresher}{0}{0}{\isa{rt{\isacharunderscore}fresher\ {\isacharequal}\ {\isasymlambda}dip\ rt\ rt{\isacharprime}{\isachardot}\ the\ {\isacharparenleft}rt\ dip{\isacharparenright}\ {\isasymsqsubseteq}\ the\ {\isacharparenleft}rt{\isacharprime}\ dip{\isacharparenright}}}
\snip{rt_fresher'}{0}{0}{\isa{{\isacharparenleft}rt\isactrlsub {\isadigit{1}}\ {\isasymsqsubseteq}\isactrlbsub i\isactrlesub \ rt\isactrlsub {\isadigit{2}}{\isacharparenright}\ {\isacharequal}\ {\isacharparenleft}nsqn\isactrlsub r\ {\isacharparenleft}the\ {\isacharparenleft}rt\isactrlsub {\isadigit{1}}\ i{\isacharparenright}{\isacharparenright}\ {\isacharless}\ nsqn\isactrlsub r\ {\isacharparenleft}the\ {\isacharparenleft}rt\isactrlsub {\isadigit{2}}\ i{\isacharparenright}{\isacharparenright}\ {\isasymor}\ nsqn\isactrlsub r\ {\isacharparenleft}the\ {\isacharparenleft}rt\isactrlsub {\isadigit{1}}\ i{\isacharparenright}{\isacharparenright}\ {\isacharequal}\ nsqn\isactrlsub r\ {\isacharparenleft}the\ {\isacharparenleft}rt\isactrlsub {\isadigit{2}}\ i{\isacharparenright}{\isacharparenright}\ {\isasymand}\ {\isasympi}\isactrlsub {\isadigit{5}}\ {\isacharparenleft}the\ {\isacharparenleft}rt\isactrlsub {\isadigit{2}}\ i{\isacharparenright}{\isacharparenright}\ {\isasymle}\ {\isasympi}\isactrlsub {\isadigit{5}}\ {\isacharparenleft}the\ {\isacharparenleft}rt\isactrlsub {\isadigit{1}}\ i{\isacharparenright}{\isacharparenright}{\isacharparenright}}}

\snip{rt_fresher2}{0}{0}{\isa{\mbox{}\inferrule{\mbox{i{\isasymin}kD\ rt\isactrlsub {\isadigit{1}}}\\\ \mbox{i{\isasymin}kD\ rt\isactrlsub {\isadigit{2}}}}{\mbox{{\isacharparenleft}rt\isactrlsub {\isadigit{1}}\ {\isasymsqsubseteq}\isactrlbsub i\isactrlesub \ rt\isactrlsub {\isadigit{2}}{\isacharparenright}\ {\isacharequal}\ {\isacharparenleft}nsqn\ rt\isactrlsub {\isadigit{1}}\ i\ {\isacharless}\ nsqn\ rt\isactrlsub {\isadigit{2}}\ i\ {\isasymor}\ nsqn\ rt\isactrlsub {\isadigit{1}}\ i\ {\isacharequal}\ nsqn\ rt\isactrlsub {\isadigit{2}}\ i\ {\isasymand}\ the\ {\isacharparenleft}dhops\ rt\isactrlsub {\isadigit{2}}\ i{\isacharparenright}\ {\isasymle}\ the\ {\isacharparenleft}dhops\ rt\isactrlsub {\isadigit{1}}\ i{\isacharparenright}{\isacharparenright}}}}}
\snip{rt_fresher2_prem_1}{0}{0}{\isa{i{\isasymin}kD\ rt\isactrlsub {\isadigit{1}}}}
\snip{rt_fresher2_prem_2}{0}{0}{\isa{i{\isasymin}kD\ rt\isactrlsub {\isadigit{2}}}}
\snip{rt_fresher2_concl}{0}{0}{\isa{{\isacharparenleft}rt\isactrlsub {\isadigit{1}}\ {\isasymsqsubseteq}\isactrlbsub i\isactrlesub \ rt\isactrlsub {\isadigit{2}}{\isacharparenright}\ {\isacharequal}\ {\isacharparenleft}nsqn\ rt\isactrlsub {\isadigit{1}}\ i\ {\isacharless}\ nsqn\ rt\isactrlsub {\isadigit{2}}\ i}\hfill\null\newline\null\hfill\isa{{\isasymor}\ {\isacharparenleft}nsqn\ rt\isactrlsub {\isadigit{1}}\ i\ {\isacharequal}\ nsqn\ rt\isactrlsub {\isadigit{2}}\ i\ {\isasymand}\ the\ {\isacharparenleft}dhops\ rt\isactrlsub {\isadigit{2}}\ i{\isacharparenright}\ {\isasymle}\ the\ {\isacharparenleft}dhops\ rt\isactrlsub {\isadigit{1}}\ i{\isacharparenright}{\isacharparenright}{\isacharparenright}}}

\snip{nsqn}{0}{0}{\isa{nsqn\ {\isacharequal}\ {\isasymlambda}rt\ dip{\isachardot}\ \textsf{case}\ rt\ dip\ \textsf{of}\ None\ {\isasymRightarrow}\ {\isadigit{0}}\ {\isacharbar}\ Some\ r\ {\isasymRightarrow}\ nsqn\isactrlsub r\ r}}
\snip{nsqn_sqn_def}{0}{0}{\isa{nsqn\ rt\ i\ {\isacharequal}\ {\isacharparenleft}\textsf{if}\ flag\ rt\ i\ {\isacharequal}\ Some\ val\ {\isasymor}\ sqn\ rt\ i\ {\isacharequal}\ {\isadigit{0}}\ \textsf{then}\ sqn\ rt\ i\ \textsf{else}\ sqn\ rt\ i\ {\isacharminus}\ {\isadigit{1}}{\isacharparenright}}}

\snip{pi2}{0}{0}{\isa{{\isasympi}\isactrlsub {\isadigit{2}}}}
\snip{pi5}{0}{0}{\isa{{\isasympi}\isactrlsub {\isadigit{5}}}}

\snip{fresher}{0}{0}{\isa{r\ {\isasymsqsubseteq}\ r{\isacharprime}\ {\isacharequal}\ nsqn\isactrlsub r\ r\ {\isacharless}\ nsqn\isactrlsub r\ r{\isacharprime}\ {\isasymor}\ nsqn\isactrlsub r\ r\ {\isacharequal}\ nsqn\isactrlsub r\ r{\isacharprime}\ {\isasymand}\ {\isasympi}\isactrlsub {\isadigit{5}}\ r{\isacharprime}\ {\isasymle}\ {\isasympi}\isactrlsub {\isadigit{5}}\ r}}
\snip{nsqnr}{0}{0}{\isa{nsqn\isactrlsub r\ r\ {\isacharequal}\ {\isacharparenleft}\textsf{if}\ {\isasympi}\isactrlsub {\isadigit{4}}\ r\ {\isacharequal}\ Aodv{\isacharunderscore}Basic{\isachardot}inv\ \textsf{then}\ {\isasympi}\isactrlsub {\isadigit{2}}\ r\ {\isacharminus}\ {\isadigit{1}}\ \textsf{else}\ {\isasympi}\isactrlsub {\isadigit{2}}\ r{\isacharparenright}}}
\newcommand{\nodes}[3]{%
    $\mbox{#2}^{\mbox{\,\isastylescript{#1}}}_{\mbox{\isastylescript{#3}}}$}
\newcommand{\subnets}[2]{{#1}\hspace{-.13em}{\isamath{\,\shortparallel\,}}{#2}}

\def\NodeS\ #1\ #2\ #3{\nodes{#1}{#2}{#3}}

\renewcommand{\isasymin}{\isamath{\,\in\,}}

\newcommand{\listconcat}{{\scriptsize\ensuremath{+\!+}}}
\newcommand{\parallelcomp}{{\scriptsize\raisebox{.19ex}{\ensuremath{\,\parallel\,}}}}

\newcommand{\gammaaodv}{{{\isasymGamma}\isactrlsub{\sf aodv}}}
\newcommand{\gammaqmsg}{{{\isasymGamma}\isactrlsub{\sf qmsg}}}
\newcommand{\sigmaaodv}{{{\isasymsigma}\isactrlsub{\sf aodv}}}

\newcommand{\selmsg}{\isa{{s}\isactrlsub {{msg}}}}
\newcommand{\updmsg}{\isa{{u}\isactrlsub {{msg}}}}
\newcommand{\selips}{\isa{{s}\isactrlsub {{ips}}}}
\newcommand{\selip}{\isa{{s}\isactrlsub {{ip}}}}
\newcommand{\seldata}{\isa{{s}\isactrlsub {{data}}}}

\usepackage{cite}
\usepackage[cmex10]{amsmath}

\newenvironment{tightcenter}{%
  \setlength\topsep{4.0pt plus 0.5pt minus 1.0pt}
  \setlength\parskip{0pt}
  \begin{center}
}{%
  \end{center}
}

\newcommand{\refsec}[1]{\hyperref[sec:#1]{Section~\ref*{sec:#1}}}
\newcommand{\reffig}[1]{\hyperref[fig:#1]{Figure~\ref*{fig:#1}}}
\newcommand{\subreffig}[1]{\hyperref[fig:#1]{Figure~\subref*{fig:#1}}}
\newcommand{\refthm}[1]{\hyperref[thm:#1]{Theorem~\ref*{thm:#1}}}
\newcommand{\reflem}[1]{\hyperref[thm:#1]{Lemma~\ref*{thm:#1}}}
\newcommand{\refdef}[1]{\hyperref[def:#1]{Definition~\ref*{def:#1}}}
\newcommand{\refeq}[1]{\hyperref[eq:#1]{(\ref*{eq:#1})}}
\newcommand{\refitem}[1]{\hyperref[item:#1]{Item~\ref*{item:#1}}}
\newcommand{\refitemr}[2]{\hyperref[item:#1]{Items~\ref*{item:#1}}
                          \hyperref[item:#2]{to~\ref*{item:#2}}}
\newcommand{\refitems}[2]{\hyperref[item:#1]{Items~\ref*{item:#1}}
                          \hyperref[item:#2]{and~\ref*{item:#2}}}
\newcommand{\seqpsos}{\isa{seqp{\isacharunderscore}sos}}
\newcommand{\parpsos}{\isa{parp{\isacharunderscore}sos}}
\newcommand{\nodesos}{\isa{node{\isacharunderscore}sos}}
\newcommand{\pnetsos}{\isa{pnet{\isacharunderscore}sos}}
\newcommand{\cnetsos}{\isa{cnet{\isacharunderscore}sos}}
\newcommand{\oseqpsos}{\isa{oseqp{\isacharunderscore}sos}}
\newcommand{\oparpsos}{\isa{oparp{\isacharunderscore}sos}}
\newcommand{\onodesos}{\isa{onode{\isacharunderscore}sos}}
\newcommand{\opnetsos}{\isa{opnet{\isacharunderscore}sos}}
\newcommand{\ocnetsos}{\isa{ocnet{\isacharunderscore}sos}}
\newcommand{\Ri}{\isa{R\isactrlsub i}}

\newcommand{\stepinv}{\isa{\isamath{\mid\!\mid\hspace{-2pt}\equiv}}}
\newcommand{\ostepinv}{\isa{\isamath{\mid\hspace{-2pt}\equiv}}}

\usepackage[printonlyused]{acronym}
\renewcommand{\acsfont}[1]{\textsc{#1}}
\newcommand{\acworkaround}[1]{{\acsfont {#1}}}
\acrodef{AODV}{Ad hoc On-demand Distance Vector}
\acrodef{AWN}{Algebra for Wireless Networks}
\acrodef{MANET}{Mobile Ad hoc Network}
\acrodef{ITP}{Interactive Theorem Prover}
\acrodef{HOL}{Higher Order Logic}
\acrodef{WMN}{Wireless Mesh Network}
\acrodef{SOS}{Structural Operational Semantics}
\acrodef{TCP}{Transmission Control Protocol}
\acrodef{rreq}{Route Request}
\acrodef{rrep}{Route Reply}
\acrodef{rerr}{Route Error}
\acrodef{rrep-ack}{Route Reply Acknowledgement}

\begin{document}

\title{A mechanized proof of loop freedom of the (untimed) AODV
routing protocol}
\titlerunning{A mechanized proof of loop freedom of AODV}

\authorrunning{T. Bourke, R.J. van Glabbeek, and P. H{\"o}fner}
  \author{
  Timothy Bourke\inst{1,2}\and
  Rob van Glabbeek\inst{3,4}\and
  Peter H\"ofner\inst{3,4}
  }
\institute{
 INRIA Paris-Rocquencourt, France\and
 Ecole normale supérieure, Paris, France\and
  NICTA, Australia\and
  Computer Science and Engineering, UNSW, Australia
  }

\maketitle
\setcounter{footnote}{0}

\begin{abstract}
The \acf{AODV} routing protocol allows the nodes in a \acf{MANET} or a 
\acf{WMN} to know where to forward data packets. Such a protocol is 
`loop free' if it never leads to routing decisions that forward packets in circles.
This paper describes the mechanization of an existing pen-and-paper proof 
of loop freedom of \ac{AODV} in the interactive theorem prover Isabelle/HOL.
The mechanization relies on a novel compositional approach for lifting 
invariants to networks of nodes. We exploit the mechanization to analyse 
several improvements of \ac{AODV} and show that Isabelle/HOL can 
re-establish most proof obligations automatically and identify exactly 
the steps that are no longer valid.
\end{abstract}

%--   }-}1%%%%%%%%%%%%%%%%%%%%%%%%%%%%%%%%%%%%%%%%%%%%%%%%%%%%%%%%%%%%
\section{Introduction}\label{sec:intro}%{-{1

\acfp{MANET} and \acfp{WMN} are self-configuring wireless networks for 
mobile devices. 
Their nodes are reactive systems that cooperate to pass data packets from 
one node to another towards each packet's ultimate destination.
This global service must satisfy certain correctness properties---for 
example, that data packets are never sent in circles.
Proofs of such properties tend to be long and complicated, often 
involving many case distinctions over possible messages sent and 
combinations of Boolean predicates over internal data structures.
For example, the only prior existing proof\footnote{Earlier, and 
simpler, proofs appear in
\cite{AODV99,BOG02} and \cite{ZYZW09}, but none of them is complete and valid
for \ac{AODV} as standardized in \cite{RFC3561}. We justify this statement 
in Section~\ref{sec:related}.} of loop freedom of the \acf{AODV} routing 
protocol---one of the four protocols currently
standardized by the IETF MANET working group, and the basis of new WMN
routing protocols such as HWMP in the  IEEE 802.11s wireless mesh network
standard~\cite{IEEE80211s}---is about 18 pages long and requires around 40 
lemmas to prove the final statement~\cite{FehnkerEtAl:AWN:2013}. This proof 
is based on a process-algebraic model.

Mechanizing process calculi and process-algebraic models in an \ac{ITP} like 
Isabelle/HOL%
~\cite{NipkowPauWen:IsabelleTut:2002} can now almost be considered routine 
\cite{
BengtsonParrow09,
Hirschkoff:picalc:1997,
GothelGle:TimedCSP:2010,
FeliachiGauWol:Circus:2012}.
However, a lot of this work focuses on 
process calculi themselves---for example, by treating variable 
binding~\cite{BengtsonParrow09} or
proving that bisimulation is a congruence \cite{Hirschkoff:picalc:1997, 
GothelGle:TimedCSP:2010}.
While the study of security 
protocols has received some attention~\cite{DutertreSch:CSP:1997}, 
comparatively
little work has been done on
mechanizing the application of such calculi to the practical 
verification of network protocols.
In this paper, however, we focus on an application and mechanize the proof 
of loop freedom of \ac{AODV}, a crucial correctness property.
Our proof uses
standard transition-system based techniques for showing safety 
properties~\cite{MannaPnu:Safety, Lynch:DistAlgo:1996},
as well as
a novel compositional technique for lifting properties from individual nodes 
to networks of nodes \cite{ITP14}.
We demonstrate these techniques on an example of significant size and 
practical interest.
\advance\textheight 1pt

\advance\textheight -1pt
The development described in this paper builds directly on the 
aforementioned model and pen-and-paper proof of loop freedom of 
\ac{AODV}~\cite{FehnkerEtAl:AWN:2013}.
While the process algebra model and the fine details of the original proof 
are already very formal, the implication that transfers statements about 
nodes to
\pagebreak[3]
 statements about networks involves coarser reasoning over execution 
sequences. Our mechanization simplifies and clarifies this aspect by explicitly stating 
the assumptions made of other nodes and by reformulating the original 
reasoning into an invariant form, that is, by reasoning over pairs of states 
rather than sequences of states.

Given that a proof already exists and that mechanization can be so 
time-consuming, why do we bother?
Besides the added confidence and credibility that come with having even the 
smallest details fastidiously checked, the real advantage in encoding model, 
proof, and framework in the logic of an \ac{ITP} is that they can then be 
analysed and manipulated \mbox{(semi-)}automatically.
\refsec{variants} describes how we exploited this fact to verify 
variations to the basic protocol, and \refsec{related} argues that such 
models aid review and repeatability.
We expect that the work described will serve as a convenient and solid base 
for showing other properties of the \ac{AODV} protocol and studying other 
protocols, and, eventually, to serve as a specification for
refinement proofs.
Finally, although any such work benefits from the accumulation of technical 
advances and engineering improvements to \acp{ITP}, we argue that it cannot 
yet be considered routine.

The paper is structured as follows. In \refsec{aodv} we informally describe the \ac{AODV} protocol.
\refsec{loopfreedom} briefly states the theorem of loop freedom 
of \ac{AODV} in the form given to Isabelle/HOL. The following three sections explain the 
meaning of this statement:
\refsec{model} describes how the model of \ac{AODV} in the process algebra 
\acs{AWN} (\aclu{AWN})~\cite{FehnkerEtAl:AWN:2013} is translated into 
Isabelle/HOL; \refsec{stating network properties} describes the 
formalization of network properties such as loop freedom; and, 
\refsec{invariance} explains our formalization of invariance of network 
properties.
\refsec{proof} summarizes how we proved the theorem 
in Isabelle/HOL\@.
\refsec{variants} describes several improvements of \ac{AODV}, proposed in 
\cite{FehnkerEtAl:AWN:2013}, and illustrates the use of Isabelle/HOL in 
proving loop freedom of these variants.
Once the original proof has been mechanized, Isabelle/HOL can re-establish 
most proof obligations for these improvements automatically, and identify 
exactly the steps that are no longer valid and that need to be 
adjusted.
A detailed discussion of related work follows in \refsec{related}, 
followed by concluding remarks.

There is only space to show the most important parts of our mechanization of 
the process algebra \ac{AWN} and of the model of \ac{AODV}\@.
Comparing these parts before \cite[\textsection\textsection 
4--6]{FehnkerEtAl:AWN:2013} and after mechanization in Isabelle/HOL, should 
give sufficient clues about the remaining parts.
By focusing mainly on the application (loop freedom of \ac{AODV}), we only 
show a glimpse of our proof method.
A companion paper \cite{ITP14} presents the technical details of the 
mechanization of \ac{AWN} and the associated framework for compositional 
proof.
Source files of the complete mechanization 
in Isabelle/HOL are available online~\cite{AODVMECH}.

%--   }-}1%%%%%%%%%%%%%%%%%%%%%%%%%%%%%%%%%%%%%%%%%%%%%%%%%%%%%%%%%%%%
\section{The \ac{AODV} routing protocol}\label{sec:aodv}%{-{1

\begin{figure*}[t]
\centering
\includegraphics[scale=.78]{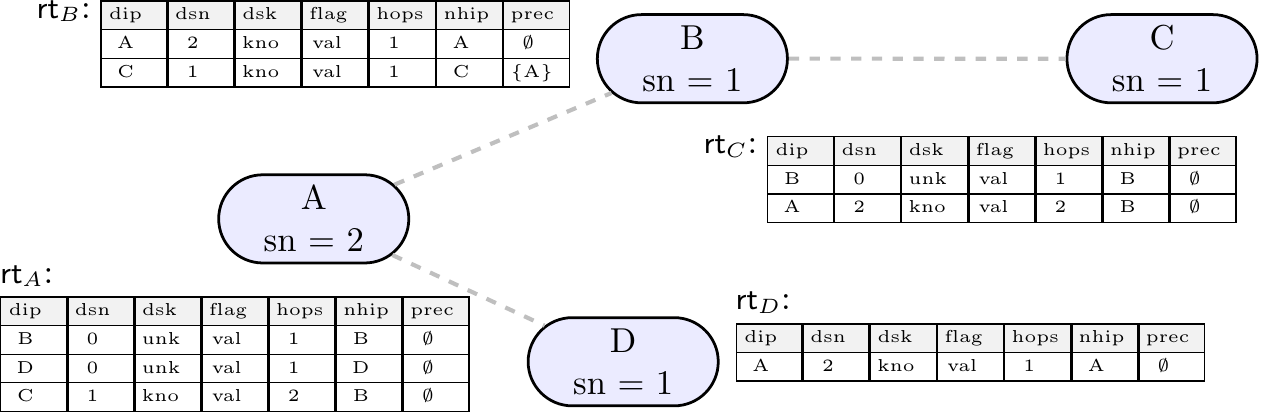}
\caption{Example \ac{AODV} instance.\label{fig:aodvnet}}
\vspace{-3mm}
\end{figure*}

\newcommand{\nA}{\isa{A}}
\newcommand{\nB}{\isa{B}}
\newcommand{\nC}{\isa{C}}
\newcommand{\nD}{\isa{D}}

The purpose of \ac{AODV}~\cite{RFC3561} is to route data packets between 
nodes.
\reffig{aodvnet} shows an example network with four nodes addressed \nA, 
\nB, \nC, and~\nD.
Each node maintains a local \emph{sequence number} (\isa{sn}) and a 
\emph{routing table} (\isa{rt}).
Imagine that the former are all set to 1, that the latter are all empty, and 
that \nA{} wants to send data to \nC.
No route is known, so \nA{} increments its \isa{sn} to 2 and broadcasts a 
\ac{rreq} message to its neighbours \nB{} and \nD.
Both neighbours immediately add a routing table entry with the 
\emph{destination address} (\isa{dip}) as \nA, the \emph{destination 
sequence number} (\isa{dsn}) as $2$, as sent by \nA, the 
\emph{destination-sequence-number status} (\isa{dsk}) as `known' 
(\isa{kno}), the \emph{route status} (\isa{flag}) as `valid' (\isa{val}), 
the \emph{number of hops to the destination} (\isa{hops}) as 1, the \emph{next hop address} (\isa{nhip}) as \nA,
and an empty set of \emph{precursors} (nodes known to be 
interested in this route).

Since neither \nB{} nor \nD{} has a routing table entry for \nC, they both 
in turn forward the \ac{rreq} message to their neighbours, which causes 
\nA{} to add entries for \nB{} and \nD{}, and \nC{} to add an entry for 
\nB{}.
Since (forwarded) \ac{rreq}s only include the \isa{sn} of the 
originating node---\nA{} in this case---the \isa{dsn} and \isa{dsk} fields 
of these new entries are set to
0 and `unknown' (\isa{unk}), respectively.
Node \nC{} also adds an entry for \nA{} to its routing table, with hop 
count~2 and next hop \nB{}. Since \nC{} is the destination, it replies to 
the request with a \ac{rrep} message, which is  destined for \nA{} and 
unicast to \nB{}.
On receipt, \nB{} updates its \isa{rt} with a route to \nC{} (adding \nA{} 
as precursor) and forwards the message to \nA.
When \nA{} receives this message it updates its \isa{rt} and starts 
forwarding data packets to \nC{} via \nB{}. All established routing table 
entries are summarized in \reffig{aodvnet}.

Besides this basic scenario, \ac{AODV} also supports early generation 
of \ac{rrep} messages by `intermediate nodes'~\cite[\textsection 6.6.2]{RFC3561}: whenever an intermediate node has 
information about a route to the destination, rather than
forwarding the \ac{rreq} message, it generates an \ac{rrep} message and 
sends it back to the originator of the \ac{rreq} message.
\ac{AODV} also supports \ac{rerr} messages for invalidating routes (sent to 
the `precursor nodes' associated with each 
entry).

\ac{AODV} features various timing requirements to expire 
packets and entries, and to limit sending rates, as well as optional 
extensions, such as `gratuitous \ac{rrep}s' to improve bi-directional 
efficiency, and
`local repair' of routes on link loss.
The model we mechanize includes the core functionality of \ac{AODV}, but 
not
timing details or optional 
features~\cite[\textsection 3]{FehnkerEtAl:AWN:2013}.

%--   }-}1%%%%%%%%%%%%%%%%%%%%%%%%%%%%%%%%%%%%%%%%%%%%%%%%%%%%%%%%%%%%
\section{Loop freedom}\label{sec:loopfreedom} %{-{1
Routing protocols must continue to correctly route data even as nodes appear, 
disappear, and move.
It is essential that they maintain \emph{loop freedom}: the absence of 
cycles across different routing tables.
For instance, the example network would have a cycle if \nD{} came into 
range of \nB{}, \nB{} updated its route for \nC{} to pass via \nD{}, and 
\nD{} added a route for \nC{} via~\nA.
Proofs of loop freedom when route replies are only generated by destination 
nodes are relatively subtle, but they become really delicate when 
intermediate nodes may also generate \ac{rrep} messages.

The main result shown in our mechanization is:
\begin{theorem}[\ac{AODV} loop freedom]\label{thm:aodv_loop_freedom}
For any well-formed network term \isa{n},\\[0mm]
\centerline{
\snippet{aodv_loop_freedom_concl}.
}
\end{theorem}

\noindent
Most of this paper is concerned with explaining the various elements of this 
statement and its proof in Isabelle/HOL\@.
In sum, the variable \isa{n} represents any well formed term 
describing a network instance---a term is well formed iff all nodes 
therein have distinct addresses.
It is mapped to an automaton by the functions \isa{pnet} and \isa{closed}.
A node with address \isa{i} comprises an instance of the protocol, 
\isa{paodv\ i}, reading messages from a queue process, \isa{qmsg}.
All reachable states of this model are shown to satisfy the formula at right 
of the `\isa{\isasymTTurnstile}', which \begin{inparaenum}
\item maps a state structured like the network term into a function from 
each node address to that node's local state (\isa{netglobal}),
\item abstracts this map into a directed graph with an arc from one node to 
another when the former has a valid routing table entry for a given 
\isa{dip} to the latter (\isa{rt-graph}), and finally,
\item claims that the transitive closure of each such graph is 
irreflexive.\end{inparaenum}

%--   }-}1%%%%%%%%%%%%%%%%%%%%%%%%%%%%%%%%%%%%%%%%%%%%%%%%%%%%%%%%%%%%
\section{Modelling \ac{AODV}}\label{sec:model} %{-{1
%{-{2

In Isabelle/HOL we formalize \ac{AODV} following the model from
\cite{FehnkerEtAl:AWN:2013}, which is expressed
in a process
algebra called \ac{AWN}~\cite[\textsection 4]{FehnkerEtAl:AWN:2013}.
In \ac{AWN}, a network instance running a protocol is modelled in five 
layers, from the bottom up: \begin{inparaenum}
\item sequential processes,
\item local parallel composition at a single network node,
\item nodes of the form ${\it ip}\mathord:P\mathord:R$ with {\it ip} the 
node's address, $R$ the set of reachable
  neighbours, and $P$ the process running on the node,
\item partial networks of nodes, and,
\item networks closed to further interaction.
\end{inparaenum}
The behaviour of each layer is defined by \ac{SOS} rules over either process 
terms or lower layers.\pagebreak[3]
By including initial states, the first layer defines an 
automaton and the others become functions 
over automata.

The four node network of \reffig{aodvnet} is, for 
example, modelled as\\[0mm]
\centerline{\snippet{eg1_all},}

\noindent where the function \isa{closed} models layer 5, closing a network,
and \isa{pnet} fabricates a partial network from a function mapping 
addresses to node processes and an expression describing the initial
topology.
For example, \snippet{eg1_node_a} becomes 
the node \msnippet{eg1_node_a'} with address \isa{A}, initial
neighbours \isa{B} and \isa{D}, and running a local composition of the 
protocol process \isa{paodv} (initialized with its address) fed by a queue 
\isa{qmsg} of incoming messages.
The communication ranges of nodes are independent of the structure of their 
composition and may change during an execution.
We now briefly describe each layer in more detail. Full details of \ac{AWN}, 
with all SOS\ rules, can be found in 
\cite{FehnkerEtAl:AWN:2013}\ and the source files~\cite{AODVMECH}.
%}-}2
\subsubsection{(1) Sequential processes} %{-{2

Both the \ac{AODV} protocol logic and the behaviour of message queues are 
specified by process terms of type \isa{('s, 'p, 'l) seqp}, parameterized by
\isa{'s}, the type
of the data state manipulated by the term, \isa{'p}, the type of process names, and \isa{'l}, the type of 
labels.
We write \isa{\isasymxi} or \isa{\isasymxi\isacharprime} for variables 
of type \isa{'s}, and \isa{p}, \isa{p'}, \isa{q}, or \isa{q'} for those of 
type \isa{('s, 'p, 'l) seqp}.
Labels are used to refer to particular control locations.

\begin{figure*}[t] %{-{2
\centering
\subfloat[Term constructors for {\sf ('s, 'p, 'l) seqp}.\label{fig:seqp:terms}]{ %{-{3
\hspace{-2pt}%
  \begin{mathpar}\!\!\!
    \msnippet{seqp_choice} \quad
        \msnippet{seqp_call} \quad
    \msnippet{lseqp_assign}   \quad
    \msnippet{lseqp_guard} \quad
    \msnippet{lseqp_ucast} \quad
    \msnippet{lseqp_bcast} \\
    \msnippet{lseqp_gcast} \quad
    \msnippet{lseqp_send} \quad
    \msnippet{lseqp_receive} \quad
    \msnippet{lseqp_deliver}\rule[-3mm]{0pt}{1mm}
  \end{mathpar}\hspace{-2pt}%
} %}-}3
\\
\subfloat[\acworkaround{SOS} rules for sequential processes: subset of
  \seqpsos.\label{fig:seqp:sos}]{ %{-{3
\hspace{-2pt}%
    \begin{mathpar}
        \msnippet{assignT'} \and
        \msnippet{choiceT1} \and
        \msnippet{callT} \and
        \msnippet{choiceT2} \and
        \msnippet{broadcastT}\rule[-2mm]{0pt}{1mm}
    \end{mathpar}\hspace{-2pt}%
    }     %}-}3
\caption{Sequential processes: terms and semantics\label{fig:seqp}}
\vspace{-4mm}
\end{figure*} %}-}2

The term constructors are summarized in \reffig{seqp:terms}:
\emph{assignment}, \snippet{lseqp_assign},
which transforms the data state deterministically (\isa{u} has type \isa{'s 
{\isasymRightarrow} 's}) and then acts as \isa{p}; \emph{guard/bind}, 
\snippet{lseqp_guard}, which returns the set of states where the guard 
evaluates to \isa{true}, one of which is chosen nondeterministically;
\emph{network synchronizations}, 
\isa{receive}/\isa{unicast}/\isa{broadcast}/\isa{groupcast}, whose 
destinations and contents depend on the data state; 
\emph{internal communications},
\isa{send}/\isa{receive}/\isa{deliver}, which do not need specified  
destinations and whose contents depend on the data~state;
\emph{choice} (\isa{\isasymoplus}), combining the possibilities of two 
subterms; and \emph{call} (\isa{call}), which jumps to a named process.
The argument \selmsg{} of \isa{broadcast} is a data
expression with variables, which evaluates to a message. It thus has type 
\isa{'s {\isasymRightarrow} msg}.
The argument \updmsg{} of \isa{receive}, on the other 
hand, is a variable that upon receipt of
a message is evaluated to the message received.
It has the type of a message-dependent state change 
(\isa{msg {\isasymRightarrow} 's {\isasymRightarrow} 's}).
A guard is of type \mbox{\isa{'s {\isasymRightarrow} 's set}}. It
can express both a construct that 
nondeterministically binds variables, giving a set of possible successor 
states, and one that returns a singleton set containing the current state 
provided it satisfies a given condition and the empty set otherwise.

The \ac{SOS} rules for sequential processes, \seqpsos{},
define a set of transitions.
A transition is a triple relating a source state, an action, and a 
destination state.
The states of sequential processes pair data components of type 
\isa{'s} with control terms of type \isa{('s, 'p, 'l) seqp}.
A set of transitions is defined relative to a \emph{(recursive) 
specification} \isa{\isasymGamma} of type \isa{'p {\isasymRightarrow} ('s, 
'p, 'l) seqp}, which maps process names to terms. 
Some of the rules are shown in \reffig{seqp:sos}.

\begin{wrapfigure}[20]{r}{0.47\textwidth}
\centering
\vspace{-7mm}
\includegraphics[width=0.4525\textwidth]{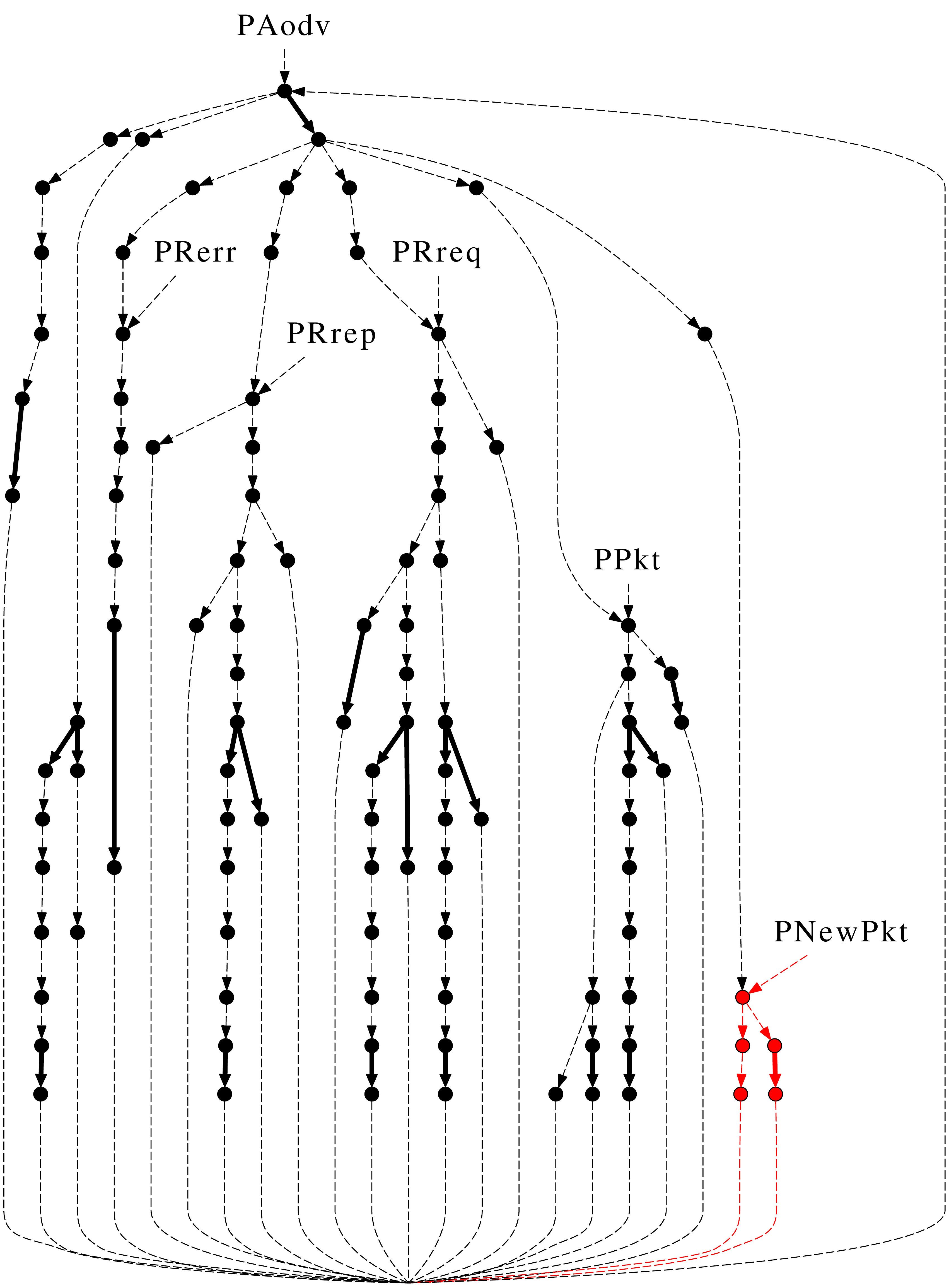}
\caption{Process term graph of \snippet{gammap}\label{fig:procterm}}
\vspace{-12ex}
\end{wrapfigure}

These elements suffice to express the control logic of 
the \ac{AODV}
protocol.
We introduce six mutually recursive processes whose names are shown in 
\reffig{procterm}.
This figure shows the control structure of the specification 
\snippet{gammap}, which maps each name to a term.
The main process is called \isa{PAodv}.
It can broadcast control packets or send data packets---the two descending 
subtrees at the very left---or on receiving a message
descend into one of the other terms depending on the message content---the 
five-pronged choice leading to the other labels.
All paths loop back to \isa{PAodv}.\vspace{1ex}
The smallest subprocess,
at bottom right,
is defined as\\
\centerline{
\begin{tabular}{r@{\,}c@{\,}l} %{-{3
\!\!\isa{\gammaaodv\ PNewPkt}
& \isa{\isacharequal}& \isa{labelled\ PNewPkt\ {\isacharparenleft}} \\
&&\isa{{\isasymlangle}{\isasymlambda}{\isasymxi}{\isachardot}\ \textsf{if}\ 
dip\ {\isasymxi}\ {\isacharequal}\ ip\ {\isasymxi}\ \textsf{then}\ 
{\isacharbraceleft}{\isasymxi}{\isacharbraceright}\ \textsf{else}\ 
{\isasymemptyset}{\isasymrangle}}
\\
& &
\isa{deliver{\isacharparenleft}data{\isacharparenright}\ {\isachardot}
\ {\isasymlbrakk}clear{\isacharunderscore}locals{\isasymrbrakk}
\ call{\isacharparenleft}PAodv{\isacharparenright}}
\\
&
\isa{{\isasymoplus}}
&
\isa{{\isasymlangle}{\isasymlambda}{\isasymxi}{\isachardot}\ \textsf{if}\ 
dip\ {\isasymxi}\ {\isasymnoteq}\ ip\ {\isasymxi}\ \textsf{then}\ 
{\isacharbraceleft}{\isasymxi}{\isacharbraceright}\ \textsf{else}\ 
{\isasymemptyset}{\isasymrangle}}
\\
& &
\isa{{\isasymlbrakk}{\isasymlambda}{\isasymxi}{\isachardot}\ 
{\isasymxi}{\isasymlparr}store\ {\isacharcolon}{\isacharequal}\ add\ 
{\isacharparenleft}data\ {\isasymxi}{\isacharparenright}\ 
{\isacharparenleft}dip\ {\isasymxi}{\isacharparenright}\ 
{\isacharparenleft}store\ 
{\isasymxi}{\isacharparenright}{\isasymrparr}{\isasymrbrakk}}
\\
& &
\isa{{\isasymlbrakk}clear{\isacharunderscore}locals{\isasymrbrakk}
\ call{\isacharparenleft}PAodv{\isacharparenright}\ 
{\isacharparenright}}\mbox{.}
\end{tabular} %}-}3
}\medskip

\noindent
It branches on whether or not the \isa{dip} and \isa{ip} variables have the 
same value in the current state, and then either delivers a message or 
updates the variable \isa{store}.
Each branch then loops back to the main process.
The \isa{labelled} function recursively labels the control states from 
\isa{PNewPkt-:0} through \mbox{\isa{PNewPkt-:4}}.

The graph of \reffig{procterm} summarizes the just over 100 control 
locations and shows that the model contains both significant branching and 
sequencing, some of which is exploited in the verification.
The thicker, solid lines are synchronizing actions.
The dashed lines are assignments or guards.
Each straight sequence of dashed lines, which correspond to a sequence of 
assignments, could in fact be replaced by a single dashed line, by nesting 
state transformations.
However, this would make the model easier to get wrong, harder to read, and 
less like implementations in programming languages.
Moreover, it is easier to verify many small steps, especially 
since typically most are dispatched automatically.

The \ac{AODV} data state is modelled in a standard way using 
records~\cite[\textsection 3.3]{SchirmerWen:Statespaces:2009}.
There are five global variables:\footnote{These variables are global in that 
their values are maintained indefinitely; however, they are local to each 
specific process running on a specific node.} \isa{ip}, the local address of 
type \isa{nat}; \isa{sn}, the current sequence number, also a \isa{nat}; 
\isa{rreqs}, a set of pairs of addresses and request identifiers that tracks 
already handled route requests; \isa{store}, a partial map from addresses to 
a status and queue of pending data packets; and \isa{rt}, the routing table, 
a partial map from addresses to entries of type \snippet{r_type}, where 
\isa{sqn} and \isa{ip} are synonyms for \isa{nat}.
Each subprocess also has its own local variables, which in the original 
process algebra~\cite{FehnkerEtAl:AWN:2013} are often initialized from 
expressions during subprocess invocation.
While it may seem `tidy' to explicitly model variable locality, it 
complicates syntactic details and the aim of working ultimately on 
transition systems where there is no notion of a subprocess.
Furthermore, the recursion through \snippet{gammap} induced by \isa{call} 
already entails one or two technical details, as we
discuss in~\cite{ITP14}, even before variable assignment is considered.
So, rather than stay faithful to these details in the mechanization, we 
simply include the union of 12 local variables in the state record.
When invoking a subprocess, these variables are set to arbitrary 
values,%
\footnote{Almost: as {\sf sip} may not equal {\sf ip}, we use {\sf 
{\isasymxi}{\isasymlparr}sip\ {\isacharcolon}{\isacharequal}\ SOME\ 
x{\isachardot}\ x\ {\isasymnoteq}\ ip\ {\isasymxi}{\isasymrparr}}.
}
by \isa{clear-locals}, before combining assignment and \isa{call}.
This pragmatic solution works well in our setting.

The routing table, \isa{rt}, is central to \ac{AODV} and thus to this 
verification.
Several functions are defined to access, add, update, and invalidate its 
entries.
For example, the function \isa{nhop} of type \snippet{nhop_type} gives the 
next hop for an address if defined; \isa{update} of type 
\snippet{update_type} encodes the rules determining if and how an entry is 
modified, \isa{invalidate} of type \snippet{invalidate_type} marks a set of 
routes as invalid (possibly individually setting sequence numbers), and 
\isa{addpreRT} of type \snippet{addpreRT_type} adds precursors to an entry.
These (partial) functions are defined in the $\lambda$-calculus of 
Isabelle/HOL.

The process state is a pair of a data record and a process term.
The (singleton) set of initial states is given by a function,
\snippet{sigmap}
where \isa{i} is the initial value of \isa{ip}.
The process term represents the control state: a sort of a symbolic program 
counter.
As the rules in \reffig{seqp} indicate, process labels have no influence on 
the flow of control (unlike in~\cite{MannaPnu:Safety}).
They exist only to aid verification.
The sets of initial states and transitions are bundled into a generic 
automaton structure with two fields to model an \ac{AODV} process:
\begin{tightcenter}
\snippet{paodv}\isa{\ i} = \snippet{paodv_term}\mbox{.}
\end{tightcenter}
The states of this automaton (an overapproximation of the set of reachable 
states) are
determined by the type of the sources and targets of the transitions.

Having considered sequential processes, we now consider the other four 
layers.

%}-}2
\subsubsection{(2) Local parallel composition} %{-{2

An instance of \ac{AODV} must always be willing to accept a message from its 
environment.
To achieve this, and to model asynchronous message transmission, the 
protocol process is combined with a simple FIFO-queue model (modelled as a 
sequential process): \snippet{paodv_qmsg_term}.
The composition operator applies to automata:
\begin{tightcenter}
\snippet{par_comp}\mbox{.}%}
\end{tightcenter}
\noindent The rules for \parpsos{}
can be found in~\cite{AODVMECH, FehnkerEtAl:AWN:2013}.

%}-}2
\subsubsection{(3--4) Nodes and partial networks} %{-{2
\newcommand{\dval}[1]{\mbox{\small\sf #1}}

Networks of \ac{AODV} instances are specified as values of an inductive 
type:
a \snippet{net_tree} is either a node \snippet{pnet_node_term} with address 
\isa{i} and a set of neighbour addresses \Ri,
or a composition of two \snippet{net_tree}s \mbox{\snippet{pnet_par_term}}.
The function \isa{pnet} maps each such value to an automaton:
\label{pnet}%
\begin{tightcenter}
\begin{tabular}{l@{\ }c@{\ }l}
\snippet{pnet1_lhs} & = & \snippet{pnet1_rhs} \\
\snippet{pnet2_lhs} & = & \snippet{pnet2_rhs}\mbox{,}
\end{tabular}
\end{tightcenter}

\noindent
where \dval{np} is a function from addresses \dval{i} to parallel 
process expressions, such as
$\lambda \dval{i}.\;\dval{paodv}\, \dval{i} 
\mathrel{\mbox{$\langle\!\langle$}} \dval{qmsg},$ and where
\begin{tightcenter}
\mbox{\snippet{node_comp}}.
\end{tightcenter}
The states of such automata mirror the structure of the original network 
term.
Node states are denoted \snippet{net_state_nodes} and composed states are 
denoted \snippet{net_state_subnets}.
During an execution of a network, the tree structure and addresses remain 
constant, but the neighbours, control states, and data states of nodes may 
evolve.

%}-}2
\subsubsection{(5) Complete networks} %{-{2

Such a network is closed to new node interactions:
\begin{tightcenter}
\snippet{closed'} = \snippet{closed_term}.
\end{tightcenter}

\myparagraph{In sum,} this section has presented the part of
\refthm{aodv_loop_freedom} to the left of the
`\isa{\isasymTTurnstile}'.

%}-}2
%--   }-}1%%%%%%%%%%%%%%%%%%%%%%%%%%%%%%%%%%%%%%%%%%%%%%%%%%%%%%%%%%%%
\section{Stating network properties}\label{sec:stating network properties} %{-{1

Our verification exploits the inductive structure of \snippet{net_tree}s and
states.
Experience taught us to keep this structure for as long as possible and only 
later to transform it into a partial mapping from addresses to data records, 
using:

\begin{tightcenter}
\begin{tabular}{lcl}
\snippet{netlift1_lhs} & = & \snippet{netlift1_rhs} \\
\snippet{netlift2_lhs} & = & \snippet{netlift2_rhs}\mbox{,}
\end{tabular}
\end{tightcenter}

\noindent
where \isa{sr} divides the state into `exposed' parts (the \isa{poadv} data 
state) and `masked' parts (the \isa{paodv} control term and the states of 
\isa{qmsg}).
The \isa{fst} then elides the latter.
The result of \isa{netlift} is a partial function from addresses to the 
exposed part of the state of the corresponding node.
It is made total by mapping missing elements to the initial (data) state 
before being passed to a property \isa{P}:
\begin{tightcenter}
   \snippet{netglobal},
\end{tightcenter}
   where \snippet{default}.\pagebreak[3]
\isa{P} is a property over a function, written $\sigma$, from addresses 
\dval{i} to data records.
It can thus speak, for
instance, of the routing table maintained by node \dval{i}.
The function \dval{netglobal} turns such a property into a
property of network states \dval{s}.

An example of \isa{P} occurs in Theorem~\ref{thm:aodv_loop_freedom}:
$\lambda\sigma.\;\forall \dval{dip}.\;
\dval{irrefl}((\dval{rt-graph} \, \sigma \, \dval{dip})^+).$
Here \dval{rt-graph} is a function that, given an address
\dval{dip} and a function $\sigma$ from addresses to data 
states, generates a routing graph: its vertices are all possible 
addresses \dval{i} and there is
an arc $({\dval{ip}},{\dval{ip}}')$ iff $\dval{ip}\mathop{\not=}\dval{dip}$ and
the entry for \dval{dip} in the routing table at node \dval{ip} has the form
$
({*},{*},\dval{val},*,{\dval{ip}'},{*}).
$
An arc in this routing graph indicates that $\dval{ip}'$ is 
the next hop on
a valid route to $\dval{dip}$ known by $\dval{ip}$; a path in a routing
graph describes a route towards $\dval{dip}$ discovered by \ac{AODV}\@.
We say $\sigma$ is \emph{loop free} if the
routing graphs $\dval{rt-graph} \, \sigma \, \dval{dip}$
are acyclic, for all destinations $\dval{dip}$, i.e.\ if $P \sigma$ holds.
A network state \dval{s} is \emph{loop free} iff the function
$\dval{netmap} \, \dval{s}$
from addresses to data records is loop free, i.e.\ if $\dval{netglobal} \, P \, \dval{s}$.
Finally, a routing protocol, such as \ac{AODV}, is \emph{loop free} iff all
reachable network expressions are loop free. This quantification over
reachable network expressions is encoded in the symbol \isa{\isasymTTurnstile}.

\myparagraph{In sum,} this section has presented the part of
\refthm{aodv_loop_freedom} to the right of the
`\isa{\isasymTTurnstile}'.

%--   }-}1%%%%%%%%%%%%%%%%%%%%%%%%%%%%%%%%%%%%%%%%%%%%%%%%%%%%%%%%%%%%
\section{Stating invariance of network properties}\label{sec:invariance} %{-{1

As \refthm{aodv_loop_freedom} is a safety property, we need only consider 
questions of invariance, that is, properties of reachable states.
The meta theory is classic~\cite[Part III]{Muller:PhD:1998}.

\begin{definition}[reachability]\label{def:reachable}
For an automaton~\isa{A} and an assumption~\isa{I} over 
actions, \isa{reachable\ A\ I} is the smallest set defined by the rules:
\begin{mathpar}
\hfill
\msnippet{reachable_init}
\hfill
\msnippet{reachable_step}
\hfill\mbox{}
\end{mathpar}
\end{definition}

\begin{definition}[invariance]\label{def:invariant}
For an automaton~\isa{A} and an assumption~\isa{I} over 
actions, a predicate~\isa{P} is \emph{invariant}, written
\snippet{invariant_lhs}, iff \snippet{invariant_rhs}.
\end{definition}

\begin{definition}[transition invariance]\label{def:transition-invariant}
For an automaton~\isa{A} and an assumption~\isa{I} over 
actions, a predicate~\isa{P} is \emph{transition invariant}, 
written
\snippet{step_invariant_lhs}, iff
\snippet{step_invariant_rhs}.
\end{definition}

\noindent
We recover the standard definition when \isa{I} is \snippet{TT_rhs} and 
write simply \snippet{invariant_TT_lhs}.

\myparagraph{In sum,}
this finishes the presentation of Theorem~\ref{thm:aodv_loop_freedom}.

%--   }-}1%%%%%%%%%%%%%%%%%%%%%%%%%%%%%%%%%%%%%%%%%%%%%%%%%%%%%%%%%%%%
\section{Our proof}\label{sec:proof} %{-{1

To show invariance, we follow the compositional strategy elucidated 
in~\cite[\textsection 1.6.2]{deRoeverEtAl:ConcVer:2001}.
That is, we take as a basic element automata with {\seqpsos\ \gammaaodv} as 
the set of transitions, show invariance using induction, and then develop 
proof rules for each of the operators defined in the previous section to 
lift the results to complete networks.
The inductive assertion method is also classic, see, for example,
Manna \& Pnueli~\cite[Rule \textsc{inv-b}]{MannaPnu:Safety}.
Its core, in Isabelle, is the induction principle associated with 
\refdef{reachable}.\pagebreak[4]

Both the original and mechanized proofs of \refthm{aodv_loop_freedom} 
involve a succession of invariants and transition invariants leading to the 
ultimate result. As an example,
\begin{equation}\label{eq:hop_count_positive}
\begin{array}{c}
  \mbox{\snippet{hop_count_positive}}
\end{array}\end{equation}
states that `all routing table entries have a hop count greater than or 
equal to one' (\isa{kD} gives the domain of a routing table; the \isa{the} 
adds an obligation to show that \isa{ip} is in the domain of \isa{dhops (rt\ 
\isasymxi)}).
This particular predicate only ranges over the data state, \isa{\isasymxi}, 
but others also range over labels, for example,
\vspace{-1mm}
\begin{equation}\label{eq:received_msg_inv}
\begin{array}{c}
  \mbox{\snippet{received_msg_inv}}
\vspace{-1mm}
\end{array}\end{equation}
states that `if for every \isa{receive m}, \isa{m} satisfies \isa{P},
then the \isa{msg} variable also satisfies \isa{P} at location 
\isa{PAodv-:1}'.
The map \snippet{onl_lhs}, defined by \snippet{onl_rhs}, 
extracts labels from control states, which obviates the need to include and maintain 
process terms in invariants.\footnote{Using labels is standard, see, for 
instance,~\cite[Chap. 1]{MannaPnu:Safety}, or the `assertion networks' of
~\cite[\textsection 2.5.1]{deRoeverEtAl:ConcVer:2001}.
Isabelle rapidly dispatches uninteresting cases.}

Invariants like these are solved by establishing them for all possible 
initial states, and showing that they are preserved by all transitions.
The soundness of this method is also formally justified
in Isabelle/HOL~\cite{ITP14}.

\renewcommand{\dval}[1]{\textit{#1}\/} % data value
\newcommand{\keyw}[1]{{\tt #1}}
\newcommand{\fnakD}{\keyw{vD}}
\newcommand{\spaces}[1]{\ #1\ }
\newcommand{\ims}{\spaces{\Rightarrow}}
\newcommand{\ans}{\spaces{\wedge}}
\newcommand{\akd}[2][N]{\fnakD_{#1}^{\ensuremath{#2}}}
\newcommand{\xiN}[2][N]{\xi_{#1}^{#2}}
\newcommand{\rtsord}[1][\dval{dip}]{\ensuremath{\sqsubset_{#1}}}
\newcommand{\rt}{\keyw{rt}}
\newcommand{\fnnhop}{\keyw{nhop}}
\newcommand{\nhp}[2][\dval{dip}]{\fnnhop_N^{\ensuremath{#2}}(\ensuremath{#1})}

This approach suffices for showing nearly 
all intermediate invariants, but not for expressing the final invariant from 
which \refthm{aodv_loop_freedom} follows.
The authors of the original proof~\cite[Theorem 7.30]{FehnkerEtAl:AWN:2013} 
introduce a notion of `quality' of routing table entries, and show that it 
strictly increases along any valid route to a destination \dval{dip}.
They formalize this as\\[1mm]
\centerline{$
\mbox{``}\dval{dip}\in\akd{\dval{ip}}\cap \akd{\dval{nhip}} 
\ans\dval{nhip}\not=\dval{dip}
\ims \xiN{\dval{ip}}(\rt)\rtsord \xiN{\dval{nhip}}(\rt)\mbox{'',}
$}\\[.5mm]
where $N$ is a ``reachable network state'', $\akd{\dval{ip}}$ are the 
addresses for which \dval{ip} has valid routing entries, and $\dval{nhip}$ 
is the address of the next hop toward \dval{dip} at \dval{ip}.
But our basic invariants, like~\refeq{hop_count_positive} or 
\refeq{received_msg_inv}, can only refer to the local model's state 
(\isa{\isasymxi}).
How can we compare the states at two nodes ($\xi^{\dval{ip}}$ and 
$\xi^{\dval{nhip}}$) without immediately introducing the whole model before 
the~`\isa{\isasymTTurnstile}'?

Our solution is to introduce `open' versions of the \ac{SOS} rules and 
operators, and of reachability and (transition) invariance.

The \emph{open} \ac{SOS} of \ac{AWN} differs from the default 
(\emph{closed}) version by modelling data states as (total) functions from 
node addresses to data records.
That is, rather than define rules over a variable \isa{\isasymxi} of type 
\snippet{state_type}, they are defined over a variable \isa{\isasymsigma} of 
type \snippet{sigma_type}.
At the level of sequential processes, the open equivalent of \seqpsos{} 
(\oseqpsos) is additionally parameterized by an address \isa{i}, and the 
\ac{SOS} rules only constrain this \isa{i}th component.
At the level of local parallel compositions within a node, we simply inherit 
the data state from the left argument (the process 
\isa{aodv}).\footnote{This suffices for our work, but a symmetric solution 
may be preferable.}
The lifting to node expressions is unproblematic.
The composition of two partial networks effectively synchronizes the state 
mappings: the only transitions that can occur are those where, given a 
source state~(\isa{\isasymsigma}), both components agree on a destination 
state~(\isa{\isasymsigma'}).

We developed a framework for stating invariants over automata with open 
transitions.
These invariants
differ from those like \refeq{hop_count_positive} and 
\refeq{received_msg_inv} in that assumptions are not stated over incoming 
messages but rather over synchronized and interleaved 
transitions---that is, over communications with an environment and over the 
independent actions of the environment---and properties are stated 
over the entire state of the network.
We show invariants at the level of a single process (\isa{paodv\ i}), with 
the additional obligation of showing their preservation under all 
interleaving transitions that satisfy the stated assumption, and then `lift' 
them to arbitrary closed networks by applying a succession of generic 
lemmas, one for each layer of \ac{AWN}.
The additional obligation is exploited as a hypothesis in the 
induction
that lifts results over partial networks 
(that is, when only one side acts, the 
property remains invariant).
The lifting rules require showing that a process satisfies the assumptions 
on synchronizations and interleavings made in the invariant statement.
These assumptions must also be lifted for each layer; care is required to 
avoid circularity in such assumption-guarantee invariants.

The framework includes a generic `transfer' lemma that infers from an 
invariant over an open model, a similar invariant over the corresponding 
closed model---in our case, the very model presented in \refsec{model}.
One need only show a relation between the \isa{np} given to \isa{pnet} (see 
page \pageref{pnet}) and a corresponding 
\isa{onp} and \isa{opnet} of the open model.
This transfer of results means that all of the definitions and lemmas 
associated with the open model are but a proof strategy: they are not needed to understand the 
statement of \refthm{aodv_loop_freedom} and their soundness is guaranteed by 
Isabelle/HOL.
The details of this proof strategy are given~in~\cite{ITP14}.

\myparagraph{Route quality.}
For completeness, we include the definition used for route quality.
We write
   \snippet{rt_strictly_fresher},
to mean that the quality of the route to address \isa{i} in route 
table~\isa{rt\isactrlsub 2} is strictly better than that 
in~\isa{rt\isactrlsub 1}, where,
\begin{tightcenter}\
   \snippet{rt_fresher2_concl},
\end{tightcenter}
provided \snippet{rt_fresher2_prem_1} and \snippet{rt_fresher2_prem_2}.
The function \isa{dhops\ rt\ i} yields the number of hops to \isa{i} 
according to \isa{rt}.
We encode the notion of \emph{net sequence numbers} from~\cite[\textsection 
7.5]{FehnkerEtAl:AWN:2013}:
\begin{tightcenter}
   \snippet{nsqn_sqn_def},
\end{tightcenter}
where \isa{flag} states whether a route is valid (\isa{val}) or invalid 
(\isa{inv}) and \isa{sqn} gives the stored sequence number.

\myparagraph{Results.}
Our mechanization of the \ac{AODV} model and the proof 
of loop freedom (not including the framework of \cite{ITP14}) involves 
$360$ lemmas, of which 40 are invariants, with a 
proof text spanning 80 printed pages.
The pen-and-paper proof \cite{FehnkerEtAl:AWN:2013} involves 
$40$ lemmas over 18 pages.
Many of the mechanized lemmas are of course trivial, for 
example, simplification rules for projections from routing tables.

The pen-and-paper proof is fastidious and we did not 
find any major errors:
\begin{inparaenum}
\item
type checking found a minor typo in the model,
\item
one proof invoked an incorrect invariant requiring the addition and proof of 
a new invariant based on an existing one,
\item
a minor flaw in another proof required the addition of a new invariant.
\end{inparaenum}
Of course, this was not known beforehand!
Nevertheless, our mechanized proof provides supplementary evidence for the 
stated property.

%--   }-}1%%%%%%%%%%%%%%%%%%%%%%%%%%%%%%%%%%%%%%%%%%%%%%%%%%%%%%%%%%%%
\section{Analysing variants of \ac{AODV}}\label{sec:variants} %{-{1
%{-{2

A mechanized model and 
proof greatly facilitates the analysis of protocol
variants, such as different interpretations of the informal text of a standard, or 
proposed improvements for future versions of a standard.
Since such variants often only differ in minor details, most proofs stay the 
same or are adapted automatically.
An \ac{ITP} tries to `replay' the 
original proof and, in case of a failure, it indicates those proof steps 
that are no longer valid.
One can thus concentrate on important changes in the proof.
This avoids the tedious, time-consuming, and error-prone manual chore of 
establishing which steps remain valid for each invariant, especially for 
long~proofs.
We support our claim by 
proving the loop freedom property of four variants of \ac{AODV}\@.

%}-}2
\subsubsection{(1) Skipping route request identifiers} %{-{2

\ac{AODV} uses route request identifiers to uniquely identify 
\ac{rreq} messages.
Since it has 
been shown~\cite[\textsection 10.1]{FehnkerEtAl:AWN:2013} that a combination 
of IP addresses and sequence numbers adequately serves the same purpose, the 
identifier can be dropped and the size of the \ac{rreq} messages reduced.
This very minor modification only requires a change in the type of 
\ac{rreq}, and related propositions and lemmas.
Implementing these changes in Isabelle/HOL only took several minutes---the 
invariant lemmas were re-proved automatically.

%}-}2
\subsubsection{(2) Forwarding route replies}%{-{2

During route discovery, an \ac{rrep} message is unicast back towards the 
originator of the triggering \ac{rreq} message.
Every intermediate node on the selected route processes the \ac{rrep} 
message and, in most cases, forwards it towards the originator.
However, intermediate nodes must discard
\ac{rrep} messages from which they cannot distil any new routing 
information.
As a consequence, the originator node will not receive a reply.\footnote{See 
\url{http://www.ietf.org/mail-archive/web/manet/current/msg05702.html}.}

An alternative is to require intermediate nodes to forward \emph{all\/} 
\ac{rrep} messages.
This behaviour is modelled by deleting three 
lines of the original specification; including one choice operator 
(\isa{\isasymoplus}) and two guards, potentially invalidating the proofs of many invariants.
To avoid the forwarding of 
outdated information, we change three lines in the 
specification (including two guards) to ensure that the best available 
information is always sent; see \cite[\textsection 
 10.2]{FehnkerEtAl:AWN:2013} for details.

Of the $360$ lemmas in the original proof, only $7$ are no longer valid.
Four of these are easily repaired: since some lines of the specification are 
deleted, the automatically generated labels change and references in the 
proofs must be adapted---a tedious, but routine find-and-replace 
chore.
So, in fact only three invariants require non-trivial
user interaction and a new proof---around three hours of manual effort.
%}}
This corresponds with the pen-and-paper proof, which requires a single 
page~\cite[pp.~106--107]{FehnkerEtAl:AWN:2013} and a new 
invariant~\cite[Prop.~7.38]{FehnkerEtAl:AWN:2013}.

%}-}2
\subsubsection{(3) From groupcast to broadcast}\label{sec:gcastbcast} %{-{2

Each routing table entry contains a set of precursors (\refsec{aodv}), a set 
of the IP addresses of all nodes that are currently known to be potential 
users of the route, and that are located one hop further away from the 
destination.
This information is recorded so that these nodes can be informed via 
\ac{rerr} message if the route becomes invalid.
However, precursor lists are incomplete: nodes not handling a route reply 
have no information about precursors for routes established while handling 
\ac{rreq} messages%
---see \cite[\textsection 
10.4]{FehnkerEtAl:AWN:2013} for examples.
As a consequence, some nodes cannot be informed of a link break and will use 
a broken route, and data packets can be lost. 

One solution is to abandon precursors and to replace \isa{groupcast}s by 
\isa{broadcast}s.
The \ac{AODV} specification is updated by dropping the precursor field of 
routing table entries, and making minor changes to related functions and function calls.
All $7$ occurrences of 
\isa{groupcast} must also be replaced by an appropriate 
\isa{broadcast}-statement; in one case this necessitates the 
introduction of a new guard.
$16$ assignments dealing with the generation and maintenance of 
precursor lists, and the calculation of groupcast destinations, are deleted.

Several changes ensued.
\begin{inparaenum}
\item
Around $30$ definitions and lemmas about precursor lists are no longer needed.
\item
About $75$ lemmas and proofs then require adjustments for typing errors and 
references to deleted lemmas.
\item
The labels in $6$ invariants must be updated.
\item
One invariant requires a careful adjustment: the removal of one case of a 
case distinction.
\end{inparaenum}

In sum, the specification changes broke many invariant proofs, but 
these were easily fixed in around three hours.

%}-}2

\subsubsection{(4) Forwarding route requests} %{-{2

During route discovery, an \ac{rreq} message is dropped if
a node (destination or intermediate) replies to the sender with an \ac{rrep} message.
This dropping of the \ac{rreq} message may inadvertently lead to non-optimal 
routes  to the originator at nodes laying `downstream' of the node 
that sent the reply~\cite{MK10}.

One possible solution is to
make nodes forward {\em all\/} \ac{rreq}s\ that they have 
not handled earlier.
The forwarded \ac{rreq} messages must be augmented with a Boolean flag to 
indicate that a reply has already been generated and sent.
The specification of this variant differs in 
only eight lines from the original~\cite[\textsection 10.5]{FehnkerEtAl:AWN:2013}.

As before, the proof is adapted in response to 
feedback from Isabelle/HOL.
\begin{inparaenum}
\item $17$ lemmas must now include the newly introduced flag.
\item The labels in $4$ invariants must be updated.
\item Only one invariant proof required major changes.
\end{inparaenum}

%}-}2
%--   }-}1%%%%%%%%%%%%%%%%%%%%%%%%%%%%%%%%%%%%%%%%%%%%%%%%%%%%%%%%%%%%
\section{Related work}\label{sec:related} %{-{1

\myparagraph{Bhargavan et al.~\cite{BOG02}.} %{-{2
Bhargavan, Obradovic, and Gunter apply a mix of manual reasoning, 
interactive theorem proving, and model checking to a preliminary draft 
(version 2) of \ac{AODV}
and show that this early draft is not loop free.
They also suggest three improvements, of which one has been incorporated 
into the standard~\cite{RFC3561}, and present a proof of loop
  freedom for the modified version.
A central role in this proof is played by an invariant stating that along a 
route either sequence numbers increase, or, 
when they stay constant, that the hop count
decreases~\cite[Theorem 17]{BOG02}. However, this is only true for valid 
routes---an assumption that is not stated.
\pagebreak[3]
Even were the assumption adopted, it is not clear how the property can be 
shown using step-by-step reasoning which must treat the case of 
invalid routes that become valid~again.
Looking at the proofs in \cite{BOG02}, it turns out that Lemma~20(1) of 
\cite{BOG02} is invalid. This failure is surprising, given that according to 
\cite{BOG02} Lemma~20 is automatically verified by SPIN\@. A possible 
explanation might be that this lemma is obviously valid for the version of 
AODV prior to the recommendations of \cite{BOG02}.

Bhargavan et al.\ concede that they do not formally prove the abstractions 
they use for model checking~\cite[p.565]{BOG02}, and 
it is not otherwise clear whether or how they ensure consistency with their 
\ac{ITP} models.
Besides the obvious question of soundness, these manual steps make it harder 
to reproduce the stated results.
If any part of the model is changed, the automatic components can be 
rerun, but both the relations between them and any other manual reasoning 
must be carefully re-evaluated.
In contrast, our development is a complete mechanization that is 
automatically validated by Isabelle/HOL in less than $15$ minutes.
\vfill

%}-}2
\myparagraph{Zhou et al.~\cite{ZYZW09}.} %{-{2
Zhou, Yang, Zhang, and Wang model \ac{AODV} as a set of finite 
traces---lists of events---defined inductively using a technique expounded 
by Paulson~\cite{Paulson:Inductive:1998}.
The model features $15$ cases for adding a new event to an existing 
trace~$\tau$; most involve conditions on `observation functions' that 
recurse over~$\tau$ to `recreate' the system state.
Zhou et al.\ show the invariant proposed by Bhargavan et al.\ but with an 
explicit assumption on route activity.
They do so using an ingenious but intricate lemma~(`l65') that exploits the 
identification of states and histories to reason across periods of route 
invalidity.

Both our mechanization and the original pen-and-paper 
proof~\cite{FehnkerEtAl:AWN:2013} were completed without access to the 
details of Zhou et al.'s work.
When we were able to examine their model in detail, we were reassured to see 
that the two models largely agree on the reading of the standard.
But there are two crucial differences.
First, we model route replies by intermediate nodes and Zhou et al.\ do not.
This feature is a core part of the \ac{AODV} standard~\cite[\textsection 
6.6.2]{RFC3561} and quite subtle---even small deviations, such as a 
  different reading of the standard, risk introducing loops~\cite{AODVloop}.
While we think it would be possible to add such replies to their model 
without introducing loops, extending the proof would not be trivial and 
would likely require new invariants similar to those that we use.
Second, Zhou et al.\ model some timing details and we do not.
But they sidestep central issues.
For example, according to the standard a 
route is expired in two steps: ``the Lifetime field in the routing 
table plays [a] dual role---for a valid route it is the expiry time, and for 
an invalid route it is the deletion time''~\cite[\textsection 
6.11]{RFC3561}.
Zhou et al.\ model the first step but not the second.
Route deletion, however, is fundamental to a timed model since 
all route information is lost; it complicates even the basic lemma that 
local sequence numbers never decrease\ (their~`l63').

Apart from content, the two models also differ in style.
While the model of Zhou et al.\ is event-based and declarative, ours 
is more operational---it states what an abstract implementation of 
the protocol does step-by-step.
The difference is important for two reasons: validation against the 
standard~\cite{RFC3561} and refinement to an implementation.
Arguably, such standards are often quite operational, perhaps because 
they are written by and for implementers.
We expect it to be easier to state 
and show some kind of simulation relation between the states and transitions 
of our model and those of an implementation (model).

As this discussion shows, in addition to guaranteeing soundness and 
facilitating repeatability and reuse, mechanized proofs aid detailed 
comparisons with related work (provided the proof scripts are available for 
study).
Since an \ac{ITP} checks the proofs, one can focus on comparing models and 
properties.
Furthermore, rather than puzzle over details omitted or unclear from 
published accounts, one can look for answers in the mechanization.
\ac{AODV} is an interesting case study since at least two  
mechanical models exist, and the protocol is of industrial relevance, 
complicated enough to be interesting, but not so large as to pose too many 
engineering problems.
\vfill

%}-}2

\myparagraph{Mechanically verifying reactive systems.}
Apart from the process-algebraic work described in the introduction, several 
other approaches for verifying reactive systems have been mechanized, namely
UNITY~\cite{HeydCre:Unity:1996},
I/O Automata~\cite{Muller:IOAuto:1998}, and
TLA${}^{+}$~\cite{ChaudhuriEtAl:TLA+:2010}.
The main difference with our approach is that they typically do not 
distinguish control and data states---specifications are essentially flat 
sets of transitions.
The last two frameworks, in particular, have focused on the verification of 
practical protocols but not, to our knowledge, on the kind of routing 
protocol exemplified by \ac{AODV}.

%}-}2
%--   }-}1%%%%%%%%%%%%%%%%%%%%%%%%%%%%%%%%%%%%%%%%%%%%%%%%%%%%%%%%%%%%
\section{Conclusion}\label{sec:conclusion} %{-{1

We have presented a mechanical proof of a model that corresponds to 
an interpretation of the current version of the \ac{AODV} 
standard~\cite{RFC3561}; the fidelity of this model is argued 
in~\cite{FehnkerEtAl:AWN:2013}.
It includes route replies from intermediate nodes but not timing features.
Such a mechanization does more than confirm the correctness of the existing 
pen-and-paper proof, it provides a computerized object that can be examined 
by others and serve as a 
foundation for analyses of variants and extensions to, and other properties 
of \ac{AODV}.
We believe that our mechanization of the process algebra \ac{AWN}, and the 
general framework for compositionally proving safety properties is also 
applicable to the study of other protocols.

A number of interesting questions remain.
\begin{inparaenum}
\item
What is the best way to manage variant models in a proof assistant?
\item
How suitable is such a model for showing refinements to more detailed 
implementation models?
\item
Can we validate our model against a real \ac{AODV} implementation as 
has been done for the \acl{TCP}~\cite{BishopEtAl:TCPinHOL:2006}?
\item
We have not modelled timing details, which is not just a question of 
modelling, but also one of which invariants are needed to show loop freedom 
when routes can be spontaneously deleted.
Ideally, timing details could also be incorporated into refinement proofs.
\end{inparaenum}

%--   }-}1%%%%%%%%%%%%%%%%%%%%%%%%%%%%%%%%%%%%%%%%%%%%%%%%%%%%%%%%%%%%
\myparagraph{Acknowledgements.} %{-{1
The authors thank G.\ Klein and M.\ Pouzet for their support and 
complaisance, M.\ Daum for his participation in early work, and L.\ Mandel 
and anonymous reviewers for their comments on a previous 
version.

NICTA is funded by the Australian Government through the Department of 
Communications and the Australian Research Council through the ICT Centre of 
Excellence Program. 
\label{lastpage}
\newpage

\bibliographystyle{splncs03}
\bibliography{paper}

%--   }-}1%%%%%%%%%%%%%%%%%%%%%%%%%%%%%%%%%%%%%%%%%%%%%%%%%%%%%%%%%%%%
\end{document}